\documentclass[iop,referee]{emulateapj-rtx4}

\usepackage{graphicx}  
\usepackage{dcolumn}   
\usepackage{bm}        
\usepackage{amsfonts,amsmath,amssymb,mathrsfs}
\usepackage{color}
\usepackage{hyperref}
\hypersetup{
  colorlinks=true,        
  linkcolor=blue,         
  citecolor=cyan,         
}

\usepackage{natbib,times}
\renewcommand{\apjl}{{ApJL}}

\newcommand{\rmn}[1]{{\mathrm{#1}}}

\shorttitle{Electromagnetic emission from long-lived BNS
  merger remnants II}

\shortauthors{D. M. Siegel \& R. Ciolfi}

\begin{document}
\title{Electromagnetic emission from long-lived binary neutron star
  merger remnants II: lightcurves and spectra}

\author{Daniel M. Siegel\altaffilmark{1} and Riccardo Ciolfi\altaffilmark{2,3}}

\altaffiltext{1}{Max Planck Institute for Gravitational Physics (Albert
  Einstein Institute), Am M\"uhlenberg 1, D-14476 Potsdam-Golm, Germany}

\altaffiltext{2}{Physics Department, University of Trento, Via
  Sommarive 14, I-38123 Trento, Italy}

\altaffiltext{3}{INFN-TIFPA, Trento Institute for Fundamental Physics
  and Applications, Via Sommarive 14, I-38123 Trento, Italy}

\email{daniel.siegel@aei.mpg.de; riccardo.ciolfi@unitn.it}

\begin{abstract}
Recent observations indicate that in a large fraction of binary
neutron star (BNS) mergers a long-lived neutron star (NS) may be
formed rather than a black hole. Unambiguous electromagnetic (EM) signatures of
such a scenario would strongly impact our knowledge on
how short gamma-ray bursts (SGRBs) and their afterglow radiation are
generated. Furthermore, such EM signals would
have profound implications for multimessenger astronomy with joint
EM and gravitational-wave (GW) observations of BNS mergers, which will
soon become reality
with the ground-based advanced LIGO/Virgo GW detector network starting
its first science run this year. 
Here we explore such EM signatures based on the model  
presented in a companion paper, which provides a self-consistent evolution of
the post-merger system and its EM emission starting from an early baryonic wind
phase and resulting in a final pulsar wind nebula that is confined by the
previously ejected material. Lightcurves
and spectra are computed for a wide range of post-merger physical properties
and particular attention is paid to the emission in the X-ray band.
In the context of SGRB afterglow modeling, we present X-ray lightcurves corresponding to the
`standard' and the recently proposed
`time-reversal' scenario (SGRB prompt emission produced at the time of
merger or at the time of collapse of the long-lived NS). 
The resulting afterglow lightcurve morphologies include, in particular, single and two-plateau
features with timescales and luminosities that are in good agreement
with the observations by the \textit{Swift} satellite. Furthermore,
we compute the X-ray signal that should precede the SGRB in the time-reversal
scenario. If found, such a signal would represent smoking-gun
evidence for this scenario. Finally, we find a bright, highly isotropic EM transient signal
peaking in the X-ray band at $\sim\!10^2-10^4\,\text{s}$ after the BNS
merger with luminosities of
$L_X\sim\!10^{46}-10^{48}\,\text{erg}\,\text{s}^{-1}$. If confirmed,
this signal would represent a very promising EM counterpart to the GW
emission of the inspiral and merger of BNSs.
\end{abstract}
		

\section{Introduction}
\label{sec:introduction}

Merging binary neutron stars (BNSs) and neutron star--black
hole (NS--BH) binaries are considered the leading scenario to explain
the phenomenology of short gamma-ray bursts
(SGRBs; e.g.,
\citealt{Paczynski1986,Eichler1989,Narayan1992,Barthelmy2005a,Fox2005,Gehrels2005,Shibata2006b,Rezzolla2011,Paschalidis2015,Tanvir2013,Berger2013,Yang2015}). The
standard paradigm for the generation of the GRB is an accretion
powered jet from a remnant BH--torus system that is formed
$\lesssim\!10-100\,\text{ms}$ after merger
(e.g., \citealt{Narayan1992,Janka1999,Rezzolla2011,Paschalidis2015}). According
to this paradigm, energy
release should cease once the torus has been accreted on a timescale of
$\lesssim\!1\,\text{s}$, which is
consistent with the duration of the SGRB prompt emission of less than $\approx\
\!2\,\text{s}$.

However, observations by the \textit{Swift} satellite
\citep{Gehrels2004} have recently revealed long-lasting
($\sim\!10^2-10^5\,\text{s}$), X-ray afterglows in a large fraction of
SGRB events that are indicative of continuous
energy release from a central engine on timescales orders of magnitude
larger than the typical torus accretion timescale (up to
$\sim\!10^4\,\text{s}$; e.g.,
\citealt{Rowlinson2010,Rowlinson2013,Gompertz2013,Gompertz2014,Lue2015}). These
afterglows are difficult to explain by accretion of the
torus or by prolonged afterglow radiation generated by the interaction
of the jet with the ambient medium (\citealt{Kumar2015} and referecnes
therein).

If a large fraction of BNS mergers result in the formation of a
stable or long-lived NS rather than a BH, extraction of rotational
energy via magnetic spin-down from such an object (typically a millisecond magnetar) could
power the observed long-lasting
afterglows (e.g.,
\citealt{Zhang2001,Metzger2008a,Bucciantini2012,Rowlinson2013,Gompertz2013,Lue2015}). If
true, this challenges the NS--BH progenitor scenario in a large
fraction of SGRB events. The
formation of a long-lived NS in a BNS merger is indeed very
likely. Recent observations of massive NSs with a mass of
$\simeq\!2\,\text{M}_\odot$ together with population synthesis models
indicate that the vast majority of BNS mergers should result in the
formation of a long-lived NS
(\citealt{Demorest2010,Antoniadis2013,Lasota1996,Belczynski2008a}; see
Section~1 of \citealt{Siegel2015b} for a more detailed discussion). However, such
magnetar models cannot readily explain how the prompt SGRB emission
should be generated. Material dynamically ejected during the merger
process as well as matter ejected subsequently in neutrino and magnetically driven
winds from the remnant NS and an accretion disk strongly pollute the
merger site with baryons (e.g.,
\citealt{Hotokezaka2013a,Oechslin2007,Bauswein2013,Kastaun2015a,Dessart2009,Siegel2014a,Metzger2014c}), which severely threatens
the generation of a relativistic outflow. Even if formed by a NS--torus
system shortly after merger, such a jet can be choked by the
surrounding material \citep{Nagakura2014,Murguia-Berthier2014}. We
note that numerical simulations of the merger process have so far not
shown any evidence for jet formation in this case \citep{Giacomazzo2013}.  

The recently proposed `time-reversal' scenario
(\citealt{Ciolfi2015a,Ciolfi2015b}; see \citealt{Rezzolla2015} for an
alternative proposal) offers a possible solution to this problem. In this scenario, the
SGRB is associated with the time of collapse of the long-lived NS,
which occurs on
the spin-down timescale $\sim\!10^2-10^4\,\text{s}$ after merger. The
long-lasting X-ray afterglow
radiation is produced by spin-down energy extracted from the NS prior
to collapse, slowly diffusing outward through the optically thick
environment composed of a pulsar wind nebula (PWN) and an outer shell
of previously ejected material (see Fig.~1 of \citealt{Siegel2015b}
and Fig.~1 of \citealt{Ciolfi2015a}). 
The problem of baryon pollution is avoided here as the NS is
surrounded by a baryon-free PWN at the time of collapse. However,
\citet{Margalit2015} recently questioned the formation of a massive torus
around the BH after the collapse of the long-lived NS and thus the formation of a jet in
this case.

While the generation of the prompt SGRB emission still remains a
matter of debate, in the present paper we focus on predicting X-ray afterglow
lightcurves and spectra for both the `standard' magnetar scenario and the
time-reversal scenario (i.e., assuming that the prompt burst occurs at the time 
of merger or at the time of collapse of the newly-formed NS) employing
a detailed dynamical evolution model
presented in a companion paper (\citealt{Siegel2015b}, henceforth
Paper I). This model should be applicable to any BNS merger leading to
the formation of a long-lived NS. It predicts the evolution of the
post-merger system and its electromagnetic (EM) emission in a
self-consistent way, given some initial data that can be extracted
from a numerical relativity simulation tens of milliseconds after
merger. It bridges the gap between the short timescales accessible to
numerical simulations of the merger process and the timescales of
interest for SGRB afterglow radiation. Our model evolves the
post-merger system through three main evolutionary phases (see Paper I
for details). During an early baryonic wind phase the
surrounding of the newly-formed NS is polluted with matter while differential
rotation is being removed (Phase I).  Once mass loss is suppressed, a pulsar
atmosphere is set up that drives a strong shock through the envelope
of previously ejected matter and sweeps up all the material into a
thin shell (Phase II). Finally, the system is composed of a radially
expanding ejecta shell that confines a PWN centered
around the NS (Phase III). The NS can collapse to a BH at any time
during Phase I-III, which gives rise to very different EM emission
scenarios. In the present paper, we apply this
model to a wide range of physically-motivated post-merger systems. We investigate
the possible morphologies of X-ray lightcurves for
both the standard and the time-reversal scenario, which can be
compared to observations of the
\textit{Swift} satellite. In particular, we also compute detailed
predictions for the X-ray radiation expected to precede the prompt
SGRB emission in the time-reversal scenario. The latter predictions
can be used to search for this X-ray signal. The presence or absence of
such radiation would have strong implications on how and when SGRBs
are produced in BNS mergers.

EM emission from BNS mergers is
also of prime importance for multimessenger astronomy.
BNS mergers are the most promising source of
gravitational waves (GW) for detection with interferometric detectors
such as advanced LIGO and Virgo
\citep{Abadie2010,Harry2010,Accadia2011}. With those detectors
starting their first science runs later this year, joint EM and GW
observations will become a routine undertaking in
the very near future \citep{Singer2014}. Such multimessenger astronomy can greatly
enhance the scientific output of GW and EM observations (see Section~1
of Paper I for a more detailed discussion), although the actual
benefit depends on the knowledge about the EM signals to be expected
in association with BNS mergers. While the prompt SGRB emission will
be observable only in a very small number of events (see
Section~\ref{sec:discussion}), (i) more isotropic EM
counterparts need to be identified that are (ii) bright and (iii) long-lasting, that
are produced in a (iv) high fraction of events, and that can (v) distinguish
between a BNS and a NS--BH merger (see also Section~1 of Paper I). We
note that a kilonova (or macronova) might be bright enough in some cases
\citep{Tanvir2013,Berger2013,Yang2015}---a near-infrared/optical
transient powered by the radioactive decay of heavy nuclei synthesized
in the material dynamically ejected during a BNS or NS--BH merger
\citep{Li1998,Kulkarni2005,Rosswog2005,Metzger2010,Barnes2013,Piran2013,Tanaka2013}. It
also fulfills the remaining criteria but the last one. While in
principle it might be possible to distinguish between BNS and NS--BH
mergers with a kilonova observation \citep{Tanaka2014}, it is
difficult in practice \citep{Yang2015}. Employing our model, we find
here an EM counterpart signal of a BNS merger that fulfills all
criteria (i)--(v). If observed, such an EM signature would represent
compelling evidence for a BNS merger and for the formation of a long-lived
NS, which, in turn, would place strong constraints on the unknown
equation of state of nuclear matter at high densities. Moreover, such
an EM counterpart can reveal important information about the evolution
of the post-merger system and the associated physical processes not
accessible to GW observations.

This paper is organized as follows. In Section~\ref{sec:observables},
we define the X-ray lightcurves and spectra the following discussion
is based on. Section~\ref{sec:params_setup} briefly describes the
parameters of our model and assigns associated ranges that define the
parameter space
for post-merger configurations to be explored in the following
sections. Further assumptions are also discussed. In
Section~\ref{sec:fiducial_model}, we present in detail the results of
a typical run with fiducial parameter values and comment on the
underlying physical processes. Section~\ref{sec:param_study} illustrates the
influence and importance of individual parameters, while
Section~\ref{sec:morphologies} explores the entire parameter space to
provide a more comprehensive overview of the
different morphologies of possible X-ray lightcurves, which are
compared to observations. Finally, Section~\ref{sec:discussion} is reserved for a
discussion of our numerical results in the context of X-ray afterglows
of SGRBs and EM counterparts to the GW signal of the inspiral and
merger of BNS systems and it presents our main conclusions. For details on the
evolution model and the underlying phenomenology, we refer to Paper I.

\section{Lightcurves and spectra: definitions}
\label{sec:observables}

The most important prediction of our model to be compared to
observations is the frequency and time-dependent lightcurve
$L_\text{obs}(t,x)$ as seen
by a distant observer (cf.~Equation~(138) of Paper
I)\footnote{Henceforth we drop the primes used in Paper I to
  distinguish between lab-frame and observer quantities. We assume
  that their respective meanings should be clear from the context.}. Here,
$x=h\nu/m_\text{e}c^2$ defines a normalized dimensionless photon
energy, where $\nu$ is the photon frequency, $h$ the Planck constant,
$m_\text{e}$ the electron
mass, and $c$ the speed of light. This lightcurve
characterizes the radiation emerging from the post-merger system and it
is reconstructed from the numerical integration of the model evolution
equations (Equations~(1)--(15) of Paper I) as described in Section~5.7
of Paper I. This reconstruction takes into account the combined
effects of relativistic beaming and the relativistic Doppler and
time-of-flight effects. In order to facilitate a comparison with
observations, we further define several luminosities restricted
to specific wavelength bands, some of which correspond to the spectral ranges of
the instruments aboard the \textit{Swift} satellite \citep{Gehrels2004}:
\begin{eqnarray}
  L_\text{obs,tot}(t) &=& \int_{x_\text{min}}^{x_\text{max}}
  L_\text{obs}(t,x)\,\rmn{d}x, \label{eq:Lobs_tot}\\
  L_\text{obs,XRT}(t) &=& \int_{x_\text{min,XRT}}^{x_\text{max,XRT}}
  L_\text{obs}(t,x)\,\rmn{d}x, \label{eq:Lobs_XRT}\\
  L_\text{obs,BAT}(t) &=& \int_{x_\text{min,BAT}}^{x_\text{max,BAT}}
  L_\text{obs}(t,x)\,\rmn{d}x,\\
  L_\text{obs,UVOT}(t) &=& \int_{x_\text{min,UVOT}}^{x_\text{max,UVOT}}
  L_\text{obs}(t,x)\,\rmn{d}x,\\
  L_\text{obs,high}(t) &=& \int_{x_\text{max,BAT}}^{x_\text{max}}
  L_\text{obs}(t,x)\,\rmn{d}x,\\
  L_\text{obs,low}(t) &=& \int_{x_\text{min}}^{x_\text{min,UVOT}}
  L_\text{obs}(t,x)\,\rmn{d}x. \label{eq:Lobs_low}
\end{eqnarray}
Here, $X=[x_\text{min},x_\text{max}]$ represents the entire spectral
range considered by our model from radio to $\gamma$-ray energies, with typically
$x_\text{min}=10^{-18}$ and $x_\text{max}=\gamma_\text{max}$
(cf.~Section~5.2 of Paper I), where $\gamma_\text{max}$ denotes the maximum
Lorentz factor of the non-thermal particles in the PWN
(cf.~Table~\ref{tab:model_parameters}). Moreover,
$X_\text{XRT}=[x_\text{min,XRT},x_\text{max,XRT}]$,
$X_\text{BAT}=[x_\text{min,BAT},x_\text{max,BAT}]$, and
$X_\text{UVOT}=[x_\text{min,UVOT},x_\text{max,UVOT}]$, with associated
energy/wavelength ranges of $0.3-10\,\text{keV}$, $15-150\,\text{keV}$,
and $170-650\,\text{nm}$, respectively, correspond to the spectral ranges of the
X-Ray Telescope (XRT; \citealt{Burrows2005}), the Burst Alert Telescope
(BAT; \citealt{Barthelmy2005b}), and the
Ultraviolet/Optical Telescope (UVOT; \citealt{Roming2005}) aboard the \textit{Swift}
spacecraft. Furthermore, the energy bands labelled as ``low'' and ``high''
contain the radiation emitted below
and above the sensitivity regimes of \textit{Swift}. The subdivision
into individual bands \eqref{eq:Lobs_tot}--\eqref{eq:Lobs_low} allows
us to make very specific predictions for the EM emission
from long-lived BNS merger remnants that falls into the sensitivity
windows of the leading instruments
for the observation of SGRBs and their afterglows. However,
predictions for any other desired spectral range can easily be generated.

\section{General input parameters \& setup}
\label{sec:params_setup}

\subsection{Parameter values}
\label{sec:params}
According to our model presented in Paper I, predictions for the
EM emission from a BNS merger remnant result from a
self-consistent numerical evolution of a set of coupled differential
equations based on several input parameters, which we
list in Table~\ref{tab:model_parameters}. In this paper, we explore
physically plausible ranges for these parameters (see ``Range'' in
Table~\ref{tab:model_parameters}) and employ a set of specific values
labelled as ``fiducial'' in Table~\ref{tab:model_parameters} as a
reference for comparison.

\begin{table*}[tbh]
\caption{Model input parameters (see Table~2 of Paper I for references
  to explicit definitions of these quantities) with corresponding
  ranges considered here and a fiducial value used for the model run
  described in Section~\ref{sec:fiducial_model}. Most of these parameters can be
  extracted from (or at least estimated/constrained using) numerical
  relativity simulations of BNS mergers.}
\label{tab:model_parameters}
\centering
\begin{tabular}{cccp{0.45\textwidth}}
\hline\hline
 Parameter & Range & Fiducial value & Description \\
\hline
$\dot{M}_\text{in}$ & $10^{-3}-10^{-1}\,\text{M}_\odot\text{s}^{-1}$ & $5\times 10^{-3}\,\text{M}_\odot\text{s}^{-1}$ &  initial mass-loss rate of the NS\\
$t_\text{dr}$ & $0.1-10\,\text{s}$ & $1\,\text{s}$ & timescale for removal of differential rotation from
                the NS\\
$\sigma_M$ & $1 - 2$ & 2 &ratio of $t_\text{dr}$ to the timescale for
                           decrease of the mass-loss rate\\
$v_\text{ej,in}$ & $0.01-0.1\,c$ & $0.01\,c$ & initial expansion speed of the baryonic ejecta
                   material\\
$\bar{B}$ & $10^{14}-10^{17}$\,G  & $10^{16}$\,G &magnetic field strength in the outer layers of the NS
            \\
$\eta_{B_\text{p}}$ & $0.01 - 0.5$ & 0.1 & dipolar magnetic field strength of
                      the pulsar in units of $\bar{B}$\\
$E_\text{rot,NS,in}$ & $(1-5)\times 10^{52}$\,erg & $5\times 10^{52}$\,erg & initial rotational energy of the NS \\
$P_\text{c}$ & $0.5-2$\,ms & 0.5\,ms & initial central spin period of the NS  
\\
       
$R_\text{e}$ & -- & 11\,km & equatorial radius of the NS
               \\
$M_\text{NS,in}$ & -- & $2.4\,\text{M}_\odot$ &initial mass of the NS
                   \\
$I_\text{pul}$ & -- & $2\times 10^{45}\,\text{g}\,\text{cm}^2$ & moment of inertia of the pulsar
                \\
$T_\text{ej,NS,in}$ & $10-30$\,MeV & 20\,MeV & initial temperature of the material
                            ejected from the NS surface\footnotemark[1]
                             \\
$\kappa$ & $0.2-10 \,\text{cm}^2\,\text{g}^{-1}$ & $0.2 \,\text{cm}^2\,\text{g}^{-1}$ & opacity of the ejecta material
          \\
$t_\nu$ & $0.3-3$\,s & 0.3\,s & neutrino-cooling timescale
         \\
$\eta_{B_\text{n}}$ & $10^{-4} - 10^{-2}$ & $10^{-3}$ &  fraction of the total pulsar wind power injected
                      as magnetic energy per unit time into the PWN\\
$\eta_\text{TS}$ & -- & 0.1 & efficiency of converting pulsar wind power into random
              kinetic energy of accelerated particles in the PWN
             \\
$\gamma_\text{max}$ & $10^4-10^{6}$ & $10^{4}$ & maximum Lorentz factor for non-thermal
                        particle injection into the PWN\\
$\Gamma_\text{e}$ & $0.5-2.5$ & 2.5 & power-law index of the
                                      non-thermal spectrum for particle
                                      injection into the PWN\\
$f_\text{coll}$& $0.1-3$ & 1 & (only in the collapse scenario) parameter specifying the time of collapse of the NS
                  in units of the spin-down timescale (collapse during Phase III) or in units of
                               $t_\text{dr}$ (``$f_\text{coll,PI}$'', collapse during Phase I)\\
\hline
\end{tabular}
\footnotetext[1]{We employ this parameter here for convenience instead of
  the initial specific internal energy of the NS material (see Table~2
  of Paper I), assuming an ideal gas equation of state with adiabatic
  index of $\Gamma = 2$.} 
\end{table*}

The majority of influential model input
parameters can be determined, estimated, or constrained
using numerical relativity simulations of the BNS merger and early
post-merger phase. In particular, $\dot{M}_\text{in}$,
$v_\text{ej,in}$, $\bar{B}$, $E_\text{rot,NS,in}$, $P_\text{c}$,
$R_\text{e}$, $M_\text{NS,in}$, and $T_\text{ej,NS,in}$
can be directly read off from a simulation (e.g.,
\citealt{Siegel2014a,Siegel2015a,Dessart2009}). Furthermore, the
neutrino-cooling timescale $t_\nu$ can be estimated from the thermal
energy content of the NS, given the total neutrino luminosity (e.g.,
\citealt{Dessart2009}). The timescale for removal of differential
rotation, $t_\text{dr}$, can be estimated by the Alv\'en timescale,
which is given in terms of the magnetic field strength $\bar{B}$, the radius
$R_\text{e}$, and the mass $M_\text{NS,in}$ of the NS
\citep{Shapiro2000}. In summary, the initial
conditions for our model can essentially be set by an appropriate
final snapshot of a BNS merger simulation.

There are further parameters, such as $\eta_{B_\text{p}}$,
$I_\text{pul}$, $\eta_{B_\text{n}}$, $\eta_\text{TS}$,
$\gamma_\text{max}$, and $\Gamma_\text{e}$, which cannot be constrained from
simulations of the merger and early post-merger process, as they refer
to properties of the pulsar and the PWN in Phase II and
III of the evolution. Fortunately, $\eta_{B_\text{p}}$ can largely be absorbed
into the parameter $\bar{B}$ (cf.~Section~\ref{sec:non_coll_models}), $I_\text{pul}$
can essentially be absorbed in $E_\text{rot,NS,in}$, varying
$\eta_\text{TS}$ results in a trivial change (just renormalizing the
lightcurves), and the numerical results are
not particularly sensitive to the remaining parameters.

Some of the parameters listed in Table~\ref{tab:model_parameters} are
not varied and, instead, set to some representative value. These are parameters
that are either well constrained and do not influence the numerical
evolution significantly, only change the results in a trivial way,
and/or they can essentially be absorbed into other parameters. One of
these parameter values merits further discussion. 

The value for
$\eta_\text{TS}$ corresponds at an order-of-magnitude level to the
efficiency of the Crab nebula in converting pulsar wind power into
particle motion, which is currently estimated to be $\gtrsim 10\%$ \citep{Kennel1984a,Buehler2014,Olmi2015}. For PWNe
formed in BNS
mergers according to our model, this efficiency factor could be
different. However, different values for $\eta_\text{TS}$ result in
different offsets for the lightcurves, i.e., in a
renormalization of the numerical results. Therefore, as long as our
model is not used to fit observational data, this parameter can be kept fixed.

Our model characterizes the ejecta material by a mean opacity
$\kappa$, which we assume to be constant over time, $\kappa=\kappa_\text{es}\approx
0.2\,\text{cm}^2\text{g}^{-1}$. This value
corresponds to Thomson electron scattering, which is the dominant source
of opacity at optical and UV wavelengths in the case of iron-rich
material (resulting from a high electron fraction). The other
important source for opacity of iron-rich material at
  optical/UV wavelengths, bound-bound absorption, is of the same order
of magnitude or less ($\sim 0.1\,\text{cm}^2\text{g}^{-1}$;
\citealt{Pinto2000,Kasen2013}). For neutron-rich
material (low electron fraction), instead, $\kappa$ can become as high
as $10\,\text{cm}^2\text{g}^{-1}$ due to bound-bound opacities
generated by transitions of the valence electrons in lanthanide
elements \citep{Kasen2013}. However, we assume here that even if the
ejecta material was neutron-rich, irradiation from the PWN is strong
enough to ensure a very high degree of ionization of the lanthanide
elements, such
that $\kappa\sim \kappa_\text{es}$. At X-ray wavelengths, which is
what we are most interested in, an important contribution to the
opacity can come from bound-free absorption, depending on the
ionization state of the material. Here we assume
again that the PWN is sufficiently luminous to ensure a very high
degree of ionization of the
ejecta material on the timescales of interest, such that
$\kappa\sim \kappa_\text{es}$. A more detailed
computation of the opacity at X-ray wavelengths based on partial
ionization of the ejecta material, e.g., along the lines of
\citet{Metzger2014a} and \citet{Metzger2014b}, can, in principle, be
included into our
model as well. Nevertheless, we take $\kappa$ as an input parameter of
our model (cf.~Table~\ref{tab:model_parameters}) and explore the
effect of higher values of $\kappa$ in Section~\ref{sec:non_coll_models}.

\subsection{Assumptions}
\label{sec:assumptions}
For the time being, we adopt two further simplifying
assumptions. First, we neglect effects of thermal Comptonization in
the PWN in Phase III, i.e., we set $\dot{n}^\text{T}_\text{C}\equiv 0$ in the
photon balance equation (cf.~Section~4.3.1 of Paper I). This reduces
the procedure of solving the coupled set of integro-differential
equations for the photon and particle spectra (Equations~(78) and (79)
of Paper I) of the PWN to the scheme outlined in Section~5.2 of Paper
I, which severely lowers the computational cost for the overall numerical
evolution of our model.

Second, for the time being, we neglect effects of further acceleration
in Phase III after $t=t_\text{shock,out}$, i.e., we set
$a_\text{ej}\equiv 0$ in Equation~(11) of Paper I. This is because we
find that for typical input parameter values,
the assumption of quasi-stationarity adopted for the
description of the radiative processes in the PWN (see Sections~4.3.1
and 5.4 of Paper I) would otherwise not be well satisfied and we
expect errors of the order of unity in this case (see also
Section~\ref{sec:monitoring_constraints}). We therefore postpone a
discussion of results for non-zero acceleration of the ejecta shell until a
time-dependent framework for the radiative physics of the PWN has been
implemented in future work. Such a time-dependent formalism is also
required to include a non-zero albedo of the ejecta material, which is
why we neglect it here throughout the entire evolution.

 \section{Results: Fiducial run}
\label{sec:fiducial_model}

In this section, we illustrate the numerical evolution of the model
equations (Equations~(1)--(15) of Paper I) using a typical set of
parameter values labelled as ``fiducial'' in
Table~\ref{tab:model_parameters}. This run also serves as a reference
for comparison with other parameter settings in the following
sections. The results for all other cases are analyzed and interpreted
in an analogous way.

\subsection{Lightcurve}
\label{sec:lightcurve}
\begin{figure}[tb]
  \includegraphics[width=0.465 \textwidth]{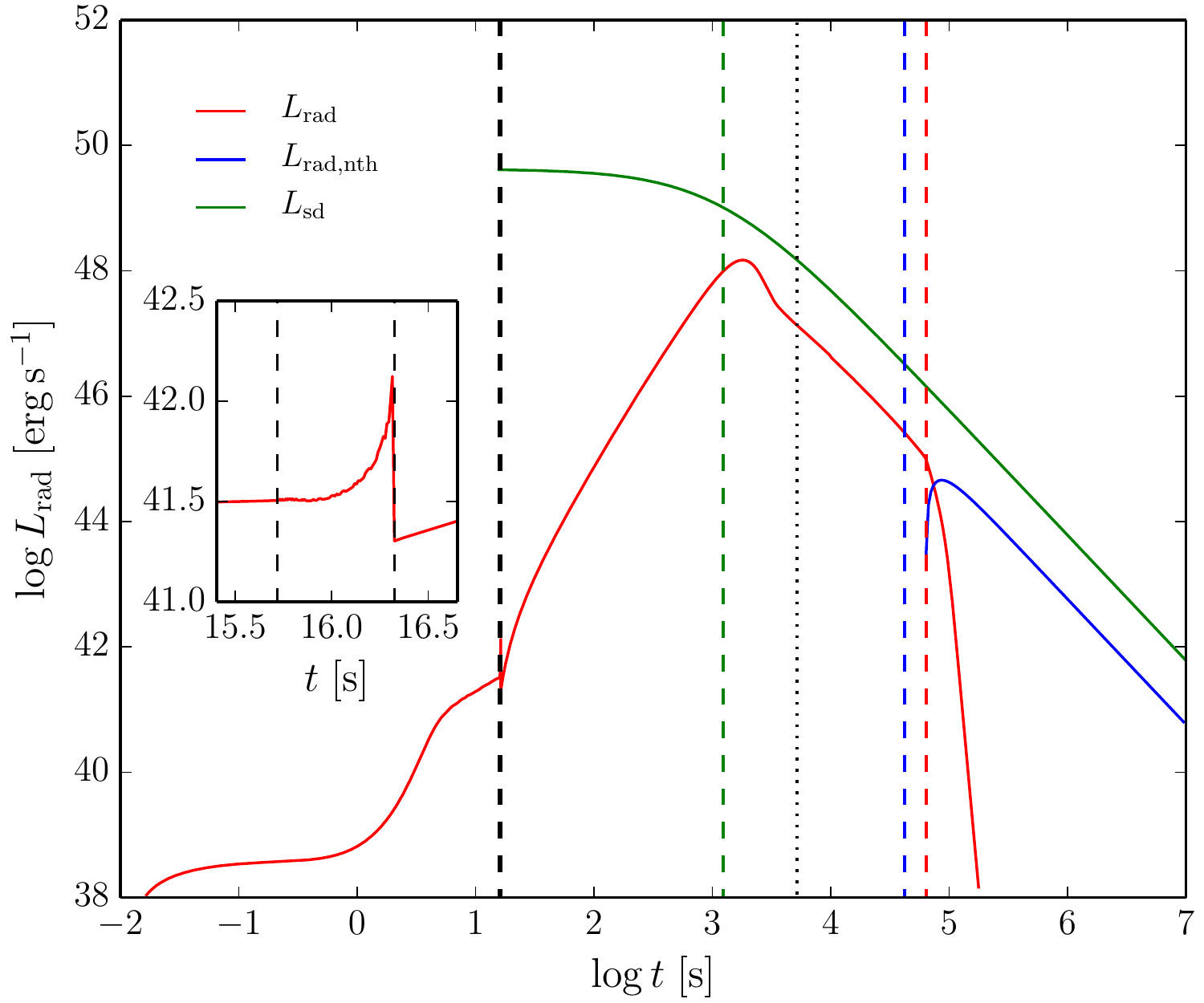}
  \includegraphics[width=0.465 \textwidth]{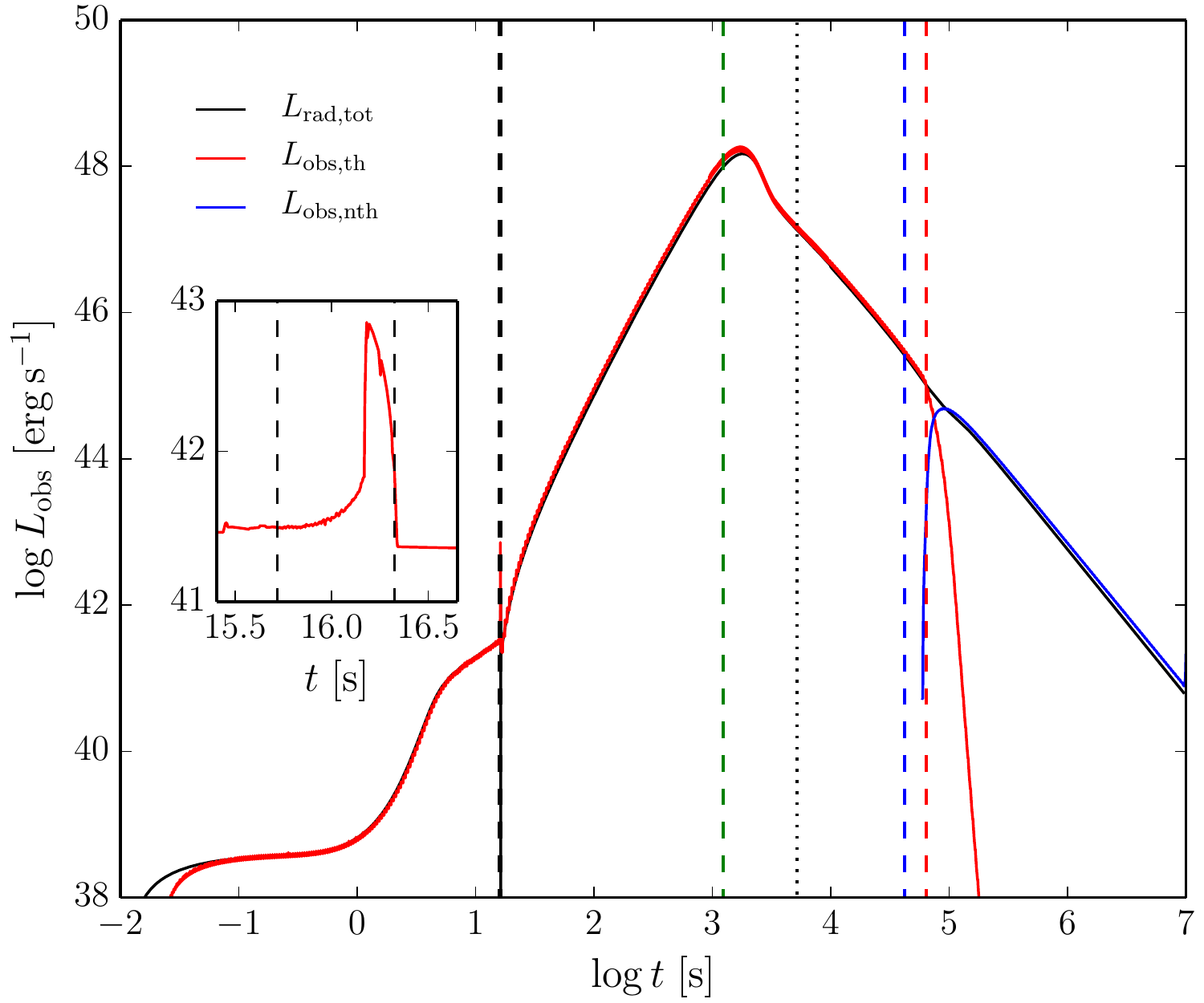}
  \includegraphics[width=0.465 \textwidth]{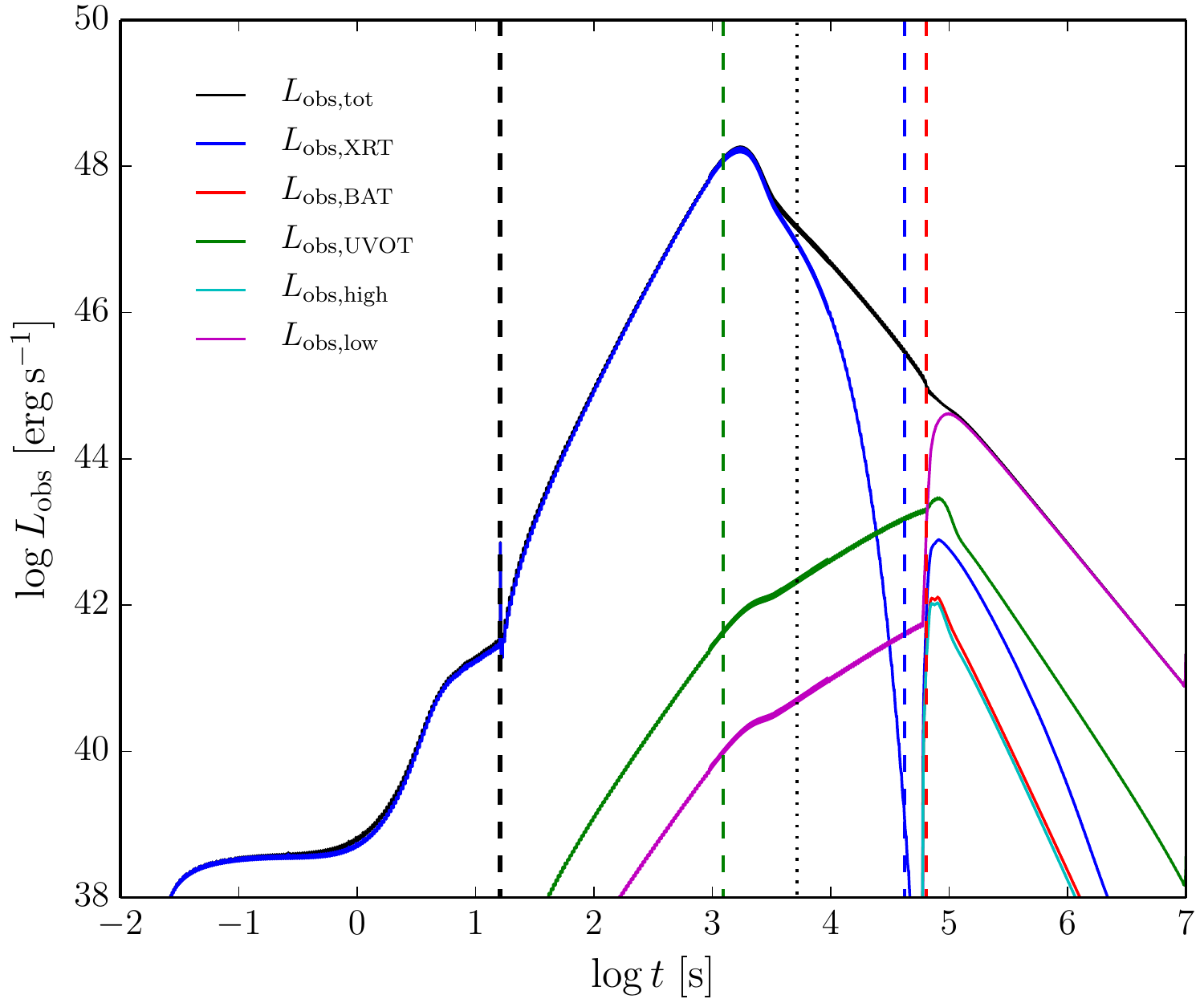}
  \caption{Luminosities throughout Phases I--III for the fiducial
    parameter setup (cf.~Table~\ref{tab:model_parameters}). Phase II
    is indicated by two black
    dashed lines (see also the inset figures), the
    black dotted line refers to $t_\text{res}$. The green, blue, and red dashed lines
    mark the spin-down timescale and the time of transition to the
    optically thin regime of the PWN and of the ejecta shell,
    respectively. Top: Lab frame thermal and non-thermal luminosities,
    together with the spin-down luminosity $L_\text{sd}$. Middle:
    Total thermal and non-thermal observer luminosities. Bottom:
    Observer luminosity in different energy bands (cf.~Section~\ref{sec:observables}).}
\label{fig:L_fiducial}
\end{figure}

Figure~\ref{fig:L_fiducial} shows the luminosity of escaping radiation
from the system throughout the entire evolution. In the following, we
discuss the evolution of the system focusing on the top panel, which reports
the thermal luminosity $L_\text{rad}$ (cf.~Equations~(34), (74), and (100) of Paper
I) and non-thermal luminosity $L_\text{rad,nth}$ (cf.~Equation~(104)
of Paper I) as computed
in the lab frame (rest frame of the NS; cf.~Appendix~B of Paper
I). The evolution of our model can be initialized by data from a numerical
relativity simulation (see Section~\ref{sec:params}). We associate the
time of merger of the BNS with
$t=0$ and start the numerical evolution of our model a few to tens of
milliseconds after merger, typically at
$t=t_\text{min}=10^{-2}\,\text{s}$, once a roughly axisymmetric state
of the NS has been reached. 

\begin{figure}[tb]
  \includegraphics[width=0.48 \textwidth]{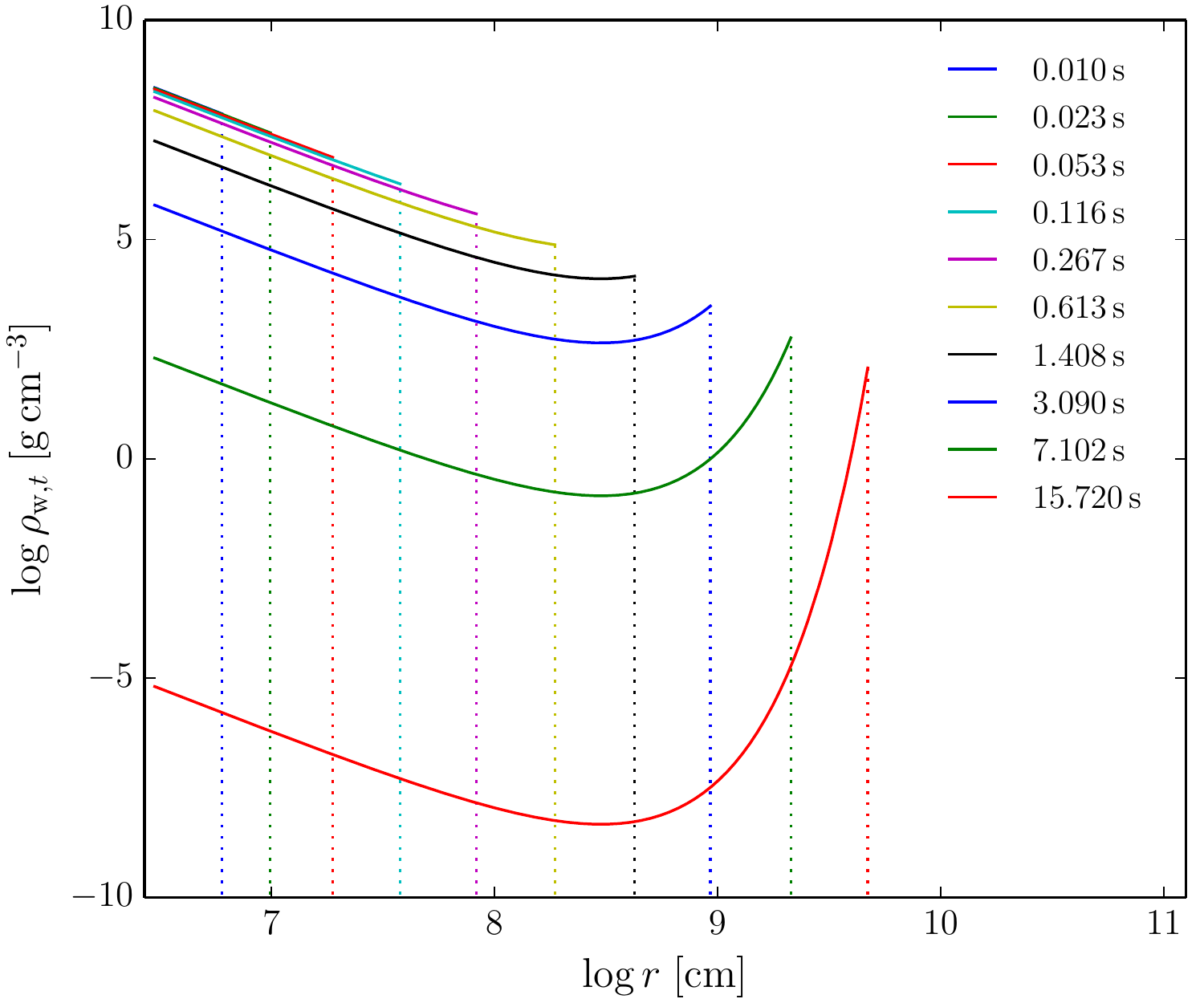}
  \caption{Snapshots of the density profiles of the baryon-loaded wind
    for selected times during Phase I for the fiducial parameter
    setup (cf.~Table~\ref{tab:model_parameters}). The shape of these
    profiles is essentially determined by the parameter $\sigma_M$
    (see Table~\ref{tab:model_parameters} and Section~4.1.1 of Paper I).}
  \label{fig:density_profiles}
\end{figure}

\begin{figure}[tb]
  \includegraphics[width=0.48 \textwidth]{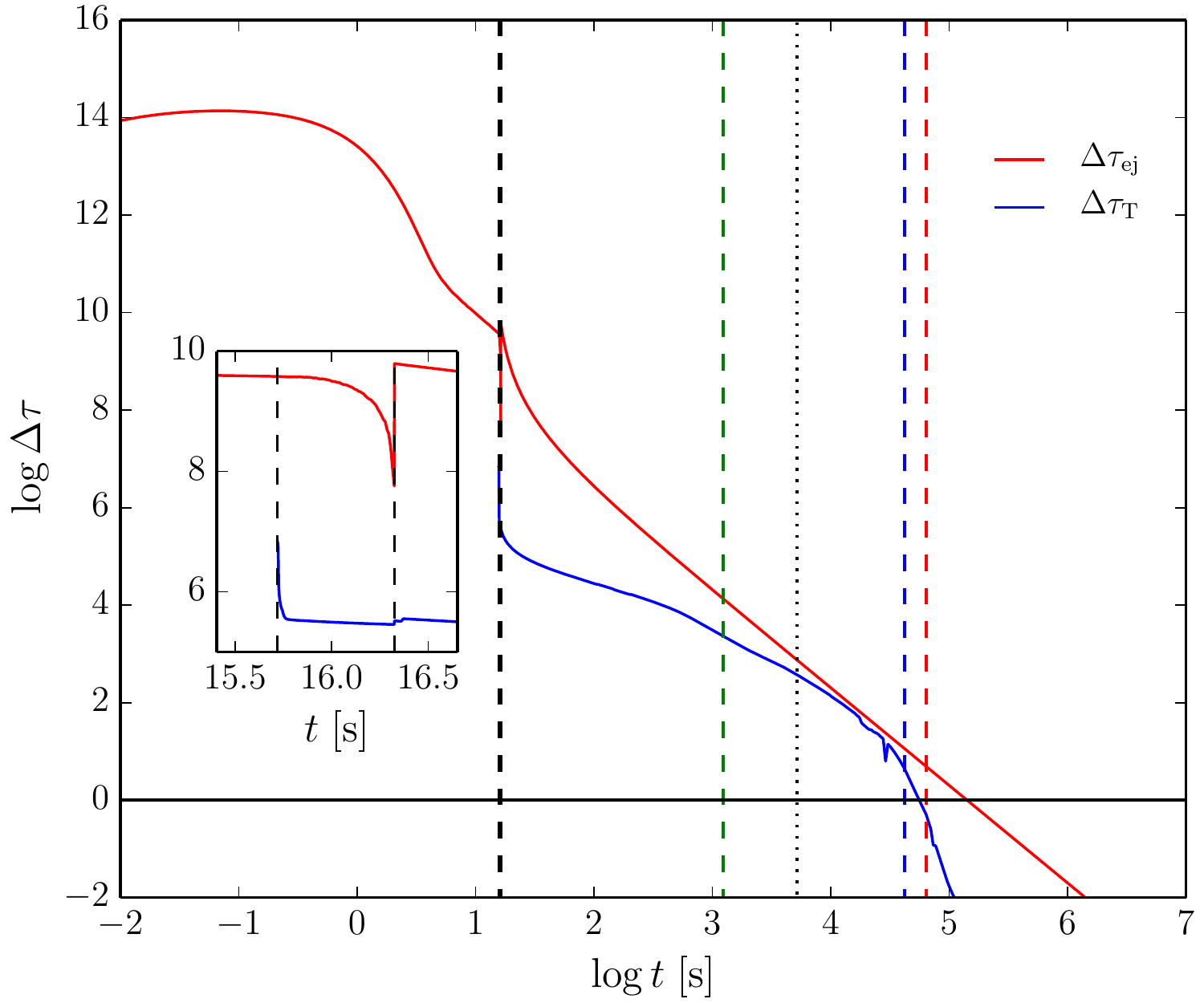}
  \caption{Evolution of the optical depth $\Delta\tau_\text{ej}$ of
    the ejected material (cf.~Equations~(36), (72), and (103) of Paper I)
    and of the optical depth $\Delta\tau_\text{T}$ of the PWN
    (cf.~Equations~(56) and Section~4.3.1 of Paper I) for the fiducial
    parameter setup (cf.~Table~\ref{tab:model_parameters}). Phase II
    is indicated by two black
    dashed lines (see also the inset figure), while the
    black dotted line refers to $t_\text{res}$ (see the text). The
    green, blue, and red dashed lines
    mark the spin-down timescale and the time of transition to the
    optically thin regime of the PWN and of the ejecta shell,
    respectively.}
  \label{fig:delta_tau}
\end{figure}

During Phase I, the NS ejects a significant amount of mass through a
baryon-loaded wind (see Section~4.1.1 of Paper I) that carries thermal
and EM energy (see Section~4.1.2 of Paper I). Density
profiles of this wind for selected times during Phase I are reported
in Figure~\ref{fig:density_profiles}. Radiative
energy loss from this wind is, however, comparatively inefficient due
to the very high optical depth of the ejecta material at these early
times (see Figure~\ref{fig:delta_tau}). The total luminosity remains
at a level of $\lesssim\!10^{39}\,\text{erg}\,\text{s}^{-1}$ and rises appreciably
only toward the end of Phase I when further mass ejection from the NS
is heavily suppressed (see Equation~(17) of Paper I) and the bulk of
the material has already moved
far away from the NS, such that the average density of the
baryon-loaded wind and thus the optical depth strongly decreases. This
steep decrease of the average density toward the end of Phase I is
evident from Figure~\ref{fig:density_profiles}.

\begin{figure}[tb]
  \includegraphics[width=0.48 \textwidth]{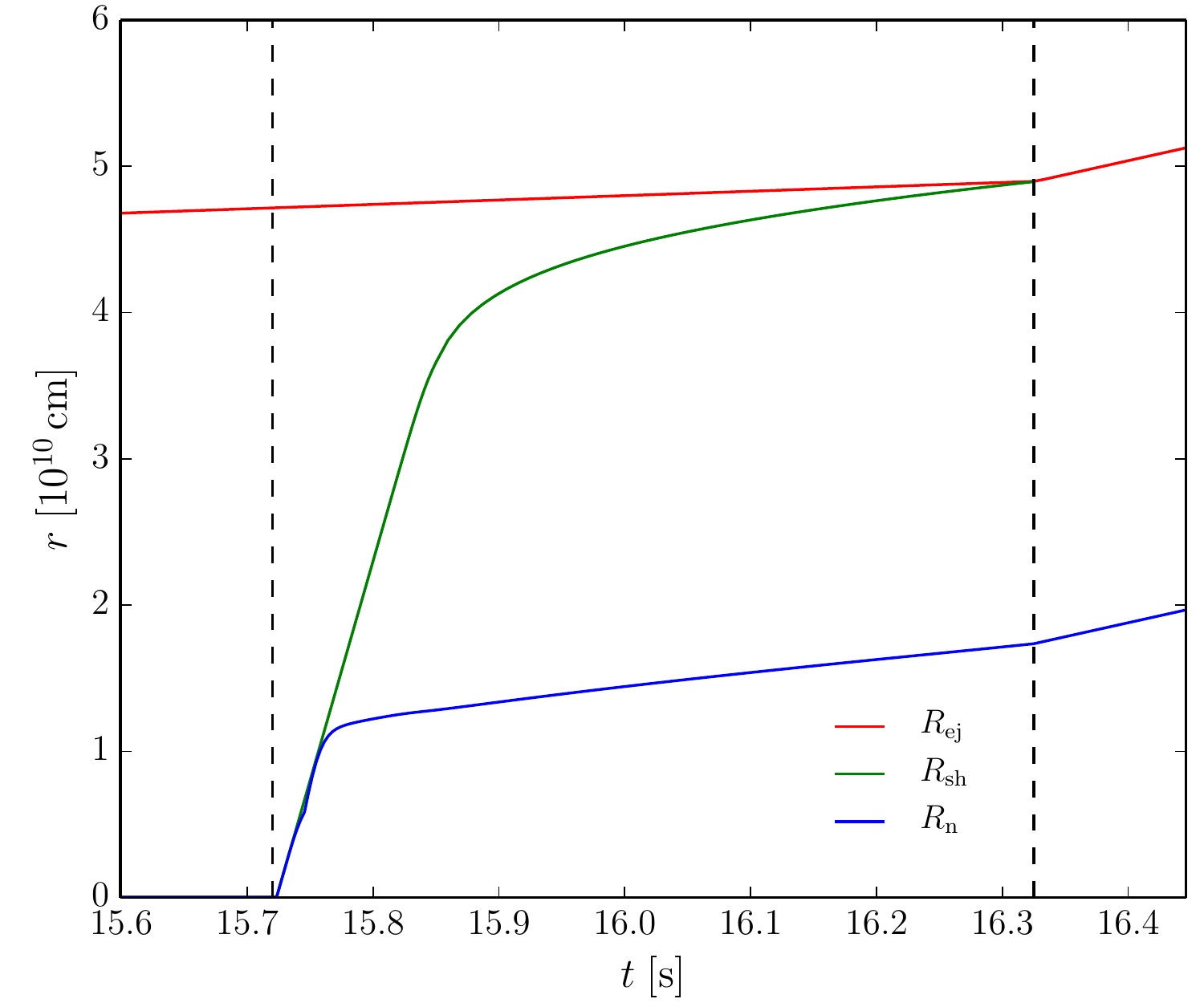}
  \caption{Evolution of three characteristic radii of the system
    during the shock propagation phase (Phase II) for the fiducial
    parameter setup
    (cf.~Table~\ref{tab:model_parameters}). The black dashed lines
    mark the beginning and the end of Phase II. $R_\text{ej}$ and
    $R_\text{n}$ denote the outer radius of the ejecta material and
    of the PWN, respectively. The shock
    front at $R_\text{sh}$ divides the ejecta layer into shocked and
    unshocked material.}
  \label{fig:Rs}
\end{figure}

At $t=t_\text{pul,in}$, once the density in the vicinity of the NS has
become sufficiently small, a pulsar magnetosphere is set up
and Phase II begins (cf.~Section~4.2.1 of Paper I). The resulting
pulsar wind inflates a PWN behind
the less rapidly expanding ejecta. This PWN is highly overpressured
with respect to
the ejecta material and thus drives a strong shock through it,
which sweeps up all the material into a layer of shock-heated ejecta
material. Initially, the shock front (denoted by $R_\text{sh}$ in
Figure~\ref{fig:Rs}) moves essentially with the speed of light and
decelerates to non-relativistic speeds when encountering higher
density material shortly before reaching
the outer ejecta radius $R_\text{ej}$ at $t=t_\text{shock,out}$ (which
terminates Phase II). At any time, this
shock front separates the
ejecta into shocked ($R_\text{n}<r<R_\text{sh}$) and unshocked
($R_\text{sh}<r<R_\text{ej}$) material, where $R_\text{n}$ denotes the
outer radius of the PWN. Phase II can be much shorter than Phase
I depending on parameters. Its duration is indicated by the black
dashed lines in Figures~\ref{fig:L_fiducial}, \ref{fig:delta_tau}, and
\ref{fig:Rs} (see the inset figures, in particular). During this phase
the lab-frame luminosity is ever increasing due to a rapidly
decreasing optical depth, which, in turn, is caused by the fact that
the thickness of the unshocked ejecta layer decreases as the shock
front is moving across the baryon-loaded wind.

When $R_\text{sh}=R_\text{ej}$ (cf.~Figure~\ref{fig:Rs}), Phase III
begins. A higher expansion speed than in Phase I and II (cf.~Section~4.3.2 of
Paper I) induces a rapid decrease of the optical depth
(cf.~Figure~\ref{fig:delta_tau}), which results in a steep rise of the
total luminosity. The luminosity reaches its maximum at
$\sim\!10^{48}\,\text{erg}\,\text{s}^{-1}$ when a
significant fraction of the energy reservoir of the ejecta material
has been consumed. This maximum brightening of the system occurs at
$\sim\!10^3\,\text{s}$, i.e., on the timescale of hours after the BNS
merger. We note that this brightening is a very robust feature of our model and it
occurs on roughly similar timescales for almost any other set of parameters (see
the following sections). If the ejecta shell is further
accelerated after $t=t_\text{shock,out}$ (not considered here for
reasons discussed in Section~\ref{sec:assumptions}), this maximum
brightening is expected to be shifted to slightly earlier times. Preliminary results
indicate that in determining this timescale, acceleration is likely to
dominate other effects such as employing different opacities.

After having radiated away most of its energy, the
ejecta shell enters an asymptotic regime, in which the internal energy
of the ejecta layer is essentially determined by the flux of inflowing
energy from
the PWN. The energy deposited and thermalized in the ejecta shell per
time step is again radiated away during the same time step, thanks to
the ever decreasing photon diffusion timescale of the ejecta shell. Hence,
there is no further energy acquisition anymore. In this
regime, the overall luminosity is determined by the luminosity of
escaping photons from the
PWN, which is again set by the spin-down luminosity
$L_\text{sd}$ of the pulsar as the PWN conserves energy (see
Appendix~D of Paper I and the discussion in Section~5.6 of Paper
I). Therefore, the total luminosity follows the spin-down luminosity (green
line in the top panel of Figure~\ref{fig:L_fiducial}; see also
Section~4.2.1 of Paper I) at a
constant ratio set by the efficiency factor $\eta_\text{TS}$
(cf.~Table~\ref{tab:model_parameters}). Hence, it also shows the
characteristic $\propto t^{-2}$ scaling at late times $t\gg
t_\text{sd}$, where $t_\text{sd}$ is the spin-down timescale
(cf.~Section~4.2.1 of Paper I; indicated
by the green dashed line in Figures~\ref{fig:L_fiducial} and
\ref{fig:delta_tau}).

In this asymptotic regime, the evolution equation for $E_\text{th}$
(Equation~(14) of Paper I) becomes stiff and an alternative scheme is
used to further evolve the set of coupled differential equations of our
model (see the discussion in Section~5.6 of Paper I). The transition
time at which we switch over to this alternative formulation is called
$t_\text{res}$, and it is indicated by a black dotted line in
Figures~\ref{fig:L_fiducial} and \ref{fig:delta_tau}.

As the optical depth approaches unity the ejecta shell
becomes transparent to the radiation from the PWN. From Equations~(99)
and (103) of Paper I, one obtains the scaling
$\Delta\tau_\text{ej}\propto R_\text{ej}^{-2}\propto t^{-2}$, where
the second proportionality assumes a constant expansion speed. This
scaling only holds once the shell thickness $\Delta_\text{ej}$ has
become much smaller than $R_\text{ej}$, which is not yet the case at
$t=t_\text{shock,out}$ for the fiducial run as illustrated by
Figure~\ref{fig:Rs}. From Figure~\ref{fig:delta_tau} it is evident,
however, that this scaling is reached at times $t\gtrsim
10^{2}\,\text{s}$. Once $\Delta\tau_\text{ej}$ has approached unity,
radiation from the PWN is not absorbed anymore by the ejecta material,
such that the thermal emission from the ejecta surface rapidly
decreases and, instead, non-thermal radiation from the PWN is emitted
toward the observer (cf.~Figure~\ref{fig:L_fiducial}). This transition
from predominantly thermal to
non-thermal spectra is discussed in more detail in
Section~\ref{sec:spectra}. The time of transition
between the optically thick and thin regime is indicated by a red
dashed line in Figures~\ref{fig:L_fiducial} and
\ref{fig:delta_tau}. We refer to Section~4.3.3 of Paper I
for a discussion of how this transition is implemented in the model
evolution equations.

As the PWN conserves energy (see Appendix~D of Paper I), the total
non-thermal luminosity $L_\text{rad,nth}$ also scales as $L_\text{sd}$
once the ejecta shell is optically thin. We note that in the fiducial
setup, the PWN becomes optically thin
prior to the ejecta matter (see
Figure~\ref{fig:delta_tau}). Hence, radiation from the deep
interior of the nebula is already being emitted toward the observer
once the PWN radiation is free to escape from the system.

The middle panel of Figure~\ref{fig:L_fiducial} shows the total
thermal and non-thermal luminosity ($L_\text{obs,th}$ and
$L_\text{obs,nth}$) as seen by a distant
observer in comparison with the total lab-frame luminosity
$L_\text{rad}$. These observer lightcurves are reconstructed using the
method presented in Section~5.7 of Paper I and take into account the
combined effects of relativistic beaming, the relativistic Doppler
effect, and the time-of-flight effect for radiation emerging from the
surface and/or the interior of a relativistically expanding sphere. As
the expansion speed of the ejecta material ($v_\text{ej,in}=0.01c$
in Phase I and II, $v_\text{ej}=0.064c$ in Phase III; cf.~Section~4.3.2 of Paper I)
remains non-relativistic for the fiducial setup (as we are neglecting
further acceleration for the
time being, see Section~\ref{sec:assumptions}), these relativistic
effects induce only small corrections
to the observer lightcurve, which closely follows the lab-frame
lightcurve.

The delayed onset of the observer luminosity with respect to
$L_\text{rad}$ at very early times is
due to the time-of-flight effect: radiation from high latitudes (small $\theta$)
of the expanding sphere reaches the observer earlier than radiation from
lower latitudes (high $\theta$, where $\theta=0$ defines the direction of
the observer; see Section~5.7 of Paper I for more details). The offset
between the time
grids of the lab frame and
the remote observer is calibrated in such a way that a photon emitted
from the spherical surface of the expanding ejecta at $\theta = 0$,
$r=R_\text{min}$, with $R_\text{min}$ being the inner spatial boundary
of our model, at a time $t=t_\text{min}$ after the BNS merger
is received by the observer at a time $t' = t_\text{min}$ (see
Section~5.7 of Paper I). This time-of-flight effect is also evident
during Phase II, where the peak observer luminosity appears to precede
the end of Phase II by $R_\text{ej}(t_\text{shock,out})/c\approx 0.16\,\text{s}$ (compare
the inset figures in the two upper panels of Figure~\ref{fig:L_fiducial}). This is
because the main contribution to the observer luminosity originates
from high latitudes, while the delayed radiation from low latitudes
only forms a tail that broadens the peak. Furthermore, we note that
the thermal observer luminosity is slightly higher than the lab-frame
luminosity in Phase III, which is due to the relativistic Doppler
effect. Finally, the time-of-flight effect can be noticed again when
the ejecta shell becomes optically thin. The onset of the non-thermal
radiation as seen by the observer precedes the corresponding onset in
the lab frame (red-dashed line in
Figure~\ref{fig:L_fiducial}) by $R_\text{n}/c$. Moreover, the observer
lightcurve starts to deviate from the lab-frame lightcurve at very
late times (barely visible in Figure~\ref{fig:L_fiducial}). This is
because the PWN is optically thin and radiation from the entire
volume of the nebula can reach the observer. As radiation
is being delayed while the overall luminosity is decreasing, this
results in a less steep slope for the observer lightcurve.

In summary, the relativistic corrections to the lightcurve are very small and
barely visible for the fiducial parameter setup. However, for higher
velocities and/or in presence of further acceleration during Phase III,
these relativistic corrections can significantly reshape the observer
lightcurve, which needs to be taken into account when comparing
the predictions of our model to observational data.

The total energy radiated away form the system is distributed over
many orders of magnitude in frequency. Depending on the energy band of
interest, very different morphologies for the
observer lightcurves are obtained (cf.~the lower panel of
Figure~\ref{fig:L_fiducial}). A remarkable
feature evident from Figure~\ref{fig:L_fiducial} is that the typical
temperature of the ejecta shell during Phase I, II, and up to the time
of maximum brightening in Phase III corresponds to an
energy in the X-ray band, such that almost all of the radiated energy
falls into the sensitivity regime of the XRT instrument. This is a
very robust feature that we also find in case of almost all other
parameter settings: the radiation from the system up to its maximal
brightness is predominantly thermal and of X-ray nature. Soon after
this, the UV and optical emission dominates, until, finally, most of
the energy is radiated away in the radio band once the ejecta shell
becomes transparent to the radiation from the nebula. This radio emission is
generated by synchrotron cooling of the electron-positron pairs in the
PWN (see also Section~\ref{sec:spectra}). Nevertheless, appreciable
X-ray luminosities of the order of $\sim
10^{42}\,\text{erg}\,\text{s}^{-1}$ may still be present at times as
late as $t\sim 10^5\,\text{s}$. For the remainder of this paper, we
focus specifically on X-ray luminosities and spectra, as they
appear to be most relevant in the context of early afterglows of a
BNS merger event.

\subsection{Monitoring the constraints}
\label{sec:monitoring_constraints}
Our approach for modeling the radiative processes in the PWN
(cf.~Section~4.3.1 of Paper I) is based
upon two important assumptions: quasi-stationarity and negligible
influence of synchrotron self-absorption. The validity of these
assumption needs
to be routinely verified during the numerical integration of the model
evolution equations. As discussed in Sections~5.4 and 5.5 of Paper I,
this can be achieved by monitoring several timescales and the optical
depth to synchrotron self-absorption, which we shall discuss here.

\begin{figure}[tb]
  \includegraphics[width=0.48 \textwidth]{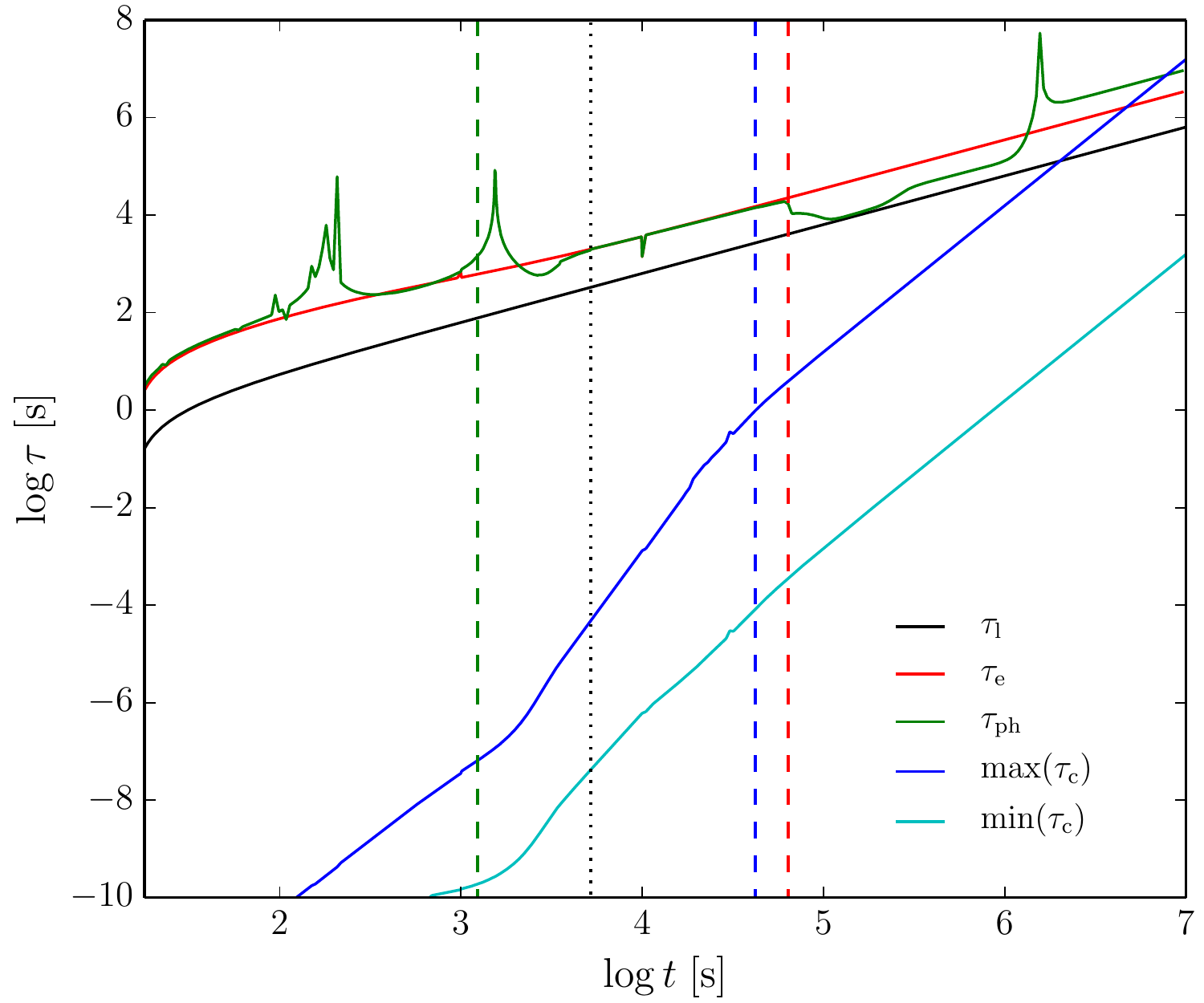}
  \caption{Evolution of several characteristic timescales related to
    the radiative processes in the PWN throughout Phase III for the
    fiducial parameter setup
    (cf.~Table~\ref{tab:model_parameters}). Shown are the light
    crossing time $\tau_\text{l}$
    of the nebula, the timescales for change of
    particle and photon injection, $\tau_\text{e}$ and
    $\tau_\text{ph}$, respectively, and the minimum and maximum value
    over frequency for the total particle cooling timescale
    $\tau_\text{c}$ in the nebula. The green, blue, and red dashed lines
    indicate the spin-down timescale and the time of transition to the
    optically thin regime of the PWN and of the ejecta shell,
    respectively. The black dotted line corresponds to $t=t_\text{res}$ (see
    Section~\ref{sec:lightcurve}).}
  \label{fig:timescales}
\end{figure}

Figure~\ref{fig:timescales} reports the evolution of the light
crossing time $\tau_\text{l}$ of the nebula, the timescales for change
of particle and photon injection into the nebula ($\tau_\text{e}$ and
$\tau_\text{ph}$, respectively), and the total particle cooling
timescale $\tau_\text{c}$ of the nebula throughout Phase III (for
definitions, see Section~5.4 of Paper I). In general we find that
\begin{equation}
  \tau_\text{c}\ll \tau_\text{ph}, \mskip40mu \tau_\text{c}\ll \tau_\text{e}, 
\end{equation}
which shows that the nebula particle distribution can
adjust fast enough to changes of the exterior conditions and thus
justifies the assumption of stationarity concerning
the particle distribution in the nebula (see Section~5.4 of Paper
I). Only at late times $t\gtrsim 10^7\,\text{s}$ these conditions are not
satisfied anymore for part of the particle energy spectrum. However, we
are not interested in the evolution of the system at such late
times. This is a general result that qualitatively holds throughout
the parameter space considered in Table~\ref{tab:model_parameters}: at
times of interest, the stationarity assumption regarding the
particle distribution is very well satisfied. Moreover,
Figure~\ref{fig:timescales} also shows that typically
\begin{equation}
  \tau_\text{l} \ll \tau_\text{ph}, \label{eq:tau_l_tau_ph}
\end{equation}
except for a short transition phase around $t\sim
10^5\,\text{s}$. This is the time when the ejecta shell becomes
optically thin and the timescale for change of the photon spectrum is
mostly determined by
the auxiliary function $f_\text{ej}$ to model this transition
(cf.~Section~4.3.3 of Paper I). In general and across the
parameter space we find, however, that Equation~\eqref{eq:tau_l_tau_ph} is
well satisfied, which justifies the stationarity assumption concerning
the photon distribution inside the PWN (see Section~5.4 of Paper
I). This picture can change when further acceleration of the
ejecta shell during Phase III is considered. We find that in this case
$\tau_\text{ph}$
can become similar to $\tau_\text{l}$ and errors of the order of unity
for the photon and particle spectra are expected. This is why
we refrain from discussing results for this case in the present
paper. To overcome this problem, a time-dependent framework for the
radiative processes in the nebula has to be developed, which we
postpone to future work.

\begin{figure}[!tb]
  \includegraphics[width=0.48 \textwidth]{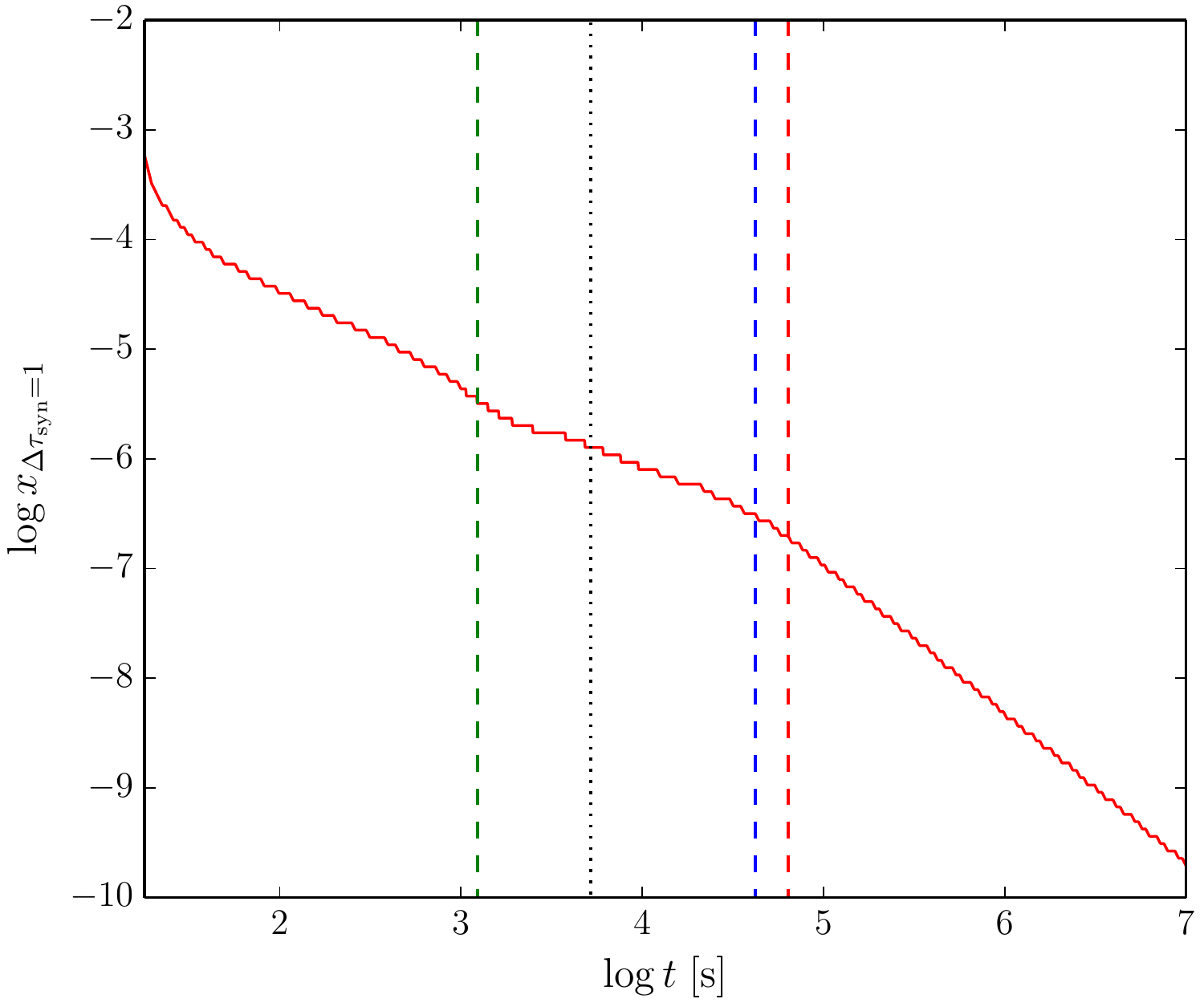}
  \caption{Evolution of the normalized dimensionless photon frequency
    $x_{\Delta\tau_\text{syn}=1}$ (cf.~Section~5.5 of
    Paper I) in Phase III for the fiducial parameter setup
    (cf.~Table~\ref{tab:model_parameters}). This frequency
    separates the optically thick ($x<x_{\Delta\tau_\text{syn}=1}$)
    from the optically thin ($x>x_{\Delta\tau_\text{syn}=1}$) parts of
    the spectrum of the PWN to synchrotron
    self-absorption. The X-ray band is always optically thin to this
    process. The green, blue, and red dashed lines
    indicate the spin-down timescale and the time of transition to the
    optically thin regime of the PWN and of the ejecta shell,
    respectively. The black dotted line corresponds to $t=t_\text{res}$ (see
    Section~\ref{sec:lightcurve}).}
  \label{fig:x_tau_syn_1}
\end{figure}

Figure~\ref{fig:x_tau_syn_1} monitors the validity of the second
central assumption: negligible influence of synchrotron
self-absorption. It
shows the normalized dimensionless frequency which separates the
optically thick and thin regimes of the photon spectrum to
synchrotron self-absorption (for a definition, see Section~5.5 of
Paper I). As we have $x_{\Delta\tau_\text{syn}=1} < 4\times 10^{-4}$
except for the first few time steps in Phase III, the spectrum in the
XRT band (which is what we are mostly interested in) is unlikely to be
affected by effects of synchrotron self-absorption. We find that this
conclusion typically also holds across the entire parameter space
considered in Table~\ref{tab:model_parameters}.

\subsection{Spectra}
\label{sec:spectra}

After having discussed the lightcurves for the fiducial parameter
setup in Section~\ref{sec:lightcurve}, this subsection presents results
for the associated spectral evolution.

\begin{figure}[tb]
  \includegraphics[width=0.48
  \textwidth]{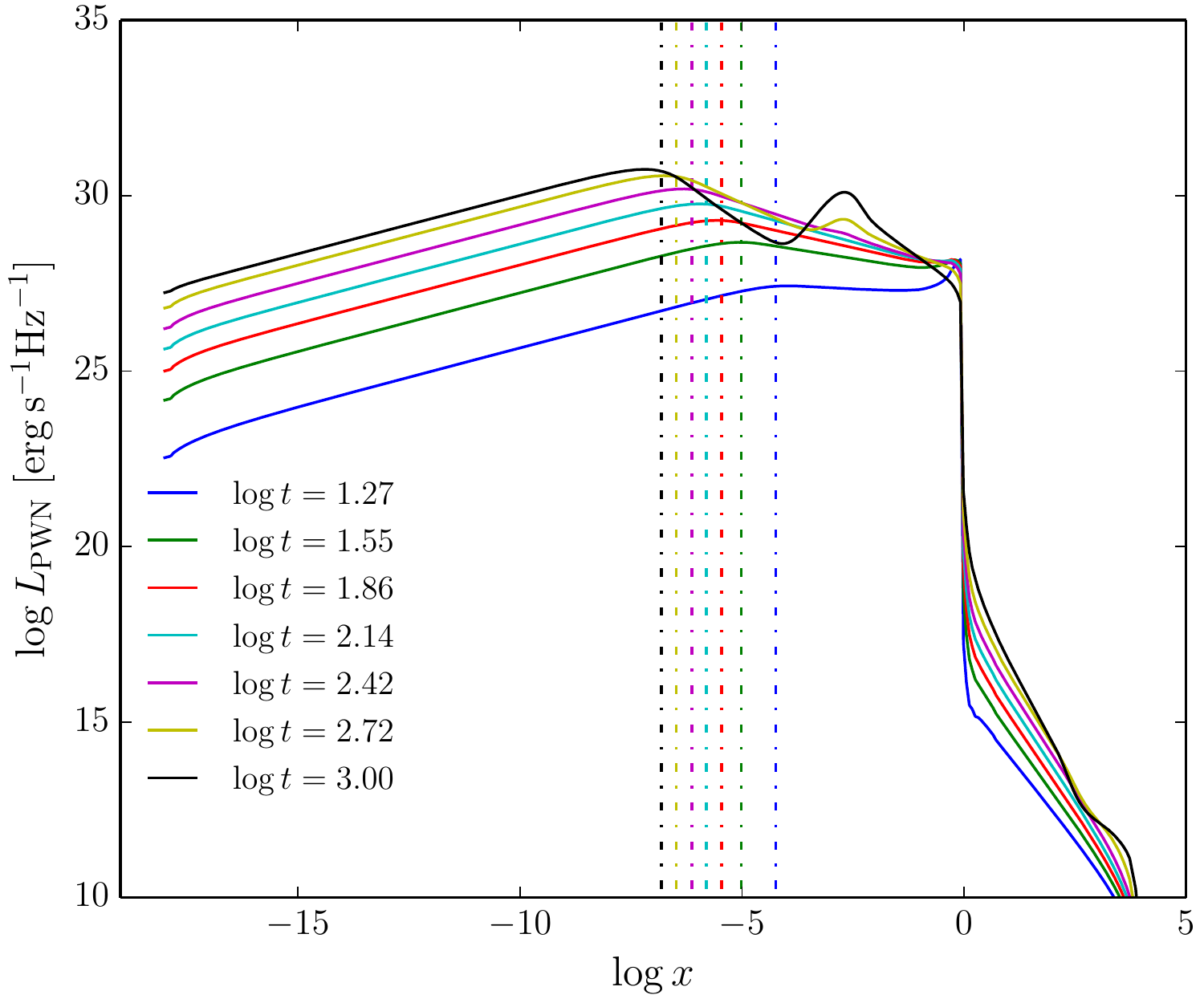}
\includegraphics[width=0.48
\textwidth]{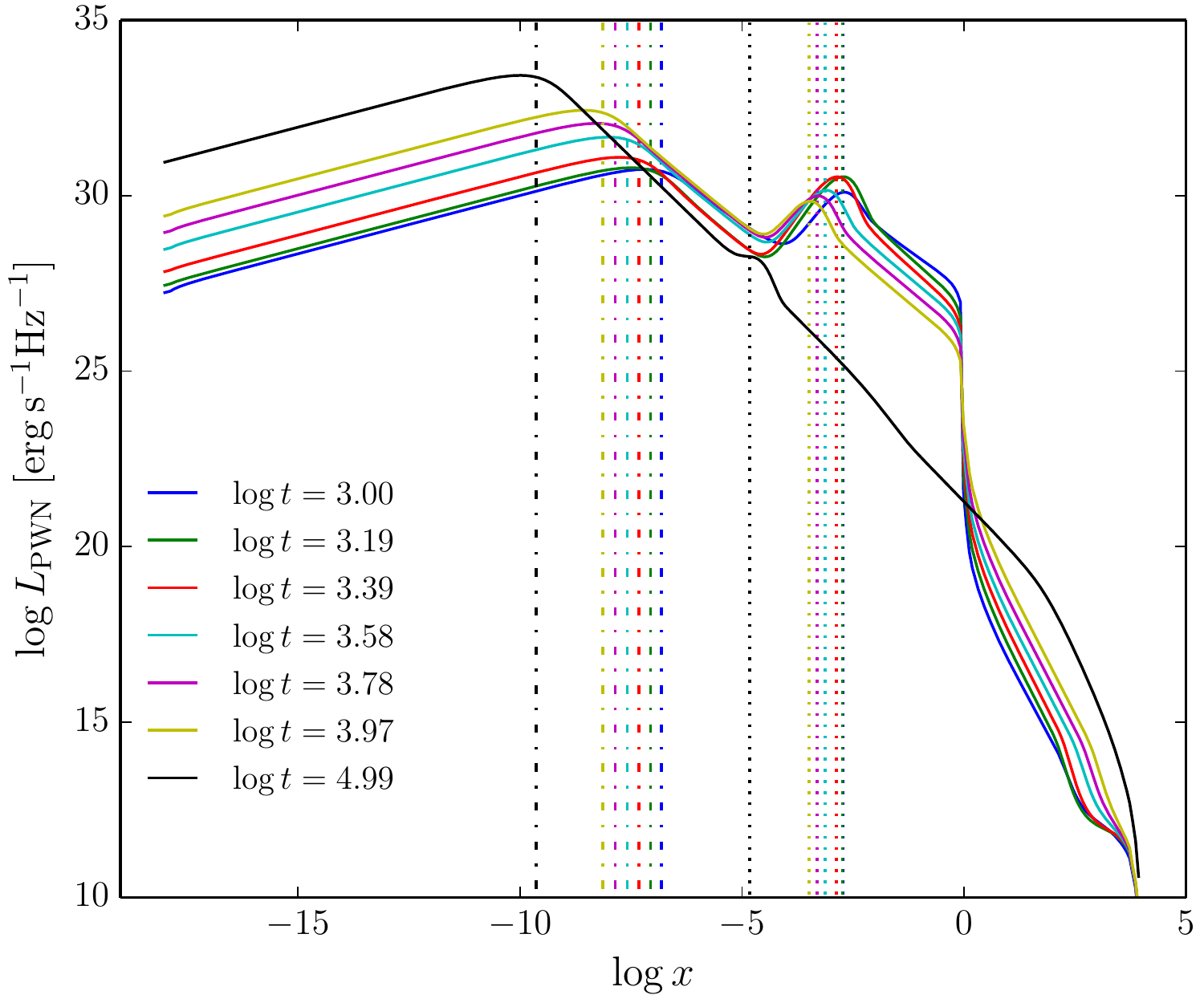}
\includegraphics[width=0.48 \textwidth]{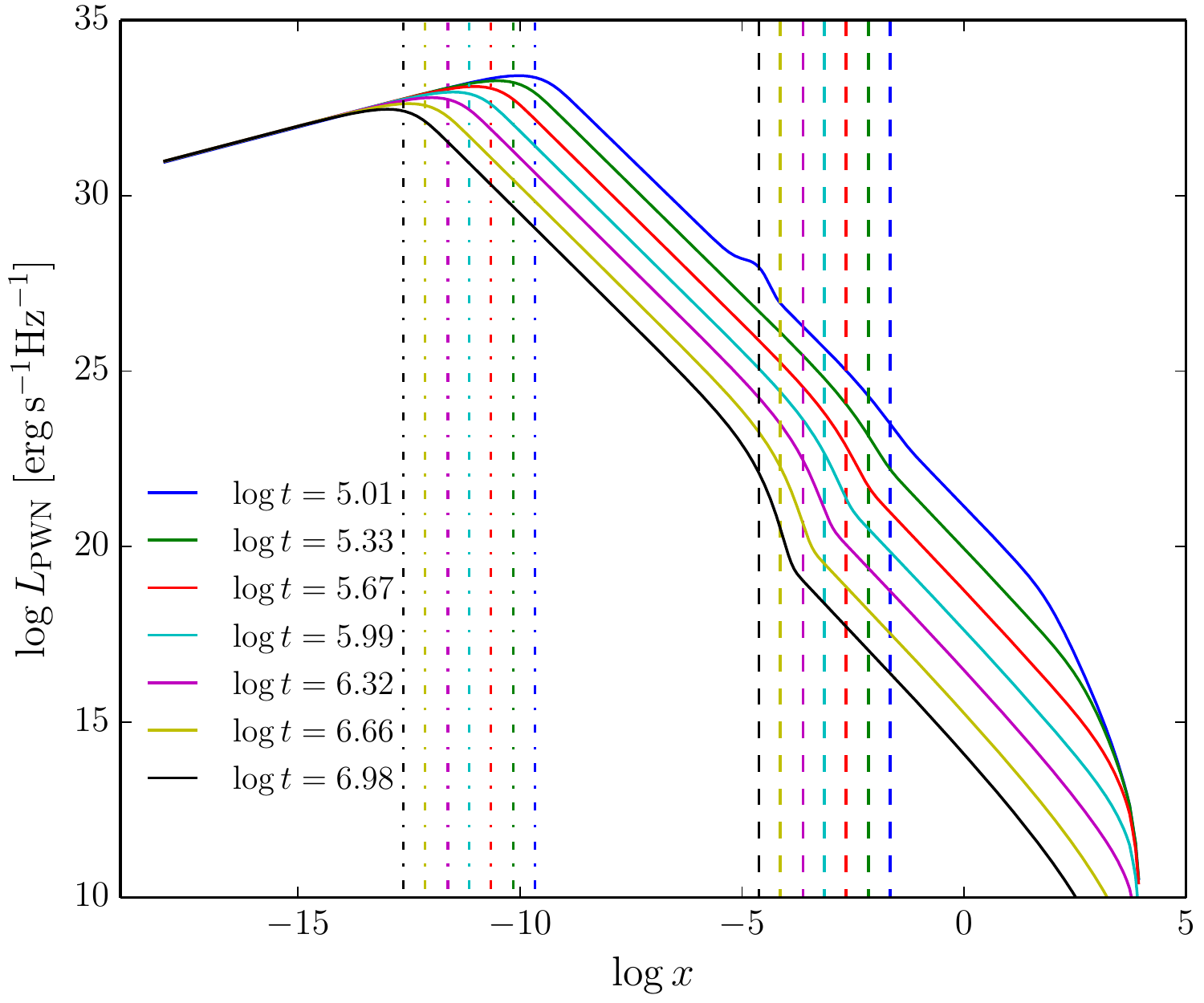}
  \caption{Luminosity per unit frequency of
    escaping radiation from the nebula as a function of
    $x=h\nu/m_\text{e}c^2$ for selected times
    during Phase III (fiducial parameters,
    cf.~Table~\ref{tab:model_parameters}). Dot-dashed and dashed lines
    refer to the critical frequencies $\nu_{\text{c},\gamma_\text{min}}$ and
    $\nu_{\text{c},\gamma_\text{max}}$, respectively (see the
    text). Dotted lines indicate the frequency at maximum of the thermal
    input spectra.}
  \label{fig:Lnth_neb}
\end{figure}

Figure~\ref{fig:Lnth_neb} shows the intrinsic spectrum
$L_\text{PWN}(\nu,t)$ of the
radiation escaping from the nebula for selected times during Phase III
(cf.~Equation~(91) of Paper I). These spectra result from the numerical
solution to the photon and particle balance equations in the PWN
(cf.~Section~4.3.1 of Paper I) and account for the combined effects of
synchrotron losses, (inverse) Compton scattering, pair
production and annihilation, Thomson scattering off thermal particles,
and photon escape.

At low frequencies, the spectra are dominated by synchrotron
radiation. The spectrum peaks at a frequency around
$\nu_{\text{c},\gamma_\text{min}}$ (cf.~the dot-dashed lines in the
three panels), where
\begin{equation}
  \nu_{\text{c},\gamma} = \frac{3}{4\pi}\frac{eB_\text{n}}{m_\text{e} c^2} \gamma^2
\end{equation}
is the critical frequency above which the synchrotron spectrum of a
single radiating particle of Lorentz factor $\gamma$ sharply decreases,
and $\gamma_\text{min}=1$ is the minimum Lorentz factor of the
particle distribution (see also Appendix~C of Paper I). The
synchrotron spectrum extends up to $\nu_{\text{c},
  \gamma_\text{max}}$ (cf., e.g., the dashed vertical lines in the bottom
panel) and shows power-law decline between
$\nu_{\text{c},\gamma_\text{min}}$ and $\nu_{\text{c},
  \gamma_\text{max}}$. In particular, this is evident at late times
(bottom panel), when the spectrum is dominated by synchrotron
radiation below $\nu_{\text{c}, \gamma_\text{max}}$. This behavior
reflects the underlying particle distribution
$N(\gamma)$ (cf.~Section~4.3.1 of Paper I), which is also of power-law
nature at late times.

At UV and higher energies other processes can determine the spectral
shape. One prominent
feature is  typically caused by the strong photon field with luminosity
$L_\text{rad,in}$ of thermal radiation from the inner surface of the
confining ejecta envelope (cf.~Equation~(101) of Paper I). A
corresponding thermal `bump' becomes visible around $\log t \approx 2.7$
(cf.~the upper panel) and persists until the ejecta material becomes
optically thin around $\log t \approx 5$. The maxima of the thermal
injection spectra at energies corresponding to $\approx 2.8\,k_\text{B}
T_\text{eff}$ are indicated by
dotted lines in the middle panel, where
$T_\text{eff}^4=T_\text{eff,com}^4/\zeta\gamma_\text{ej}$ defines the
effective temperature of the ejecta material as measured in the lab
frame (see Equation~(101) of Paper I). It is the full non-linear
interaction of these injected photons with the particle distribution
via (inverse) Compton scattering and pair creation/annihilation that
essentially determines the spectral shape from UV to gamma-ray
energies up to $x_\text{max}=\gamma_\text{max}$.

\begin{figure}[tb]
  \includegraphics[width=0.48
  \textwidth]{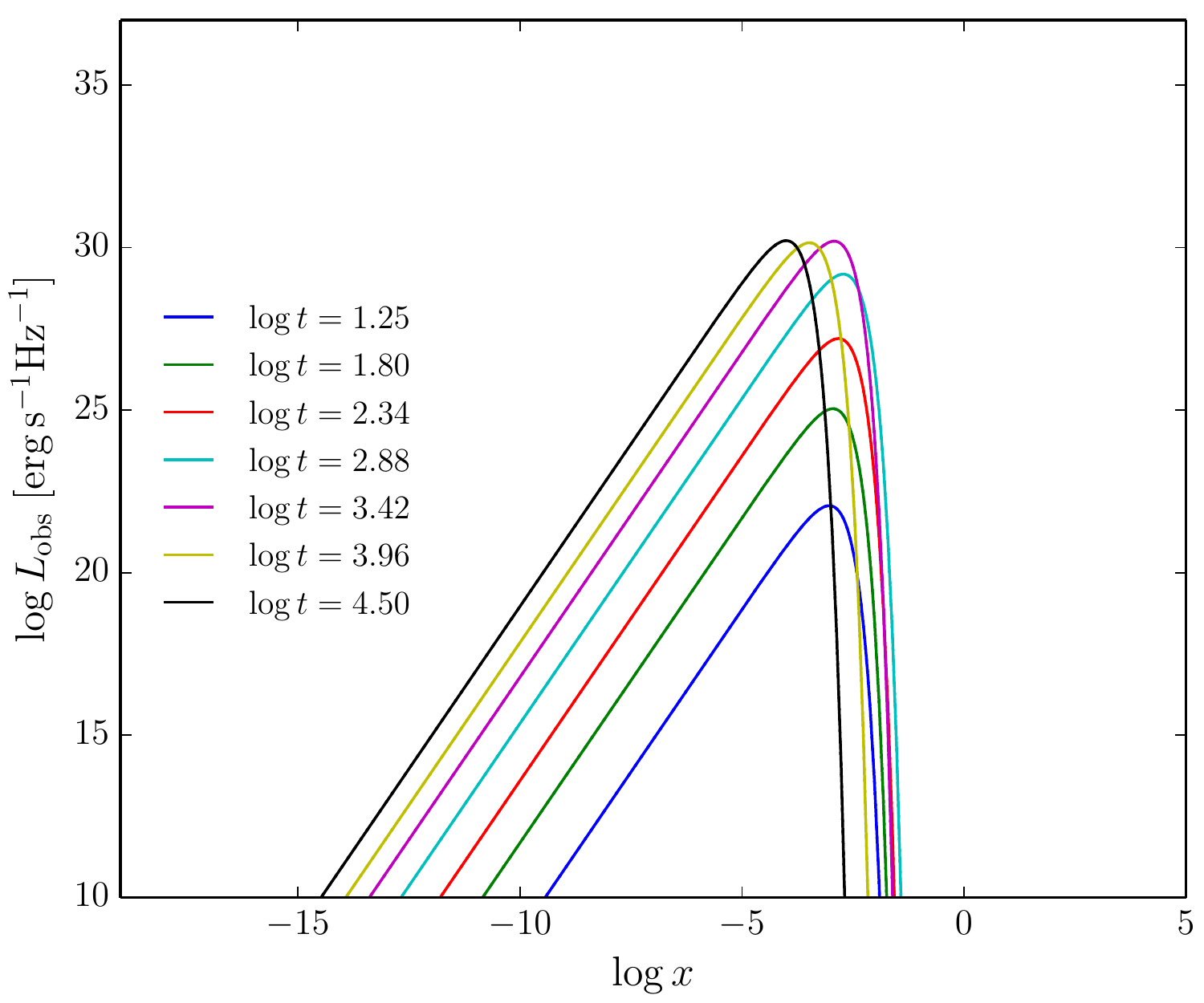}
\includegraphics[width=0.48
\textwidth]{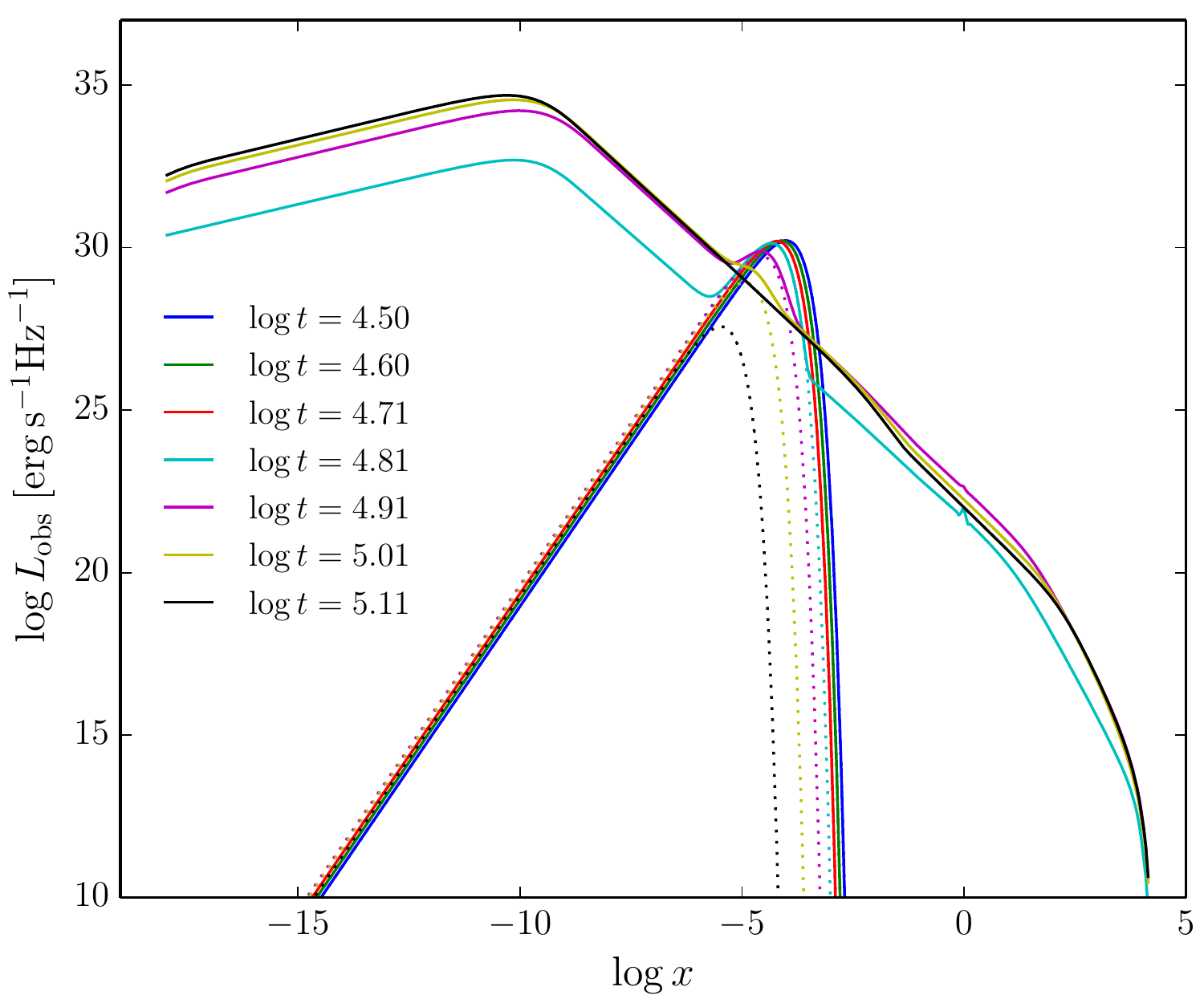}
\includegraphics[width=0.48 \textwidth]{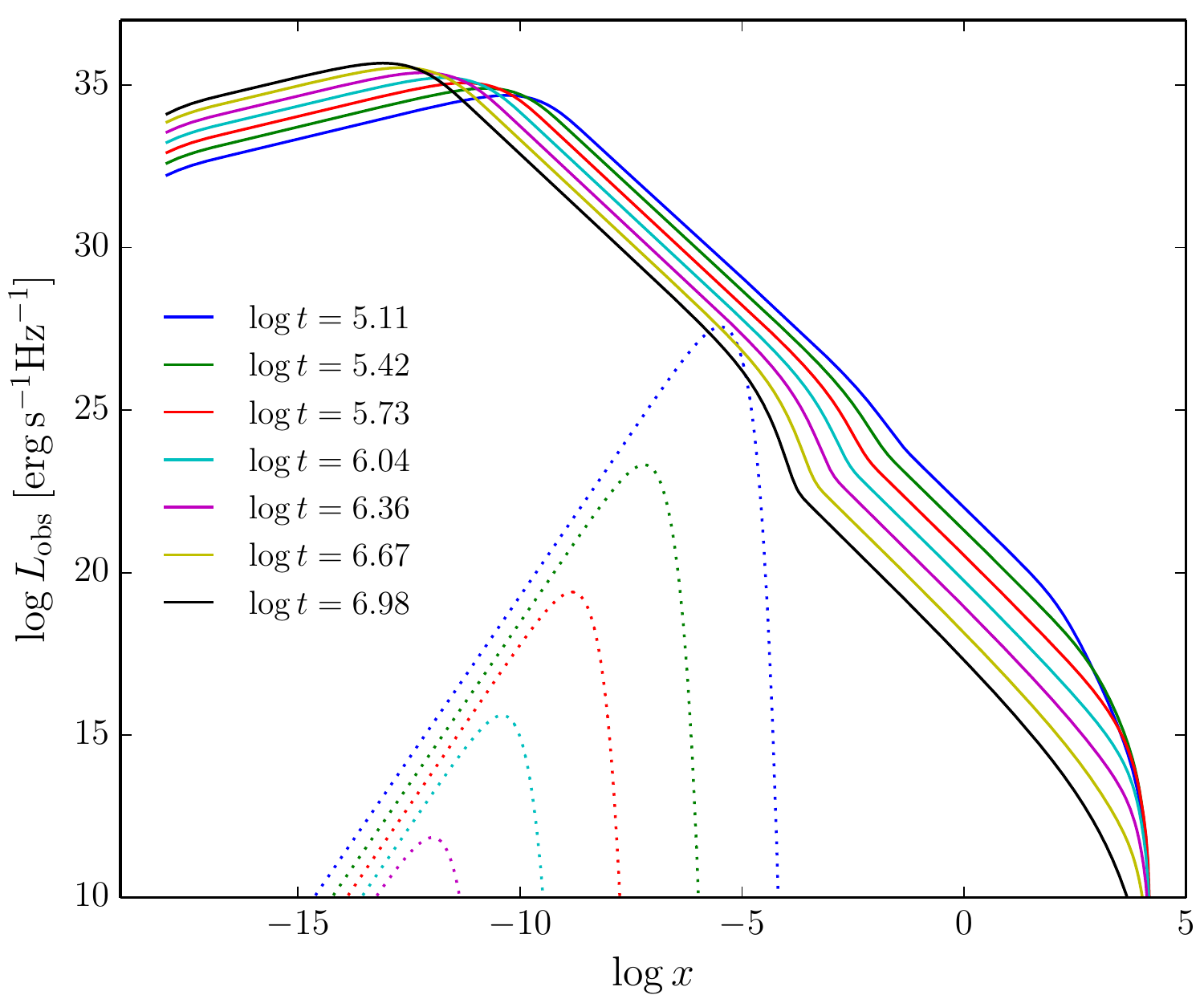}
  \caption{Luminosity per unit frequency of escaping radiation from
    the system as seen
    by a remote observer (cf.~Section~\ref{sec:observables}) as a
    function of $x=h\nu/m_\text{e}c^2$ for selected times
    during Phase III and the fiducial parameter setup
    (cf.~Table~\ref{tab:model_parameters}). Dotted curves indicate
    the thermal contribution separately. The middle panel shows the
    transition from thermal to non-thermal emission.}
  \label{fig:Lnth_obs}
\end{figure}

The non-thermal radiation escaping from the
PWN is reabsorbed and thermalized by the surrounding ejecta layer as
long as the confining envelope is optically thick. The spectra as seen
by a remote observer are thus characterized by thermal spectra at
early times (cf.~Figure~\ref{fig:Lnth_obs}, upper panel), a transition
from thermal to non-thermal spectra
when the ejecta material becomes optically thin
(cf.~Figure~\ref{fig:Lnth_obs}, middle panel), 
and non-thermal spectra at late times (cf.~Figure~\ref{fig:Lnth_obs},
bottom panel). We refer to Section~5.7 of Paper I for a discussion of
how these observer spectra are reconstructed. The thermal
part of the spectrum is plotted as dotted curves in the lower two
panels and shows how quickly this contribution fades away when the
optical depth of the ejecta approaches unity. We note that the
effective temperature of the ejecta material at early times falls into
the XRT band, which
typically also holds for all other runs across the entire parameter space
(see also Section~\ref{sec:lightcurve}). This is a remarkable feature
that makes XRT an ideal instrument to
observe and analyze the radiation escaping from the system at
$\lesssim 10^4-10^5\,\text{s}$ after the BNS merger.

According to our model, the observable radiation at early times is purely
thermal. However, due to the high degree of ionization of the
material, there are numerous free electrons upon which photons can
inverse Compton-scatter to form a non-thermal high-energy
tail of the spectrum. Furthermore, we have employed a single effective
temperature to characterize the ejecta layer. The ejecta shell
will, however, be characterized by strong temperature gradients, which
will give rise to a superposition of individual Planck spectra of
different temperatures. Hence, the actual observable spectrum at early
times might
be different from the simple Planck spectrum employed here and could
even appear as a power law in the XRT band.
The transition from this `thermal' spectrum at early times to
an intrinsically non-thermal spectrum at later times, which carries
the signatures of the
PWN, is a characteristic prediction of our model that is common to
essentially all parameter settings. The timescale for this transition
depends on the optical thickness of the ejecta material, which depends
on a number of parameters, such as the initial density of the ejecta
shell (i.e., $t_\text{dr}$,
$\dot{M}_\text{in}$, $\sigma_M$), its opacity $\kappa$, and the expansion
velocity $v_\text{ej}$. Depending on these parameters, it can occur
significantly earlier than for the fiducial parameter set.

 \section{Results: Influence of parameters}
\label{sec:param_study}

Most of the energy radiated away from the system at timescales of
interest is contained in the X-ray band (cf.~the previous
section). This section is devoted to study the influence of individual model
input parameters (see Table~\ref{tab:model_parameters}) on the
numerical result for the X-ray lightcurves as seen by a remote
observer. We discuss and compare results for the
XRT band varying only one parameter at a time. We distinguish between
``non-collapsing'' models, for which we arbitrarily set
$f_\text{coll}=\infty$, and ``collapsing
models'', in which the NS collapses to a black hole at a fraction or
multiple of the spin-down timescale (set by $f_\text{coll}$, see
Table~\ref{tab:model_parameters}).

\subsection{Non-collapsing models}
\label{sec:non_coll_models}

\begin{figure}[tb]
  \includegraphics[width=0.48
  \textwidth]{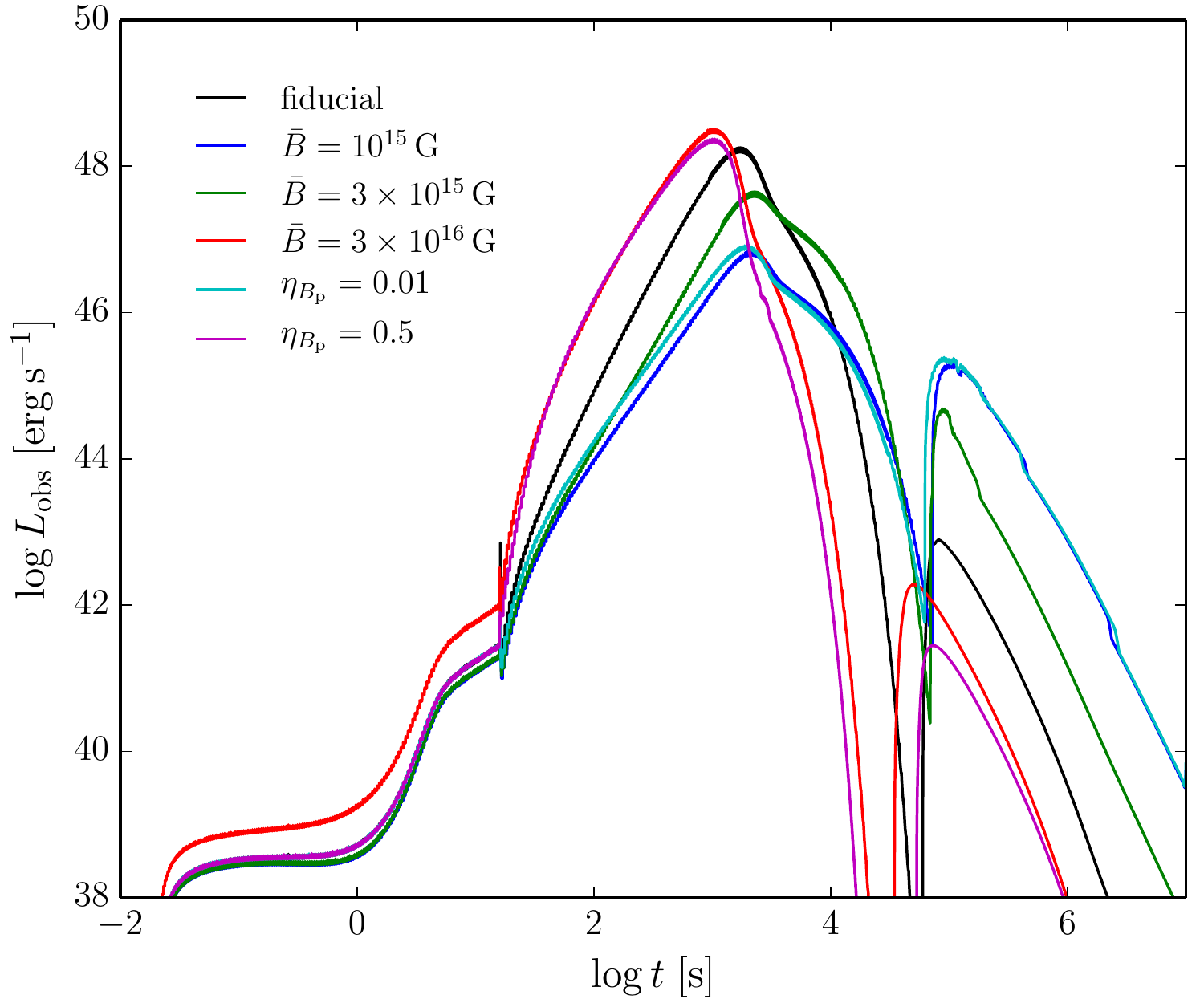}
\includegraphics[width=0.48
\textwidth]{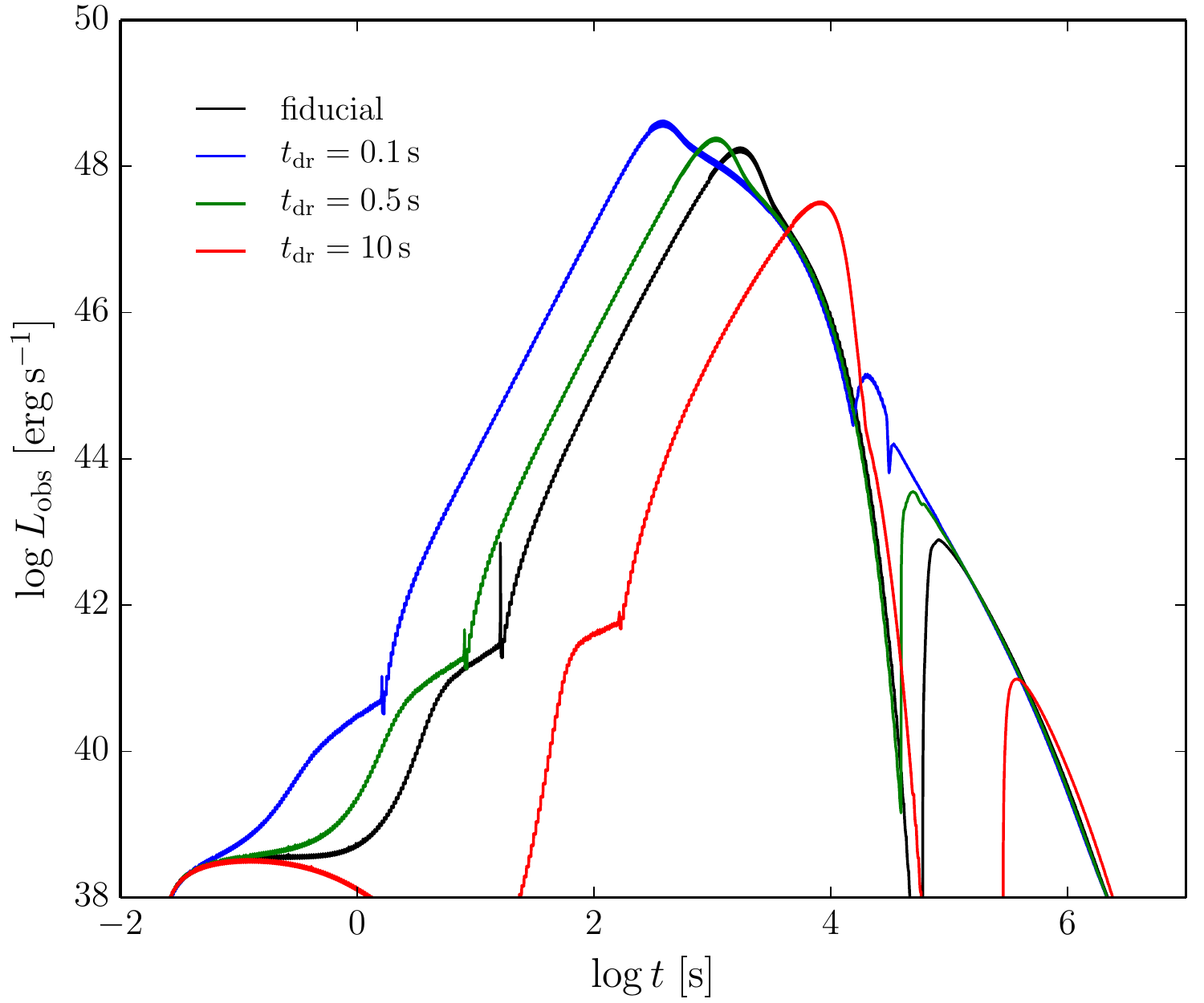}
\includegraphics[width=0.48 \textwidth]{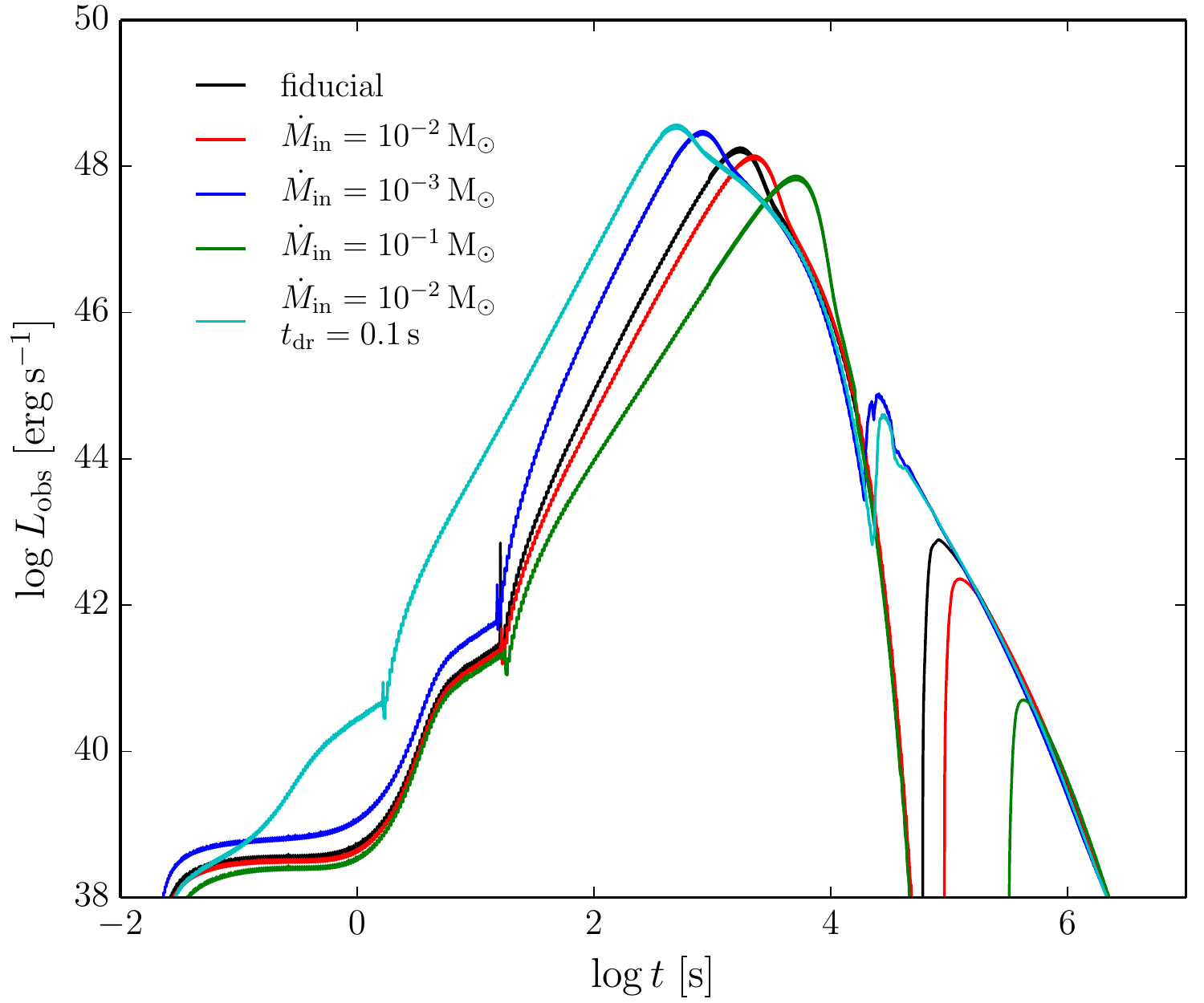}
  \caption{Luminosity $L_\text{obs,XRT}$  of radiation in the XRT band as seen
    by a remote observer (cf.~Section~\ref{sec:observables}) for the
    fiducial parameter setup  (cf.~Table~\ref{tab:model_parameters})
    and different values of $\bar{B}$ and
    $\eta_{B_\text{p}}$ (top), different values of $t_\text{dr}$
    (middle) and different values of $\dot{M}_\text{in}$
    (bottom). Phase II of the evolution is comparatively short and can
  be noticed by the kink in the lightcurves between 1 and $10^2$\,s.}
  \label{fig:comp_BtdrMdot}
\end{figure}

Non-collapsing model runs correspond to a scenario in which the merger
results in a NS that is indefinitely stable against gravitational
collapse. We restrict the discussion here to those parameters that
influence the model results significantly, i.e., $\bar{B}$, $\eta_{B_\text{p}}$,
$t_\text{dr}$, $\dot{M}_\text{in}$, $E_\text{rot,NS,in}$, and
$\kappa$. We find that the predictions of our model
concerning the dynamics of the system and the
observer lightcurves are not significantly influenced by the
remaining parameters.

Figure~\ref{fig:comp_BtdrMdot} compares the prediction for the X-ray
lightcurve in the XRT band and the fiducial parameter setup
(cf.~Table~\ref{tab:model_parameters}) with runs employing different
values for $\bar{B}$, $\eta_{B_\text{p}}$, $t_\text{dr}$, and
$\dot{M}_\text{in}$. The top panel shows that a higher total initial magnetic
field strength in the outer layers of the newly formed NS already
leads to an appreciably higher X-ray luminosity in Phase I and II. This
is because the baryon-loaded wind is endowed with a much higher
Poynting flux for higher values of $\bar{B}$, which is dissipated in
the ejecta material ($L_\text{EM}\propto
\bar{B}^2$; cf.~Equation~(32) of Paper I). At the beginning of Phase
III, higher values of $\bar{B}$ lead to a significantly steeper rise
of the luminosity, because the dipolar magnetic field strength
at the pole of the NS, $B_\text{p}$, is also higher (cf.~Equation~(41)
of Paper I). As the spin-down luminosity scales as $L_\text{sd}\propto
B_\text{p}^2$ (cf.~Section~4.2.1 of Paper I) and the PWN conserves
energy (cf.~Appendix~D of Paper
I), more energy is deposited in the ejecta material at earlier
times. Furthermore, the maximum value for the luminosity is
essentially determined by
$L_\text{sd}$\footnote{A rough estimate for the maximum possible X-ray observer
  luminosity in absence of strong relativistic effects is given by
  $\sim\!\eta_\text{TS} L_\text{sd,in}$, were
  $L_\text{sd,in}$ is the initial spin-down luminosity at
  $t=t_\text{pul,in}$ (cf.~Section~4.2.1 of Paper I).}, which is why
the maxima of the runs spread over two orders of magnitude. The time
of this maximal
brightening of the system is roughly
$\sim\!10^3\,\text{s}$ and varies only marginally among the different
runs. This timescale is essentially determined by the optical depth
of the ejecta shell, which does not directly depend on $\bar{B}$ (see
also below). After the maximum in the
luminosity, the lightcurve decays steeper for higher values of
$\bar{B}$, which is due to the fact that the spin-down timescale
scales as $t_\text{sd}\propto B_\text{p}^{-2}$ and, hence, energy
injection into the ejecta material fades away more rapidly. Moreover,
it is important to note that varying either $\bar{B}$ or
$\eta_{B_\text{p}}$ by the same factor yields roughly identical
results in Phase III (compare the cyan and blue and the red and purple
lines, respectively). This is because with all other parameters being
identical, those runs also share the same value
of $B_\text{p}$, which determines the spin-down luminosity and
therefore the energy output in Phase III. In Phase I, however, the
luminosity differs, as $\bar{B}$ directly enters the source term
$L_\text{EM}$ (see above). Consequently, as far as Phase III is
concerned, the parameter $\eta_{B_\text{p}}$ can be absorbed into
$\bar{B}$.

The parameter $t_\text{dr}$ is highly influential in determining
timescales, as shown by the middle panel of
Figure~\ref{fig:comp_BtdrMdot}. With all other parameters fixed, it
does not only set the duration of
Phase I, which is directly proportional to $t_\text{dr}$ (the
timescale for the density in the surrounding of the NS to decrease is
$t_\text{dr}/\sigma_\rho$; see
Sections~4.1.1 and 4.2.1 of Paper I). It also impacts the time of maximum
brightness, which lies between $\sim\!10^2-10^4\,\text{s}$ for the
cases considered here. This is because with higher values of $t_\text{dr}$, the
total ejected mass $M_\text{ej} \approx \dot{M}_\text{in}
t_\text{dr}/\sigma_M$ (cf.~Equation~(31) of Paper I) increases and
thus the ejecta shell needs more time to expand in order to reach a
comparable average density and thus a comparable optical depth. At the
same time, however, the material cools down, which leads to a lower
peak brightness as $t_\text{dr}$ increases.

\begin{figure}[tb]
  \includegraphics[width=0.48
  \textwidth]{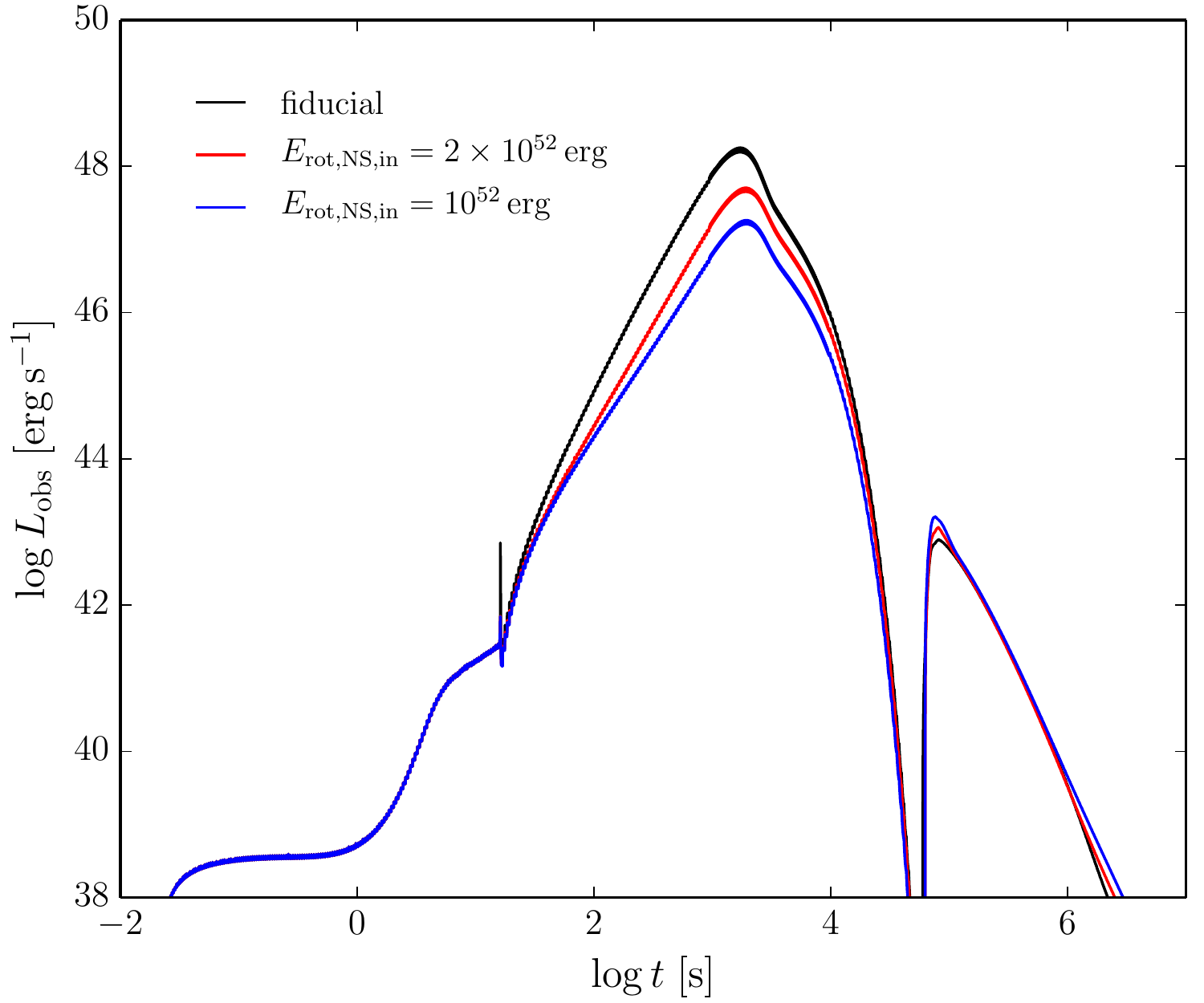}
\includegraphics[width=0.48 \textwidth]{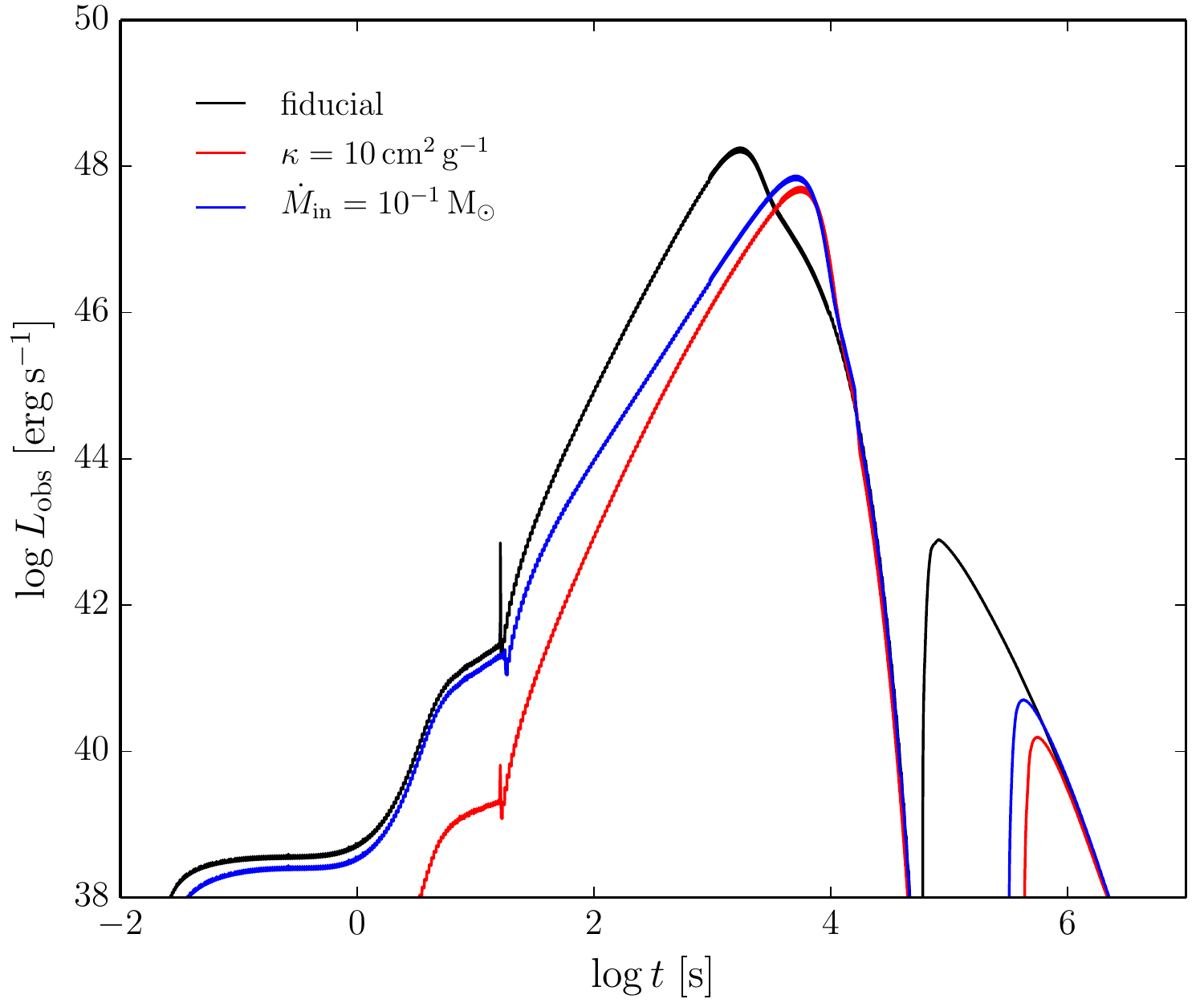}
  \caption{Luminosity $L_\text{obs,XRT}$ of radiation in the XRT band as seen
    by a remote observer (cf.~Section~\ref{sec:observables}) for the
    fiducial parameter setup  (cf.~Table~\ref{tab:model_parameters})
    and different values of $E_\text{rot,NS,in}$ (top) and different
    values of the opacity $\kappa$ (bottom). Phase II of the evolution
    is comparatively short and can
  be noticed by the kink in the lightcurves between 1 and $10^2$\,s.}
  \label{fig:comp_ErotAkappa}
\end{figure}

Smaller initial mass loss rates $\dot{M}_\text{in}$ induce moderately higher
luminosities in Phase I, II, and up to the global maximum of
the X-ray lightcurve in Phase III (see bottom panel of
Figure~\ref{fig:comp_BtdrMdot}), thereby also shifting the time of
maximum brightness to earlier times. This is because lower values of
$\dot{M}_\text{in}$ result in a smaller amount of ejected mass
$M_\text{ej}(t)$ up to time $t$ in Phase I (cf.~Equation~(31) of Paper
I) and, hence, in a lower average density and optical depth of the
ejecta material. As the total amount of ejected material $M_\text{ej} \approx \dot{M}_\text{in}
t_\text{dr}/\sigma_M$ is also reduced, the ejecta shell
starts off with a lower average density at $t=t_\text{shock,out}$ and
is thus characterized by a lower optical depth throughout
Phase III. The red line in the lower panel of
Figure~\ref{fig:comp_BtdrMdot} corresponds to a case in which
both $t_\text{dr}$ and $\dot{M}_\text{in}$ differ from the
fiducial value, but combine to give the same total ejected mass as in
the $\dot{M}_\text{in}=10^{-3}\,\text{M}_\odot$ case (blue curve). The
evolution up to the global maximum is different in the two
cases, showing that the two parameters cannot be reduced to a single
one (e.g., to $M_\text{ej}$). This is because $t_\text{dr}$ has a much
stronger effect on shifting the onset of Phase II. However, on much
longer timescales this shift becomes irrelevant and the two curves
essentially agree.

Different initial rotational energies of the newly-born
NS after merger affect the X-ray lightcurve only by rescaling the
global maximum (see the top panel of
Figure~\ref{fig:comp_ErotAkappa}). This scaling can roughly be
explained by the fact that $L_\text{sd,in}\propto
E_\text{rot,NS,in}^2$ (assuming that the rotational energy of the pulsar
is approximately given by $E_\text{rot,NS,in}$; cf.~Section~4.2.1 of
Paper I). This is similar to the scaling found for
$\bar{B}$ and $\eta_{B_\text{p}}$ (see above).

Figure~\ref{fig:comp_ErotAkappa} (bottom panel) shows the effect of a
much higher  (by a
factor of 50) mean opacity $\kappa$ than in the fiducial case (cf.~the
discussion in Section~\ref{sec:params}). This
increases the optical depth, which thus results in
a dimmer X-ray lightcurve and a delayed global maximum (as it takes
more time to radiate away the acquired energy). The cooling
timescale of the ejecta material after the global maximum in the
luminosity, however, depends mostly on the
energy input, which is similar in the two runs. Therefore, both
lightcurves fall off in a very similar way after
$\sim\!10^4\,\text{s}$ as the temperature of
the Planck spectrum moves out of the XRT band. Furthermore, we note
that concerning the behavior in Phase III, higher opacities have a very
similar effect as a higher initial mass-loss rate for the baryonic
wind in Phase I, as shown by the blue curve. This shows that part of
the uncertainty in the opacities can effectively be absorbed into,
e.g., the initial mass-loss rate.

\subsection{Collapsing models}

Collapsing models correspond to the most frequent case in which the NS
is supramassive (or hypermassive) at birth (cf.~Section~1 of Paper I)
and it is thus doomed to
collapse to a BH. Everything that has been
noted in Section~\ref{sec:fiducial_model} and in the previous
subsection still applies (at least up to the time of collapse), with
the additional complication that the NS collapses to a BH at
$t=t_\text{coll}$, which alters the subsequent evolution of the
system. After $t=t_\text{coll}$, the evolution
equations are numerically integrated as described in Section~4.4 of
Paper I. We first concentrate on the more likely case of a
gravitational collapse during Phase III (Section~\ref{sec:col_p3}),
and return to the possibility
of a collapse during Phase I in Section~\ref{sec:col_p1}.

\subsubsection{Collapse during Phase III}
\label{sec:col_p3}

\begin{figure}[tb]
  \includegraphics[width=0.48
  \textwidth]{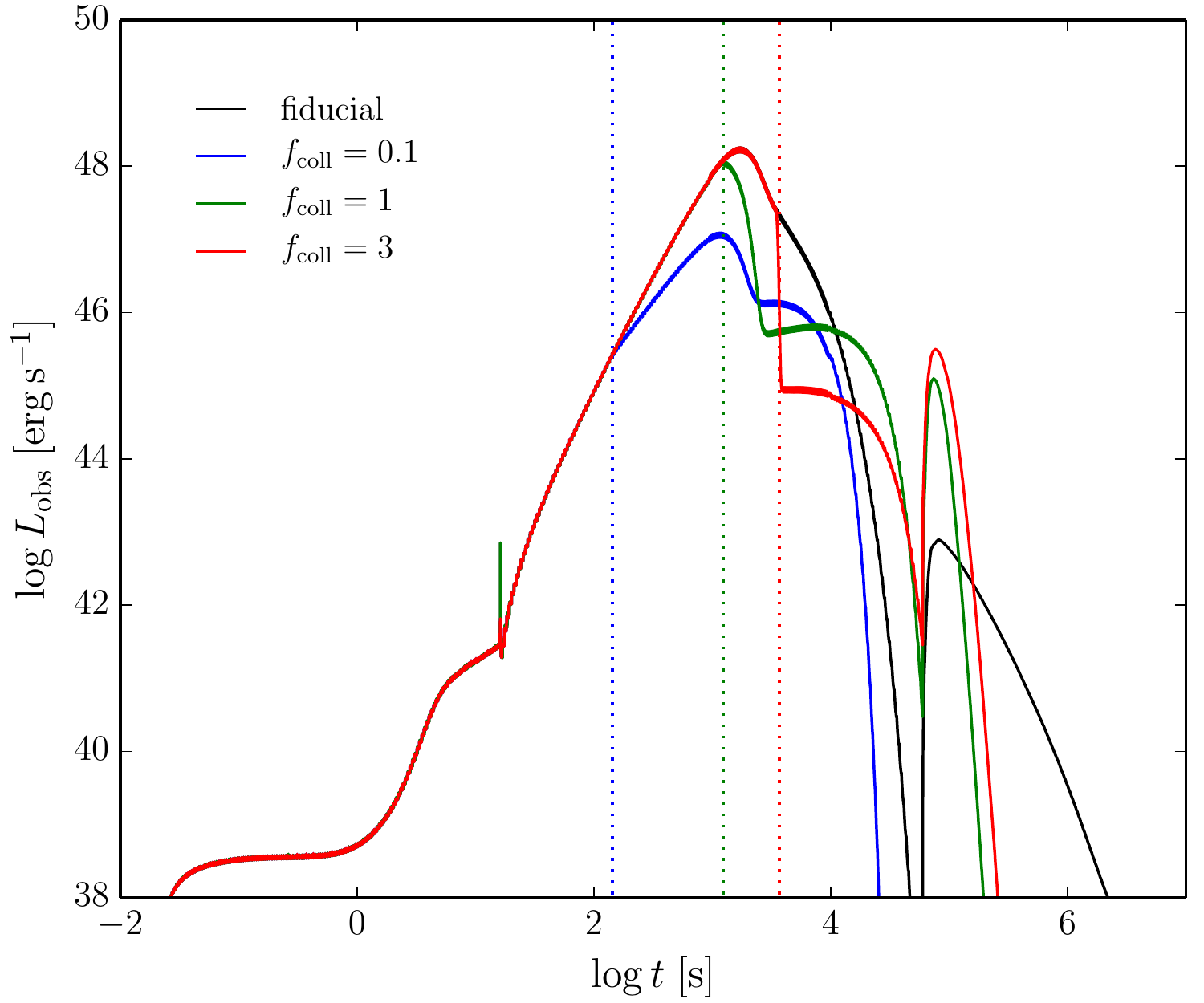}
\includegraphics[width=0.48
\textwidth]{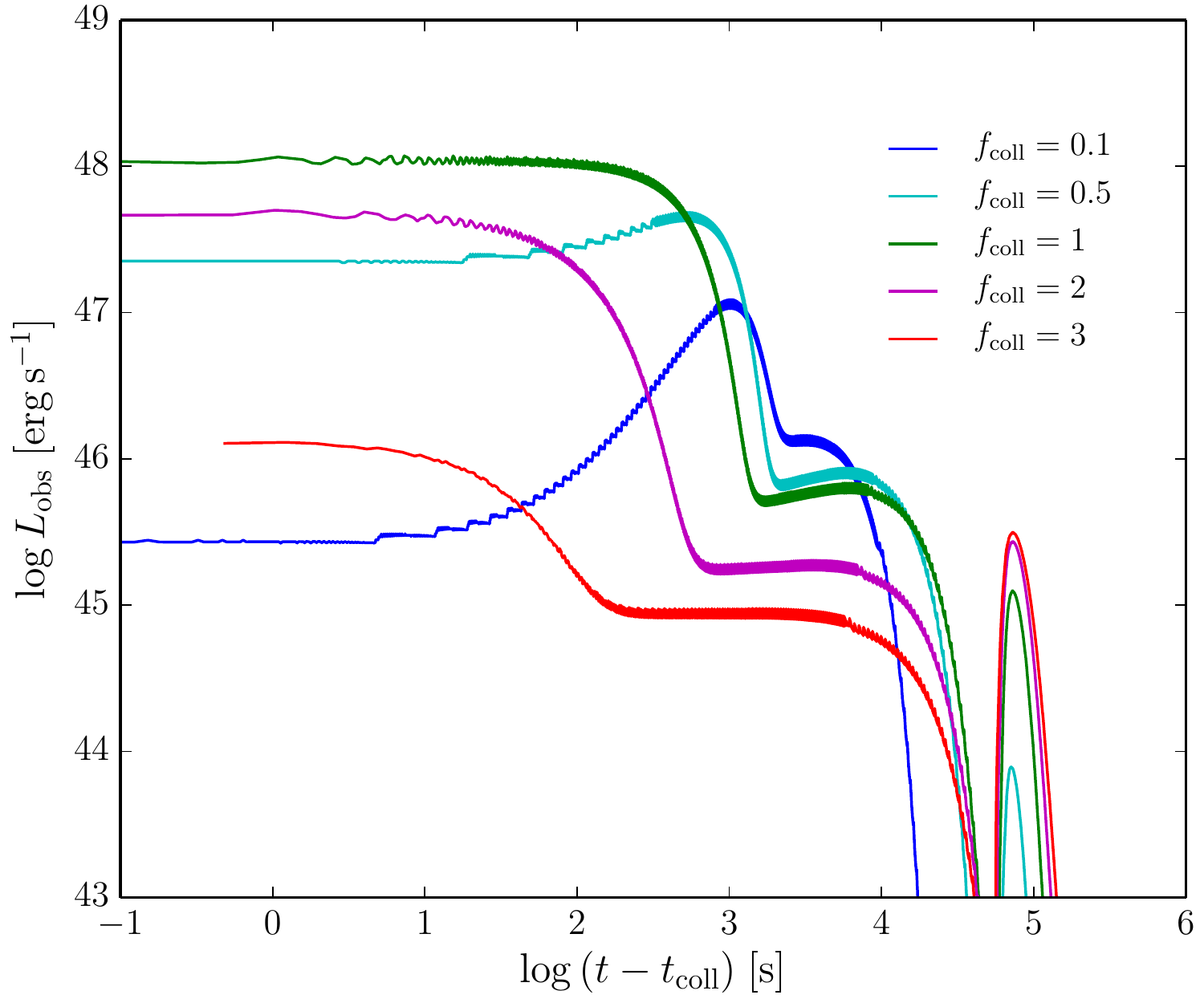}
  \caption{Top: luminosity $L_\text{obs,XRT}$ of radiation escaping
    from the system in the XRT band as seen
    by a remote observer (cf.~Section~\ref{sec:observables}) for the
    fiducial parameter setup  (cf.~Table~\ref{tab:model_parameters})
    and different values of $f_\text{coll}$. Dotted lines indicate the
    time of collapse $t_\text{coll}$ of the NS in the lab
    frame. Bottom: same lightcurves as in the top panel (plus
    additional ones), but as a function of the time after the
    respective collapse. The latter lightcurves represent the ``X-ray
    afterglow'' in the time-reversal scenario.}
  \label{fig:comp_fcoll}
\end{figure}

If the NS is supramassive at birth it can survive until a substantial
fraction of its rotational energy has been dissipated. Its lifetime is
therefore expected to be of the order of the spin-down timescale
$t_\text{sd}$. The top panel of
Figure~\ref{fig:comp_fcoll} shows the evolution of the system for the
fiducial parameter set, but different times of collapse. Dotted lines
indicate $t=t_\text{coll}$, which is set in units of $t_\text{sd}$ by
the parameter $f_\text{coll}$ (see
Table~\ref{tab:model_parameters}). 

If the collapse occurs prior to the
time of maximal brightening, it
manifests itself only as a `kink' in the X-ray lightcurve (cf.,
e.g., the blue curve). The
latter still increases up to a maximum after the time of collapse due
to the fact that there is still a substantial amount of internal energy
(stored in the ejecta material) yet to be radiated away. The sooner
the time of collapse, the dimmer is this global maximum of the
lightcurve, since further energy injection by $L_\text{PWN}$ is
substantially reduced at
$t=t_\text{coll}$ (see below). After the maximum the lightcurve
gradually decreases
(cf.~blue and green curves) until a plateau-like phase is reached. 

In contrast, if the NS collapses after the time of maximal brightening,
the lightcurve directly decreases to a plateau-like phase. This is
because the ejecta shell enters an `asymptotic' phase shortly after having
reached the global maximum (cf.~Section~5.6 of Paper I). In this
phase, most of the acquired internal energy has already been radiated
away and the present luminosity depends on
the present influx of energy from the PWN. This influx is severely
reduced after $t=t_\text{coll}$, as only gradual cooling of the PWN
can still supply the ejecta shell with further energy (as described in
Section~4.4 of Paper I).

\begin{figure}[tb]
  \includegraphics[width=0.48
  \textwidth]{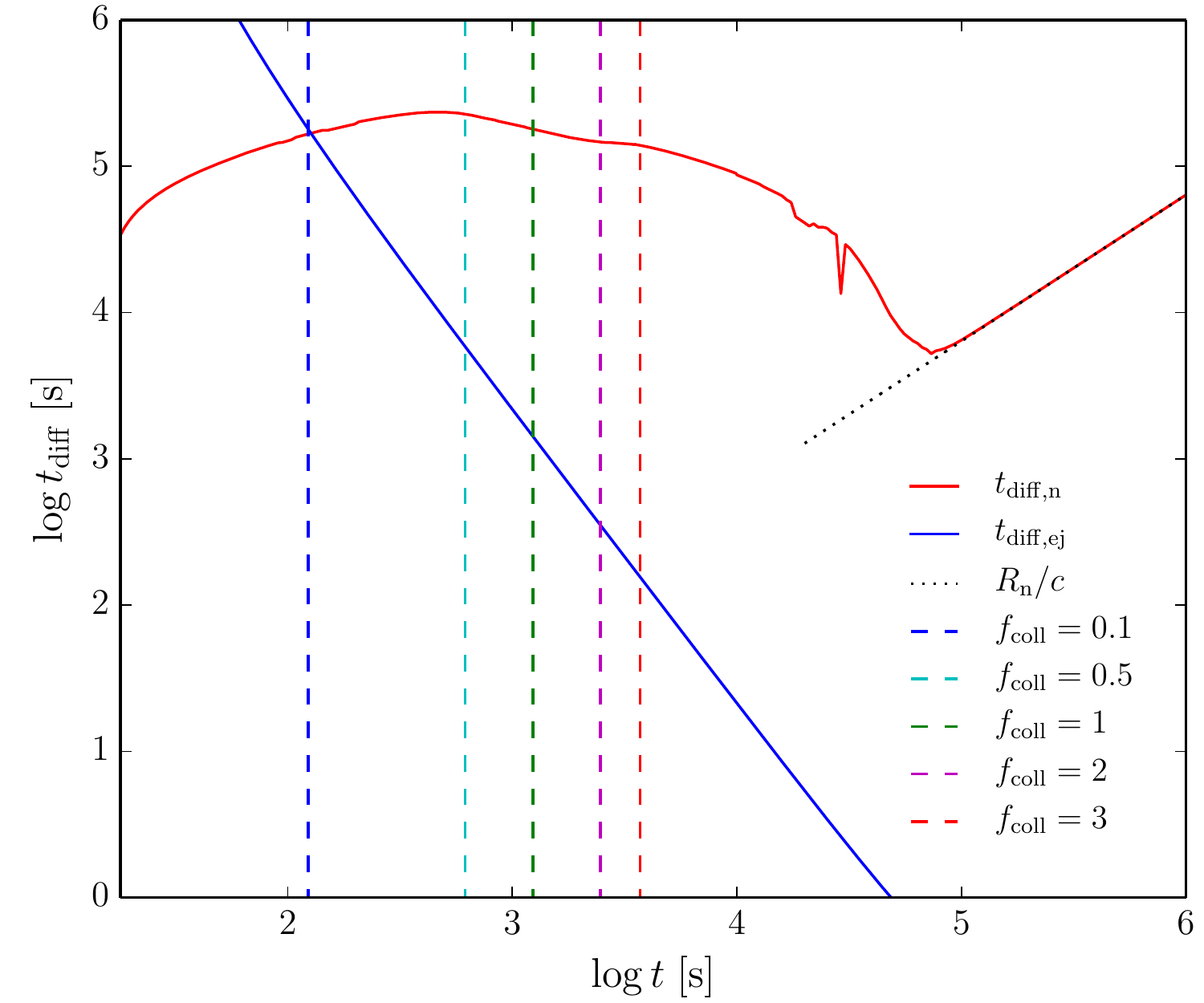}
  \includegraphics[width=0.48
  \textwidth]{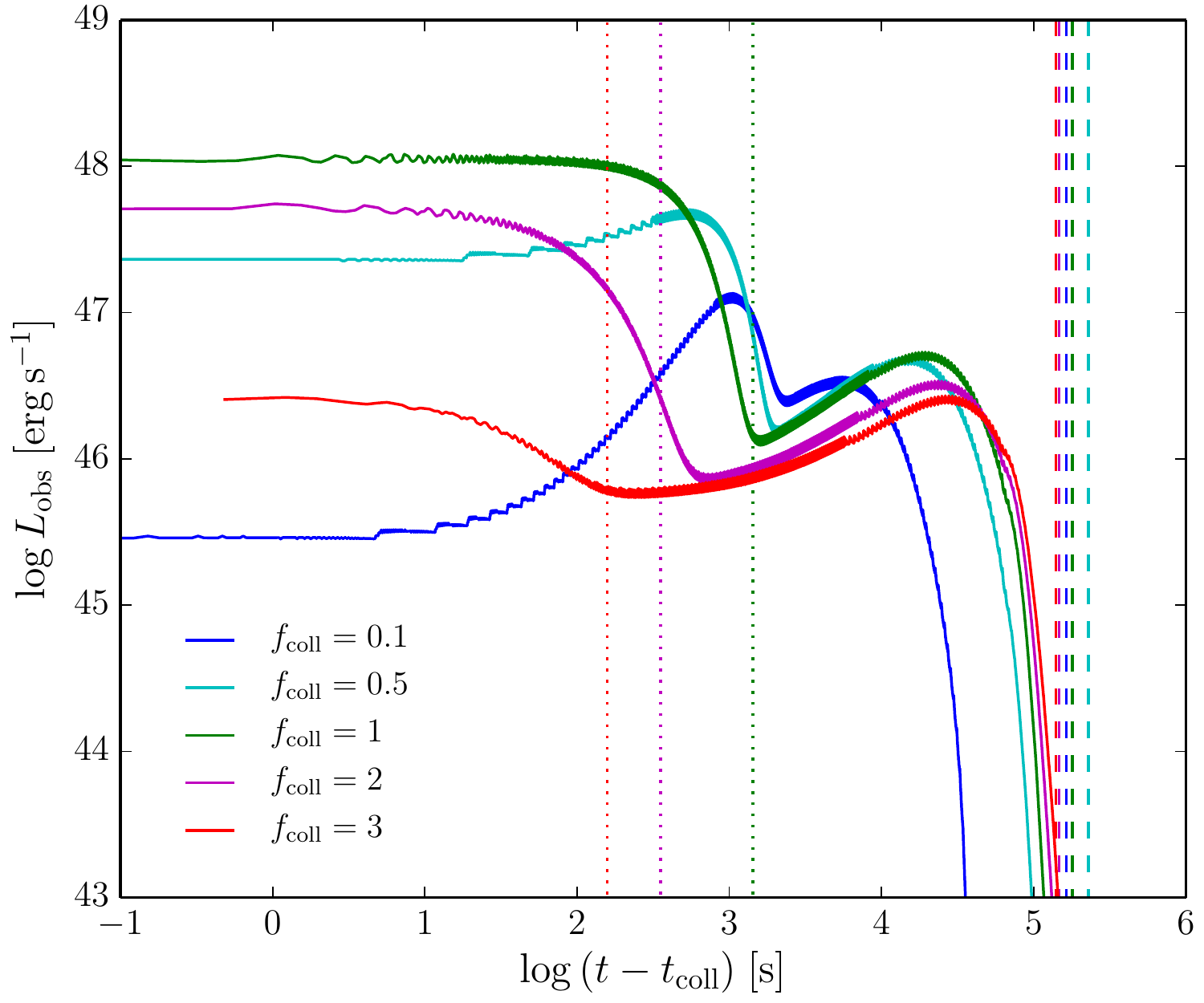}
  \caption{Top: photon diffusion time of the PWN ($t_\text{diff,n}$)
    and of the ejecta shell ($t_\text{diff,ej}$) for the
    fiducial parameter setup (cf.~Table~\ref{tab:model_parameters})
    and $f_\text{coll}=\infty$, i.e., without a collapse of the NS to
    a black hole. Dashed lines indicate the different times of collapse
    considered in Figure~\ref{fig:comp_fcoll}. Bottom: total luminosity $L_\text{obs,tot}$
    (cf.~Section~\ref{sec:observables}) as seen
    by a remote observer after the respective collapse, corresponding
    to the runs depicted in the bottom panel of
    Figure~\ref{fig:comp_fcoll}. Dashed lines represent
    $t_\text{diff,n}$ at the time of collapse (cf.~the upper panel), which represent upper
    limits on the duration of the second `plateau' evident in these
    lightcurves. Dotted lines indicate $t_\text{diff,ej}$ at the time
    of collapse for the runs with $f_\text{coll}\ge 1$, which provide an estimate
    for the duration of the first plateau for these runs.}
  \label{fig:tdiff_plateaus}
\end{figure}

It is this supply with residual energy from the PWN that
creates the plateau phase characteristic of all runs across the entire
range for $f_\text{coll}$ (cf.~the upper panel of
Figure~\ref{fig:comp_fcoll}). The characteristics of the lightcurves after
$t=t_\text{coll}$ are, however, better
illustrated by plotting the luminosity as a function of the time
after collapse (cf.~the bottom panel of Figures~\ref{fig:comp_fcoll}
and \ref{fig:tdiff_plateaus}). 

Depending on the collapse time, the
lightcurves show a characteristic two-plateau structure after
$t=t_\text{coll}$ (cf.~the bottom panel of Figures~\ref{fig:comp_fcoll}
and \ref{fig:tdiff_plateaus}). Only for
early collapse times, the lightcurve essentially shows a single rising
plateau plus a `bump' of radiation (cf., e.g., the case
$f_\text{coll}=0.1$) caused by the release of energy stored in the
ejecta shell (see above). In this case, the lightcurve after $t=t_\text{coll}$
strongly depends on the previous history as the radiation escaping
from the system is still sourced by energy acquired by the ejecta material
prior to the collapse. However, for runs in which
the ejecta shell has already entered
or is about to enter the asymptotic regime (see above; cf., e.g., the
cases $f_\text{coll}=1,2,3$) its properties strongly depend
on the present state of the nebula and the total observer
luminosity $L_\text{obs,tot}$ clearly shows a two-plateau structure
(see the bottom panel of Figure~\ref{fig:tdiff_plateaus}). The duration of
the first plateau in these runs can be roughly estimated
by the photon diffusion timescale $t_\text{diff,ej}$ of the ejecta
layer at the time of
collapse (cf.~Equation~(122) of Paper I and the top panel of
Figure~\ref{fig:tdiff_plateaus}). This timescale represents
the time needed for the ejecta material to adjust to the sudden
(instantaneous) change
of the PWN conditions at $t=t_\text{coll}$.\footnote{We note that the
  collapse of the NS proceeds on the dynamical timescale, which is of
  the order of milliseconds. At the times of interest here,
  the collapse can thus be implemented as an instantaneous
  modification of the numerical integration of the evolution equations
at $t=t_\text{coll}$ (see also Section~4.4 of Paper I).}
The latter process is nicely reflected by
the aforementioned lightcurves depicted in the lower panel of
Figure~\ref{fig:tdiff_plateaus}, in which case $t_\text{diff,ej}$ at
the time of collapse (indicated by dotted lines) indeed provides reliable
predictions for the duration of the first plateau.

The duration of the second plateau phase in the
total observer luminosity $L_\text{obs,tot}$
(cf.~Section~\ref{sec:observables}) is essentially determined by the photon
diffusion time $t_\text{diff,n}$ of the PWN (cf.~Section~4.4 of Paper
I and the top panel of Figure~\ref{fig:tdiff_plateaus}), which sets
the cooling timescale of the nebula after the
collapse. As $t_\text{diff,n}$ is monotonically decreasing after
$t=t_\text{coll}$ (cf.~Equation~(105) and (106) of Paper I), the
diffusion time at $t=t_\text{coll}$ provides an upper bound on the
plateau duration (see the bottom panel of
Figure~\ref{fig:tdiff_plateaus}). This fact has also been noted by
\citet{Ciolfi2015a} based on a simplified analysis. As $t_\text{diff,n}$
varies only slightly across the different collapse times considered
here (cf.~upper panel of Figure~\ref{fig:tdiff_plateaus}), the plateau
durations are remarkably similar
($\sim\!10^5\,\text{s}$; cf.~the bottom panel of
Figure~\ref{fig:tdiff_plateaus}). Moreover, the absolute luminosity
levels $L_\text{obs,tot}$ during the second plateau phase are also
very similar. This is because the total luminosity during this phase
is essentially determined by the ratio of the internal
energy of the nebula $E_\text{nth}$ (which is very similar among the
different runs) and $t_\text{diff,n}$ (cf.~Section~4.4 of Paper
I). 

It is important to note that the ejecta material cools down and that the
maxima of the thermal spectra drift out of the XRT band on the timescales
considered here. Therefore, after
$t=t_\text{coll}$ the X-ray lightcurves $L_\text{obs,XRT}$ differ
somewhat from the total luminosity
$L_\text{obs,tot}$ (cf.~the bottom panel in
Figures~\ref{fig:comp_fcoll} and \ref{fig:tdiff_plateaus}). In
particular, the second plateaus of the X-ray lightcurves are
characterized by lower
luminosity levels and shorter durations of the order of
$\sim\!10^4\,\text{s}$. Furthermore, as the PWN
becomes optically thin before $t=10^5\,\text{s}$ (cf.~also
Figure~\ref{fig:delta_tau}), a short breakout
of luminous, non-thermal radiation from the interior of the PWN
becomes visible in
the X-ray band shortly after the second plateau
(cf.~Figure~\ref{fig:comp_fcoll}). This breakout is only
short-lived as further cooling of the nebula prevents such high
X-ray luminosities soon after the onset. In the case of $f_\text{coll}=0.1$,
the PWN has even cooled down to such an extent that this
transition to the optically thin regime of the PWN does not lead to
appreciable X-ray luminosities at all.

Finally, we note that in the recently proposed time-reversal scenario
\citep{Ciolfi2015a,Ciolfi2015b}, the X-ray lightcurves depicted in the
lower panel of Figure~\ref{fig:comp_fcoll} correspond to the
observable X-ray afterglow of the SGRB as seen in the XRT band. We
note that the two-plateau structure found here is remarkably similar
to the structure of X-ray afterglows observed in a number
of SGRB events (e.g., \citealt{Gompertz2014}). Furthermore, as the
prompt SGRB emission is associated with the
collapse of the NS in the time-reversal scenario, the X-ray emission
predicted by our model prior
to the collapse (as, e.g., depicted in the upper panel of Figure~\ref{fig:comp_fcoll})
corresponds to a precursor signal that can be searched for by future
X-ray missions. If found, it would constitute a remarkable piece of
evidence in favor of the time-reversal scenario. The essence of `time
reversal' in this scenario is nicely illustrated by
Figures~\ref{fig:comp_fcoll} and \ref{fig:tdiff_plateaus}: the energy
radiated away from the system after $t=t_\text{coll}$ has been
extracted from the NS
prior to the collapse. This energy diffuses outward on the respective
diffusion timescales and produces a characteristic two-plateau
structured X-ray lightcurve.

\subsubsection{Collapse during Phase I}
\label{sec:col_p1}

In this section, we return to the problem of
predicting the EM transient signal associated with an
early collapse of the NS during Phase I. Such an early collapse is
expected to occur either if the NS is hypermassive at birth or if it
is supramassive but collapses due
to (magneto-) hydrodynamic instabilities. In either case we
numerically integrate the corresponding evolution equations
(Equations~(1)--(2) of Paper I) after the collapse as described in
Section~4.4 of Paper I. In this case, the time of collapse,
$t=t_\text{coll}$, is parametrized in units of $t_\text{dr}$, $f_\text{coll,PI} =
t_\text{coll}/t_\text{dr}$.

\begin{figure}[tb]
  \includegraphics[width=0.48
  \textwidth]{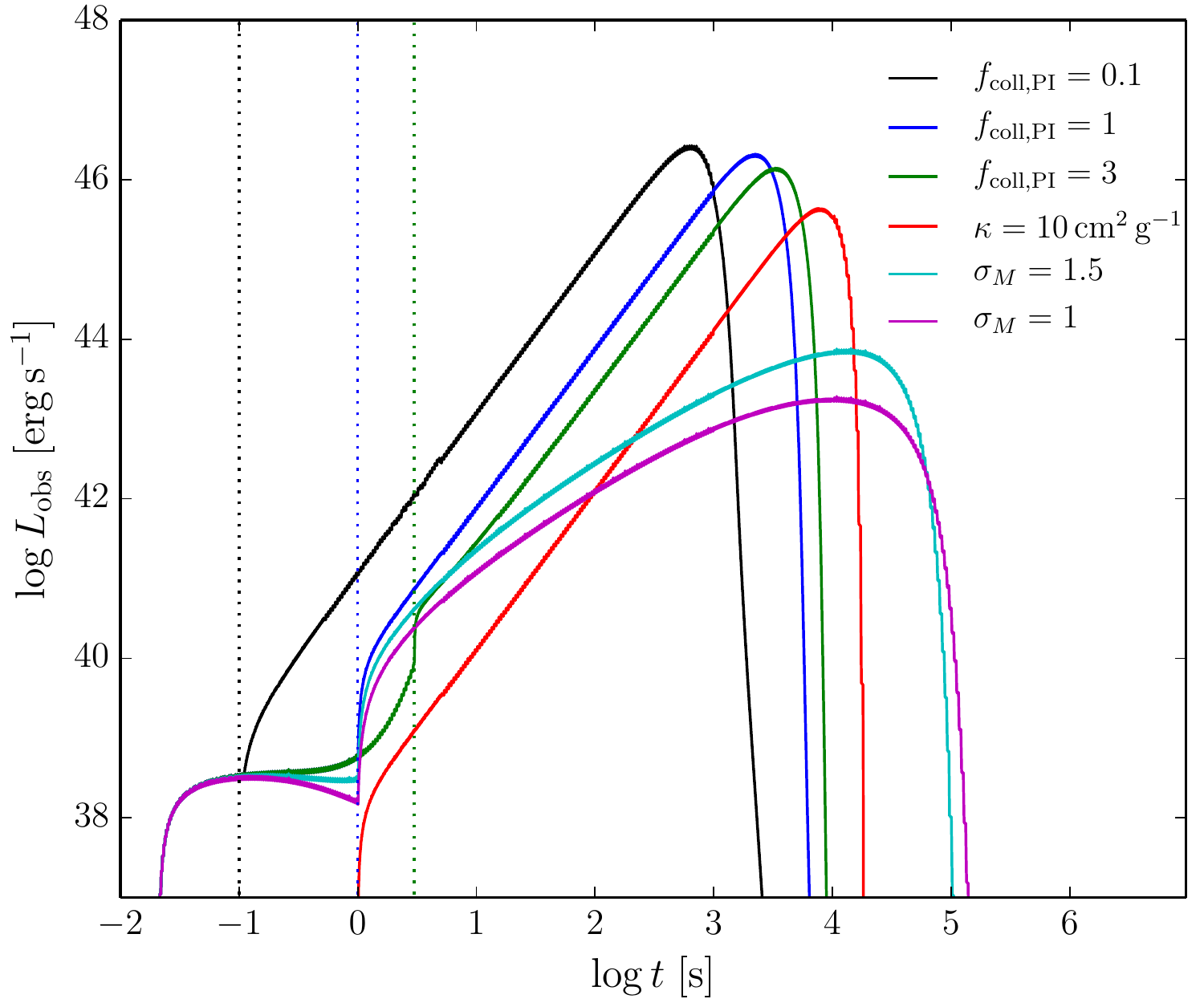}
  \includegraphics[width=0.48
  \textwidth]{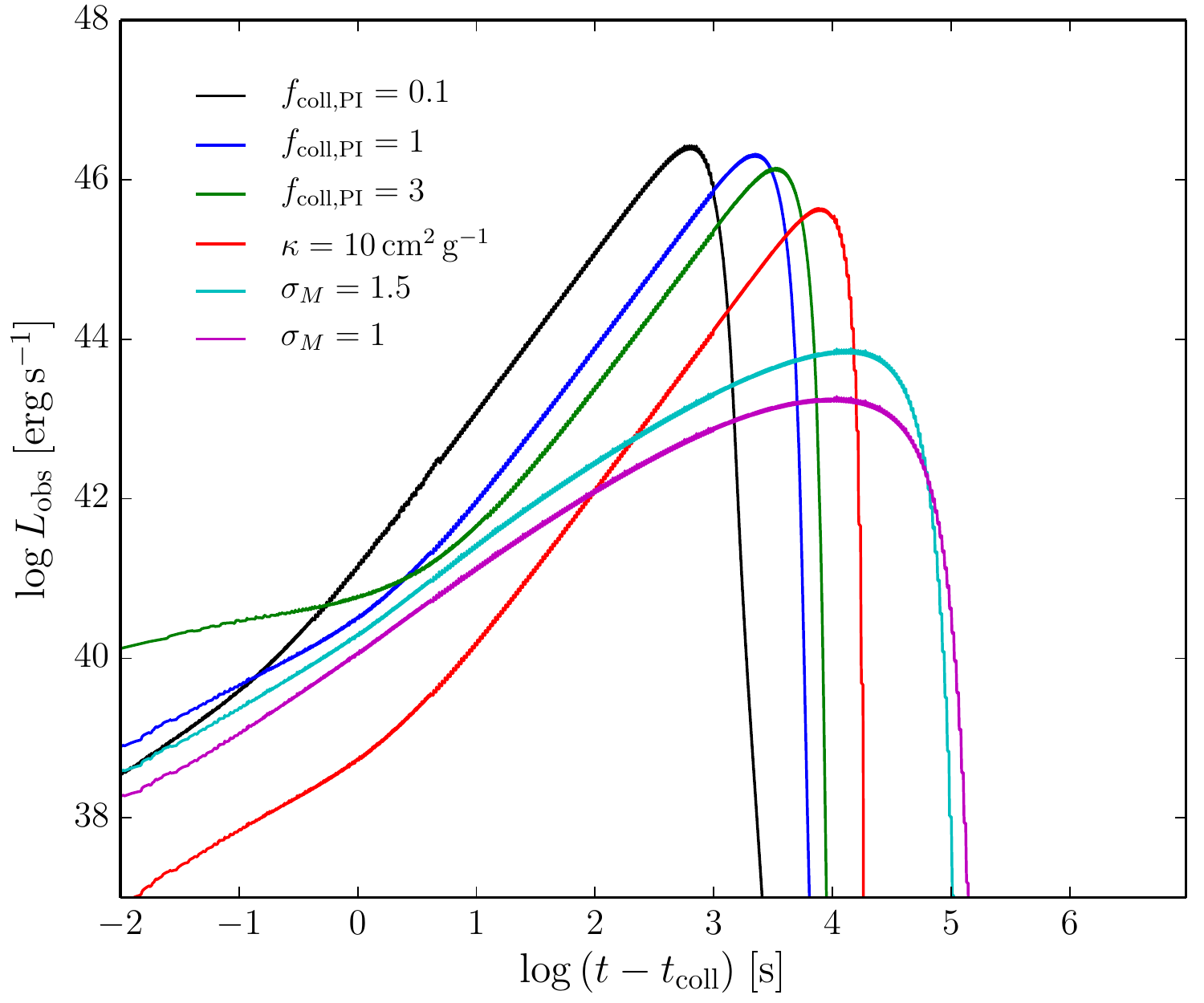}
  \caption{Early collapse of the NS during Phase I. Top: X-ray
    luminosity $L_\text{obs,XRT}$
    (cf.~Section~\ref{sec:observables}) as seen
    by a remote observer for the fiducial parameter setup
    (cf.~Table~\ref{tab:model_parameters}), but different values of
    $f_\text{coll,PI}=t_\text{coll}/t_\text{dr}$, $\sigma_M$, and
    $\kappa$. Dashed lines indicate
    the different times of collapse $t_\text{coll}$. Bottom: same
    lightcurves, but plotted as a function of the time after the
    respective collapse.}
  \label{fig:coll_p1}
\end{figure}

\begin{figure}[tb]
  \includegraphics[width=0.48
  \textwidth]{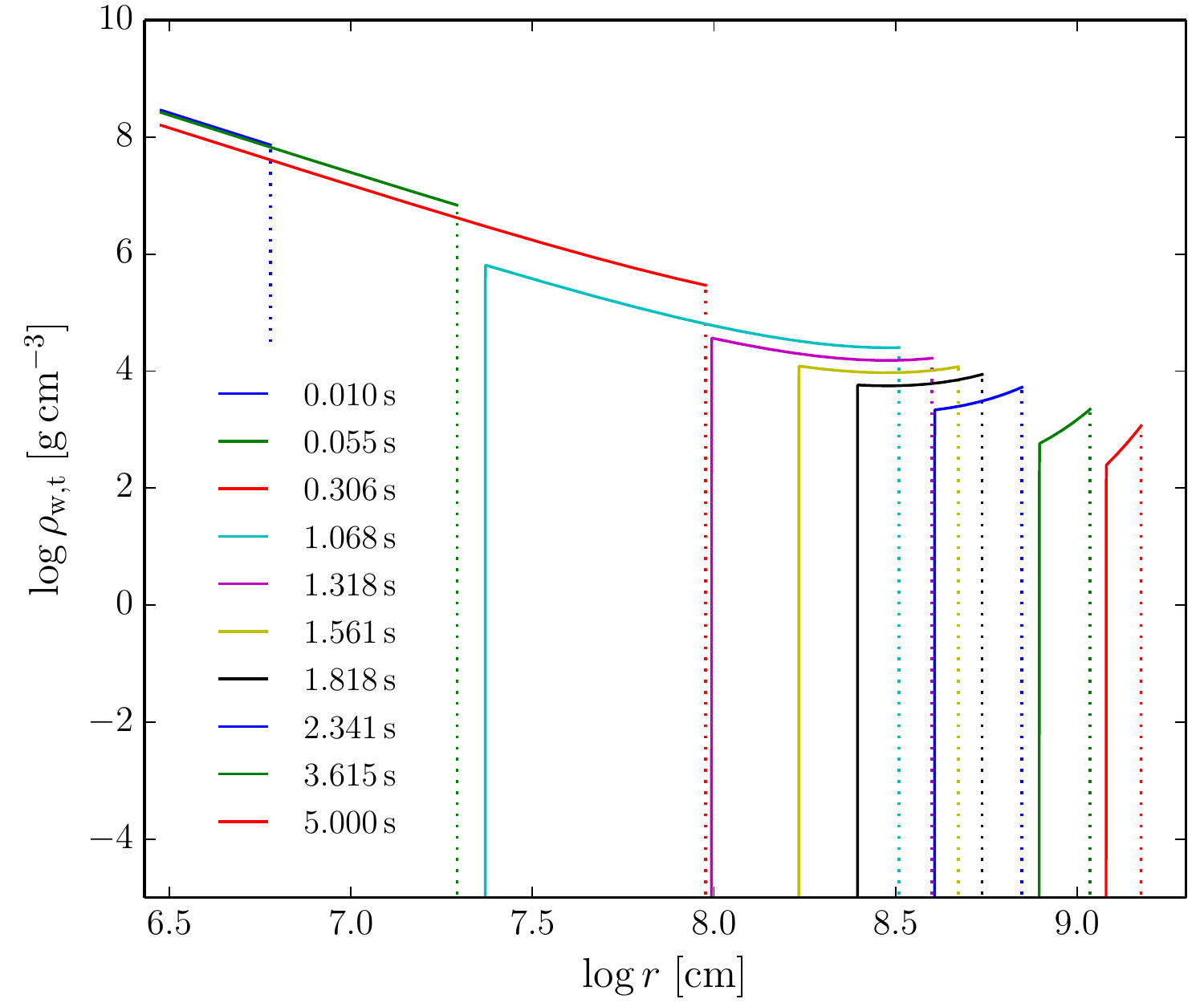}
  \caption{Snapshots of the density profiles of the baryon-loaded wind
    for selected times (fiducial parameter
    setup with $t_\text{coll}=1\,\text{s}$,
    cf.~Table~\ref{tab:model_parameters}; cf.~also Sections~4.4 and
    4.1.1 of Paper I). After $t=t_\text{coll}$ the
    wind forms a spherical shell. }
  \label{fig:coll_p1_rho}
\end{figure}

Figure~\ref{fig:coll_p1} shows X-ray lightcurves of
radiation in the XRT band (cf.~Section~\ref{sec:observables}) for the
fiducial parameter setup varying the time of collapse, $\sigma_M$, and
$\kappa$. These are the most influential parameters concerning this early
collapse scenario. At $t=t_\text{coll}$ mass ejection from the NS
suddenly stops and the ejected material starts to form a spherical
shell of thickness $R_\text{ej}-R_\text{in}$ (cf.~Section~4.1.2 of
Paper I) that moves outward with a constant head
speed $v_\text{ej,in}$ (see Figure~\ref{fig:coll_p1_rho}). After
$t=t_\text{coll}$ the total luminosity
$L_\text{obs,tot}$ typically increases as a power law in
time\footnote{As both $R_\text{ej}$ and $R_\text{in}$ scale linearly
  with time, all other quantities show power-law behavior as well.} up
to a maximum at a timescale $\sim\!10^{3}-10^{4}\,\text{s}$ after the
BNS merger. For the
fiducial setup (cf.~Table~\ref{tab:model_parameters}), the shell
thickness remains constant (since $\sigma_v =
0$; cf.~Section~4.1.1 of Paper I) and the peak emission frequency remains in the
XRT band until late times. This is why the corresponding X-ray
lightcurves show this power-law behavior and follow the total
luminosity until shortly after the global maximum (cf.~the
$f_\text{coll,PI}=0.1,1,3$ curves in Figure~\ref{fig:coll_p1}). When
$\sigma_M<2$, $\sigma_v > 0$ and the shell thickness increases
$\propto t$. Hence, the temperature decreases more rapidly and the
corresponding X-ray lightcurves deviate from a power-law increase soon
after $t=t_\text{coll}$ as the peak emission frequency moves out of
the XRT band (cf.~the $\sigma_M = 1,1.5$ cases in
Figure~\ref{fig:coll_p1}). Finally, we note that the
higher opacity of $\kappa=10\,\text{cm}^2\,\text{g}^{-1}$ leads to an
overall dimmer lightcurve, with the
power-law increase and global maximum being shifted to larger
timescales.

In this early-collapse scenario, the maximum of the lightcurve is
always followed by a monotonic and abrupt decrease in contrast to the
typical (two-)plateau morphology found in the case of a collapse
during Phase III. This is due to the absence of a PWN, which could further
supply the ejecta matter with energy once the energy reservoir has
been depleted.

\section{Lightcurve morphologies}
\label{sec:morphologies}

\begin{figure*}[tb]
\centering  
\includegraphics[width=0.33
  \textwidth]{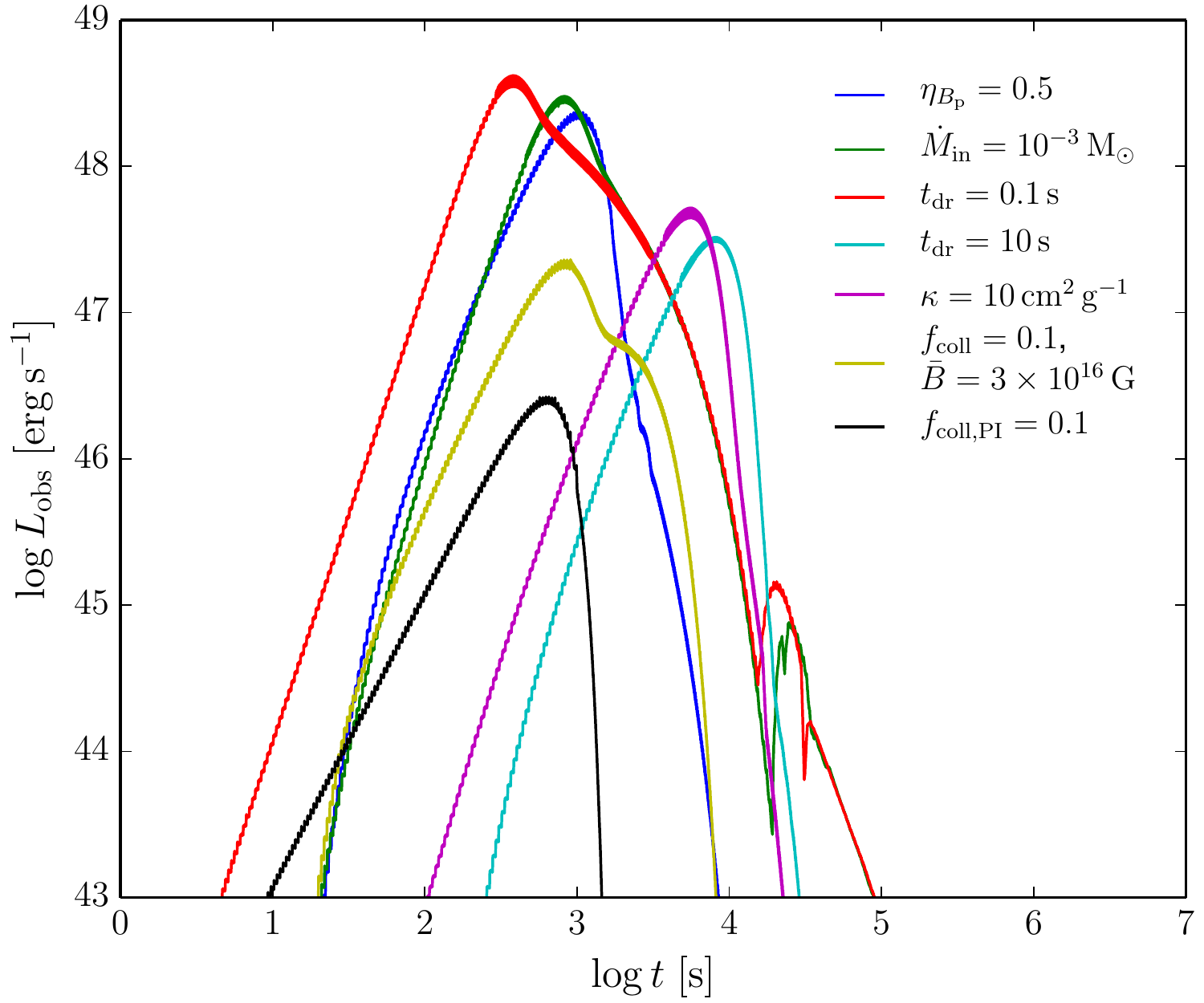}
\includegraphics[width=0.33
  \textwidth]{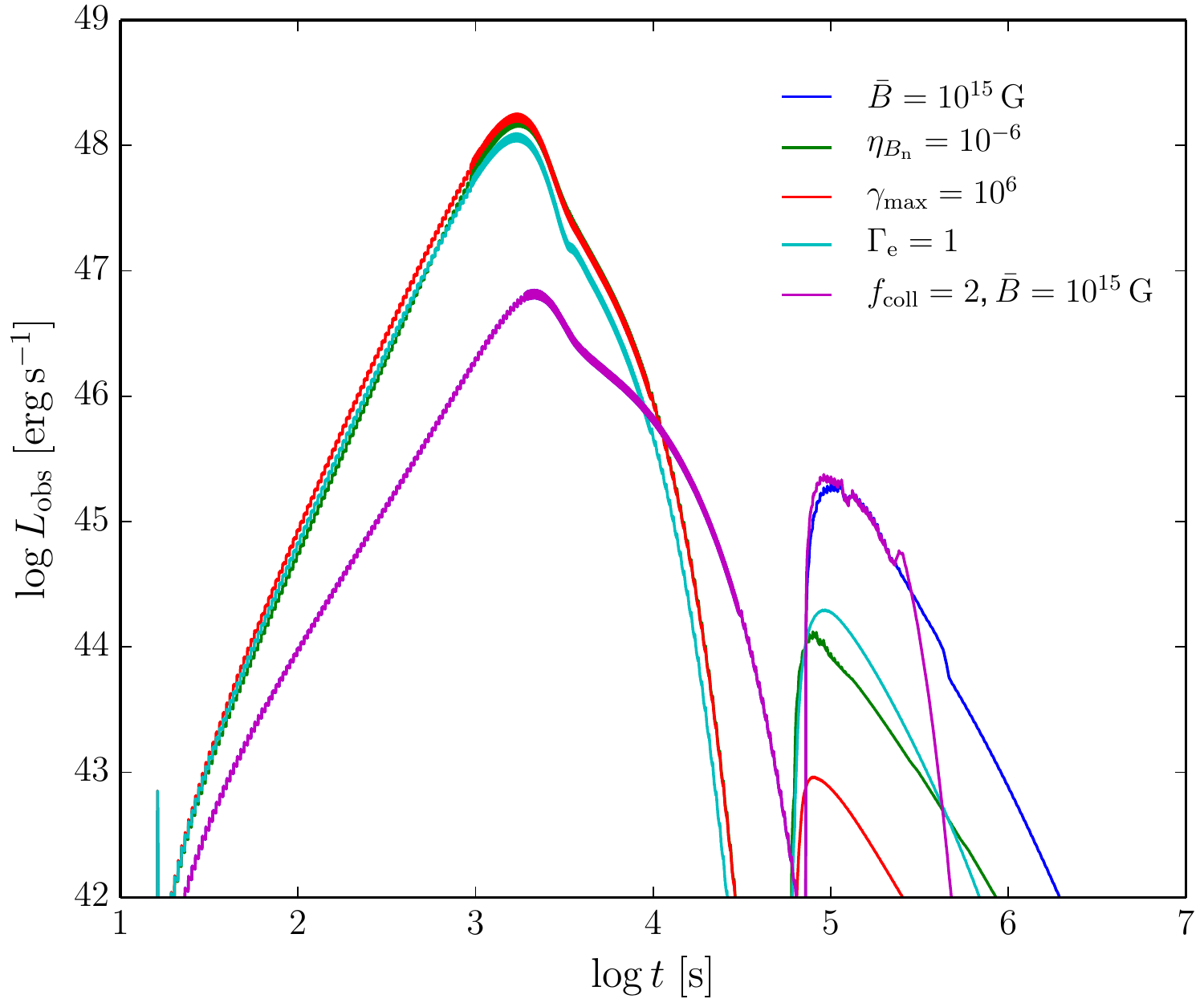}\\
\includegraphics[width=0.33
  \textwidth]{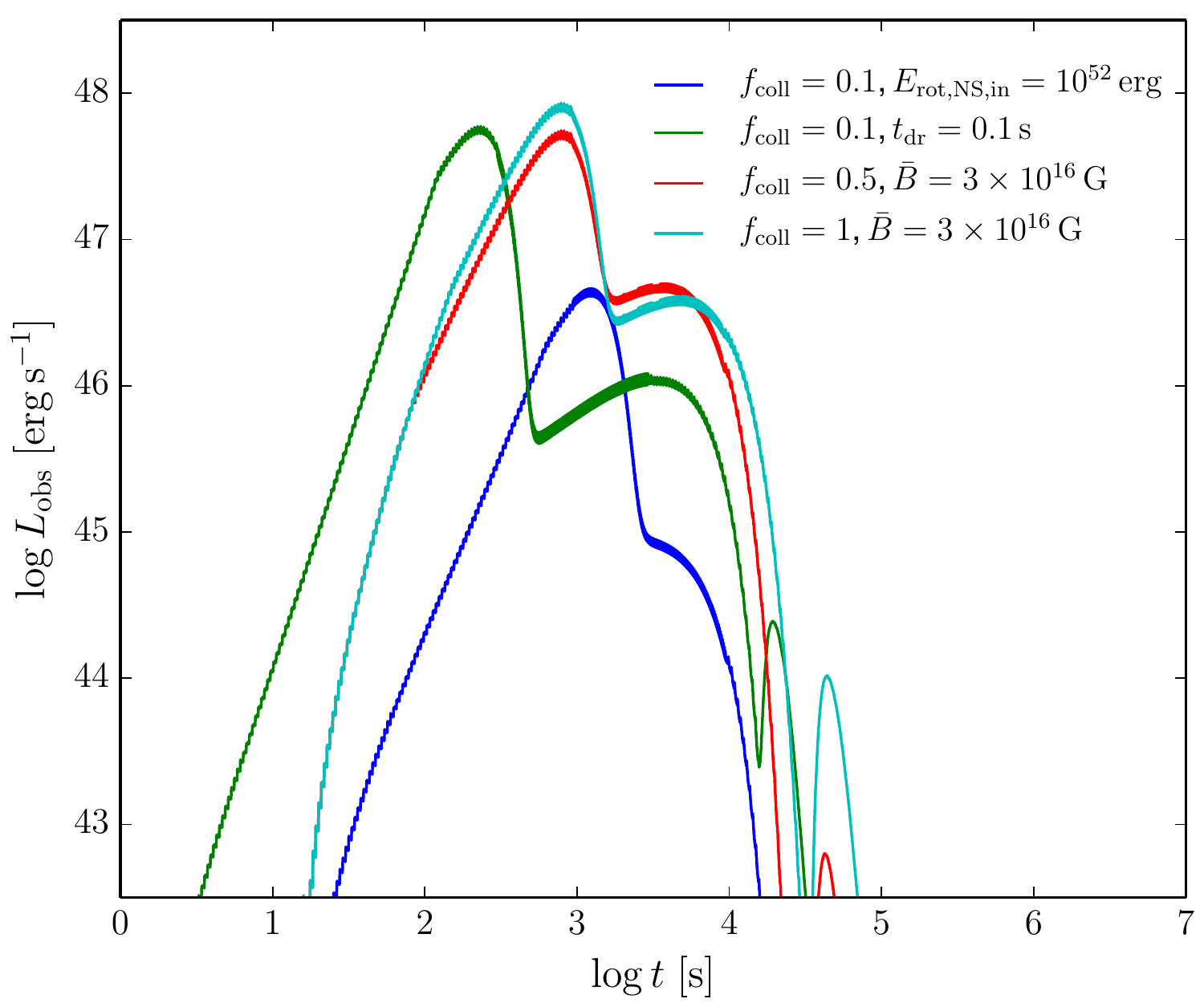}
\includegraphics[width=0.33
  \textwidth]{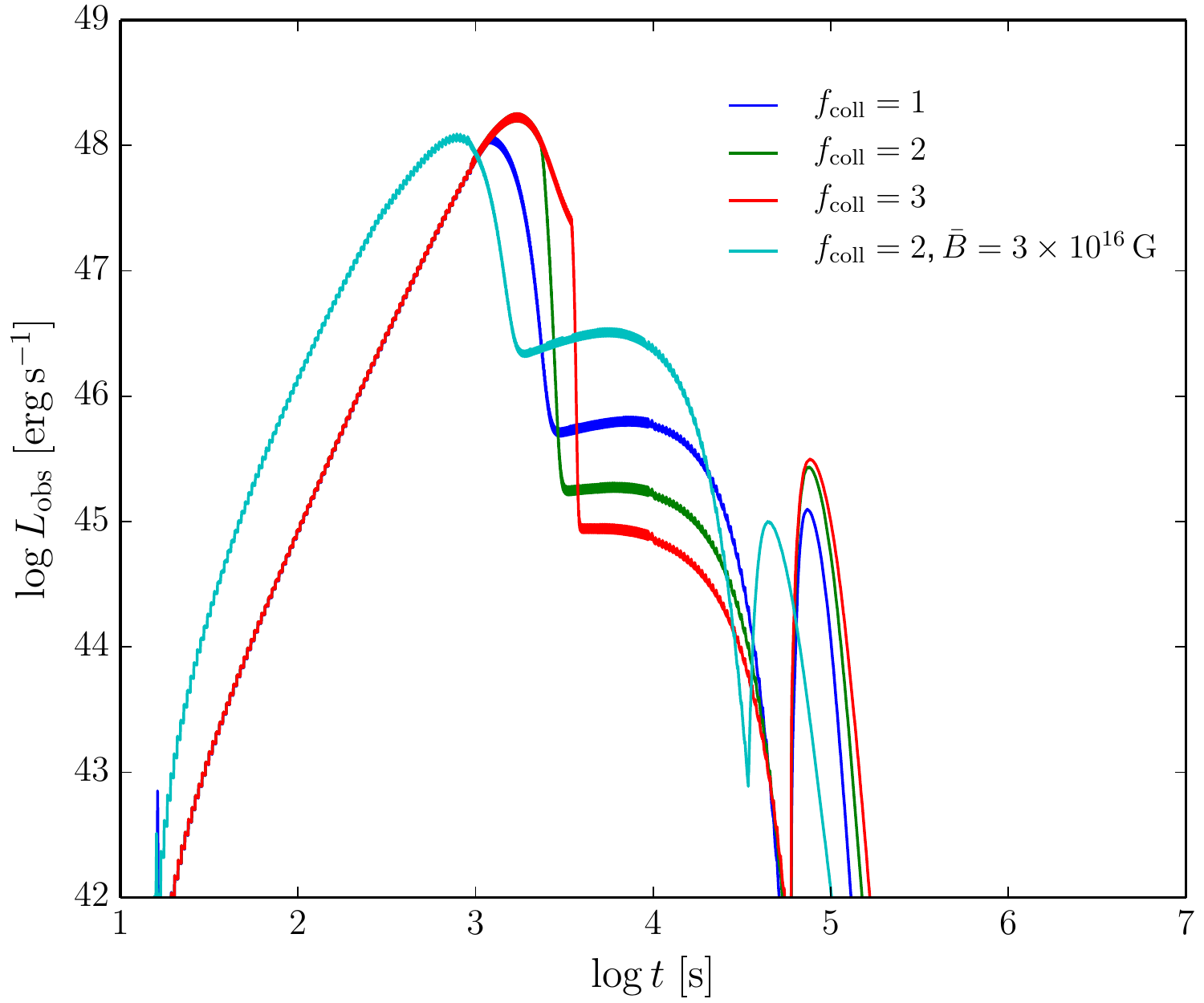}
\includegraphics[width=0.33
  \textwidth]{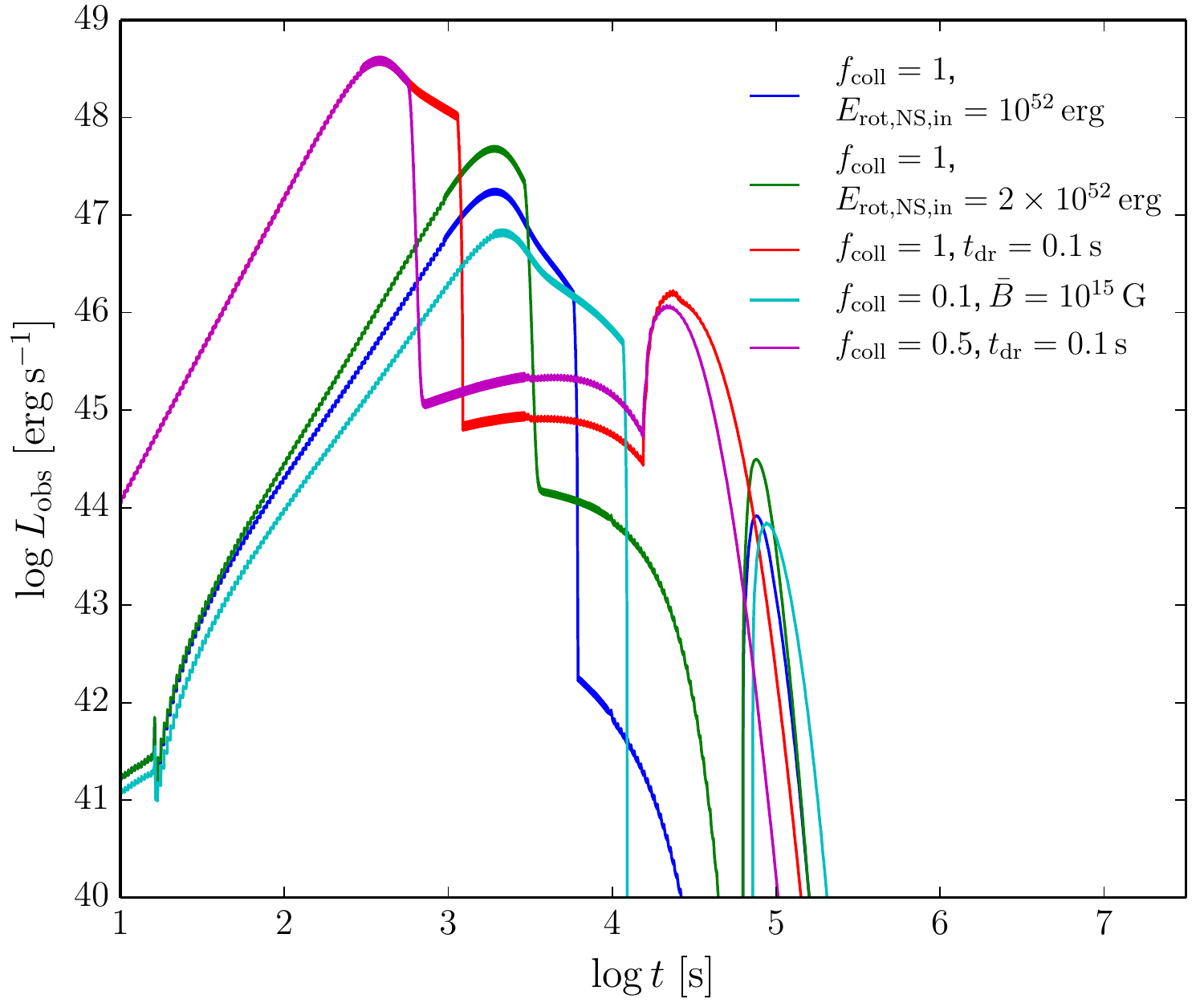}
  \caption{X-ray lightcurve morphologies (standard scenario). Upper row (left to right):
    single-maximum lightcurves with steep decay (M), single-maximum
    lightcurves with secondary maximum due to non-thermal radiation at
    late times (MM). Bottom row
    (left to right): plateau lightcurves (P), plateau lightcurves with a late-time
    flare of non-thermal radiation (PF), lightcurves with a late-time flare of non-thermal
    radiation showing absence
    of strong X-ray radiation at intermediate times (GF). All parameters are
    as in the fiducial setup
(cf.~Table~\ref{tab:model_parameters}), except for the ones specified
in the respective legend.}
  \label{fig:morphologies_noTR}
\end{figure*}

After having discussed the influence of individual model parameters on the
X-ray lightcurves in the previous section, we now explore the entire
parameter space and classify the resulting lightcurves according to
their morphology into different categories. For each of these
categories, we show a few representative examples (see
Figures~\ref{fig:morphologies_noTR} and
\ref{fig:morphologies_TR}) and point to some candidate
SGRBs\footnote{XRT data of all SGRB events mentioned in this paper are
  available at \url{http://www.swift.ac.uk/xrt_curves/}.} that are
potentially consistent with the
respective morphology. This division into categories is somewhat
arbitrary and it is not necessarily exhaustive. We stress that a
comparison with observational
data is only intended to be on a qualitative level at the present
stage. Fitting observational data with our model will be the subject
of future work. Overall, we also distinguish
between the standard scenario in which the SGRB itself is
associated with the time of merger of the BNS system and the
time-reversal scenario \citep{Ciolfi2015a,Ciolfi2015b} in which
the SGRB is associated with the
collapse of the remnant NS. In both cases, the lightcurves shown in
Figures~\ref{fig:morphologies_noTR} and \ref{fig:morphologies_TR}
correspond to the X-ray afterglows as seen in the XRT band of the
\textit{Swift} satellite. General issues regarding, e.g., the
thermal/non-thermal nature of the spectra and late-time power-law
decays observed in some SGRBs are discussed in
Section~\ref{sec:discussion}.

\subsection{Standard scenario}
\label{sec:morphologies_noTR}

Figure~\ref{fig:morphologies_noTR} shows five different categories of
X-ray lightcurve morphologies, which we briefly discuss below. In
contrast to the time-reversal case
(Section~\ref{sec:morphologies_TR}), these
lightcurves are characterized by a `delayed onset': it typically takes
$\sim\!1-10\,\text{s}$ after the SGRB to reach appreciable X-ray
luminosities of the order of $\gtrsim 10^{42}\,\text{erg}$. This is
mainly due to the very high optical depths of the ejecta matter at
early times (cf.~Section~\ref{sec:lightcurve}). The time of peak
brightness is typically reached between $\sim\!10^2-10^4\,\text{s}$
after the SGRB with luminosities of
$\sim\!10^{46}-10^{49}\,\text{erg}\,\text{s}^{-1}$.

\paragraph{Cat.~M} These lightcurves show a single maximum followed
by a steep decay (upper left panel of
Figure~\ref{fig:morphologies_noTR}). They correspond to thermal radiation
from the ejecta shell, with a non-thermal `tail' of radiation from
the nebula at late times in some cases (cf.~$t_\text{dr}=0.1$ and
$\dot{M}_\text{in}=10^{-3}\,\text{M}_\odot$). Because of the single
maximum and abrupt decay these lightcurves are qualitatively similar
to the shapes of the observed X-ray afterglows of, e.g., GRB 080905A, GRB 090515,
GRB 090607, GRB 100117A.

\paragraph{Cat.~MM} These lightcurves show a global maximum due to
thermal radiation from the ejecta shell, followed
by a steep decay and a separated second `bump' of non-thermal radiation
from the nebula (upper right panel of
Figure~\ref{fig:morphologies_noTR}). The non-thermal radiation
typically decreases as a power law $\propto t^{-a}$, where
$a\gtrsim 2$. This could explain some more moderate declines of
X-ray luminosity at late times in some observed cases. Characteristic
of these lightcurves is also the absence of high X-ray luminosity at
intermediate times. Potential
candidates for such a morphology could be, e.g., GRB
051227, GRB 061201, GRB 060313, GRB 070724A, GRB 071227, GRB 110112A, GRB
121226A, GRB 131004A. Furthermore, regarding timescales and
luminosity levels, the second maximum of
non-thermal radiation is consistent with the late-time X-ray rebrightening
observed in some SGRB events, such as GRB 050724, GRB 080503,
and the r-process powered kilonova candidate GRB 130603B
\citep{Grupe2006,Perley2009,Fong2014}. Based on their models,
\citealt{Metzger2014b} and
\citealt{Gao2015} argued that such a
rebrightening for GRB 080503 and GRB 130603B could be caused by
radiation from a stable magnetar
surrounded by a PWN (similar to our setup here)
and thus called those GRBs candidate events for a
``magnetar driven transient'' or ``magnetar-driven merger-novae''. We
note that, in particular, \citet{Fong2014} reported a late-time
($t\gtrsim\text{day}$) X-ray
excess of $L\simeq 4\times 10^{43}
(t/\text{day})^\alpha\,\text{erg}\,\text{s}^{-1}$ for GRB 130603B,
where $\alpha =
-1.88\pm 0.15$. As noted above,
some of our lightcurves also show a power-law behavior with an
exponent close to $-2$. Furthermore, the timescales and luminosity
levels of some of our models are consistent with such an X-ray excess
at late times.

\paragraph{Cat.~P} These lightcurves are characterized by a global
maximum (as in the previous cases), followed by a plateau that is
caused by the collapse of the NS (cf.~Section~\ref{sec:col_p3}) and a
final steep decay (lower left panel of
Figure~\ref{fig:morphologies_noTR}). The radiation originates from the
ejecta shell throughout the evolution and is thus thermal. The
plateaus typically extend up to $\sim\!10^4\,\text{s}$ and the
luminosity levels are typically roughly two orders of magnitude lower
than the preceding peak at $\sim\!10^2-10^3\,\text{s}$. This
morphology together with the timescales
and luminosity levels seems to resemble some of the SGRBs considered
by \citet{Gompertz2013,Gompertz2014}. More specifically, potential
candidates for such a morphology could be, e.g., GRB 051227, GRB
070714B, GRB 070724A, GRB 071227, GRB 080123, GRB 111121A.

\paragraph{Cat.~PF} These X-ray afterglow lightcurves are very
similar to Cat.~P, except for the fact that there is a noticeable flare-like
event of non-thermal radiation from the nebula after the plateau
(bottom row of Figure~\ref{fig:morphologies_noTR}, middle panel). This
is due to the fact that the ejecta shell becomes optically thin during
the plateau phase of the total luminosity
(cf.~Section~\ref{sec:col_p3}). A potential candidate for this
morphology could be, e.g., GRB 050724, which shows a flare-like event
just after the plateau phase. However, as such flares
are not always well separated from the plateau itself, this morphology
can also be confused with Cat.~P in some cases. Therefore, some of
the GRB candidates for Cat.~P could also apply to this category and
vice versa.

\paragraph{Cat.~GF} These are more peculiar lightcurves shown in the
lower right panel of Figure~\ref{fig:morphologies_noTR}. They are
similar to Cat.~PF, but characterized by an abrupt decrease after the
global maximum and by absence of strong X-ray radiation (a `gap') at
intermediate times. The final flare dominates the luminosity
at intermediate times.

\subsection{Time-reversal scenario}
\label{sec:morphologies_TR}

\begin{figure*}[tb]
\centering  
\includegraphics[width=0.33
  \textwidth]{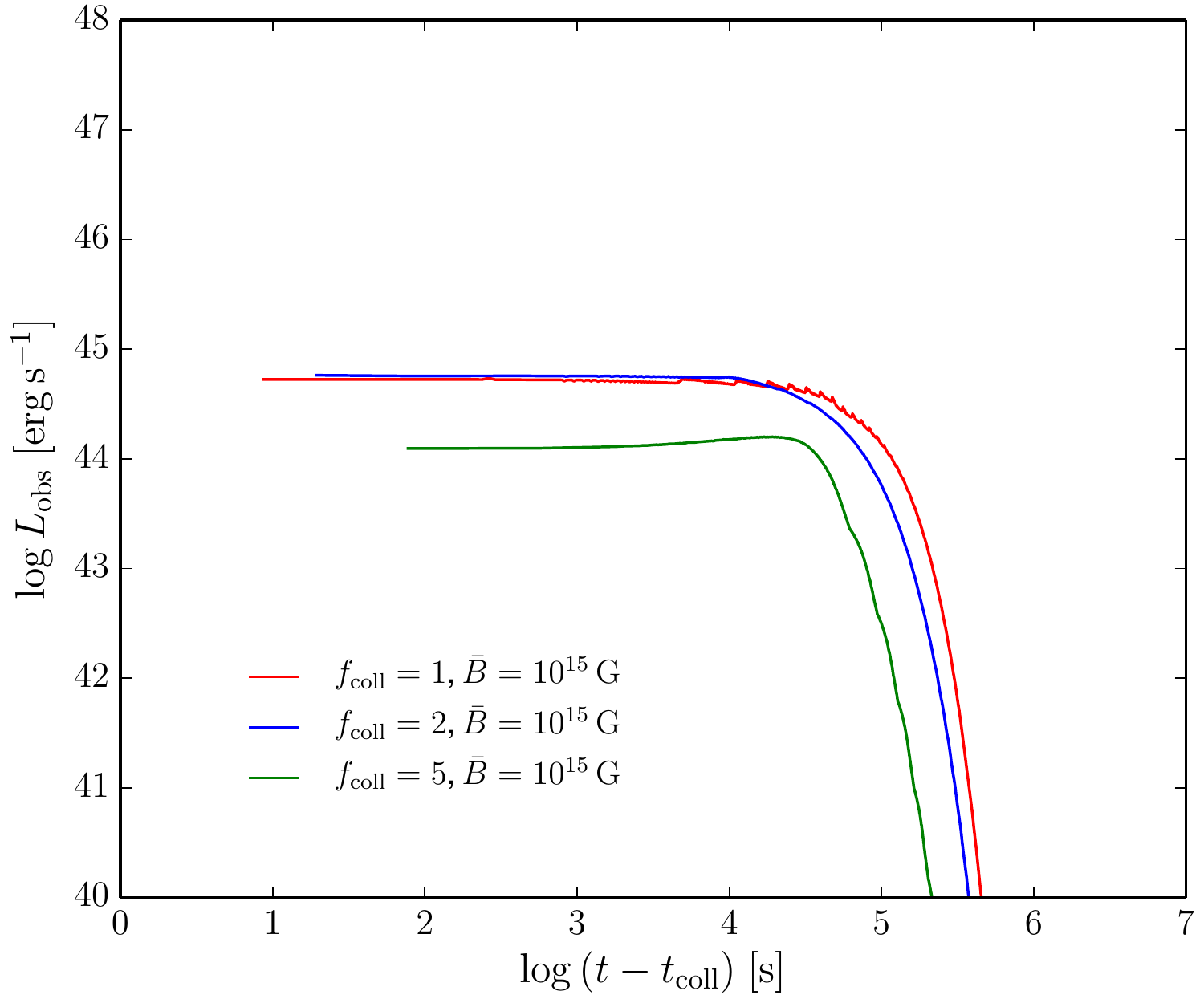}
\includegraphics[width=0.33
  \textwidth]{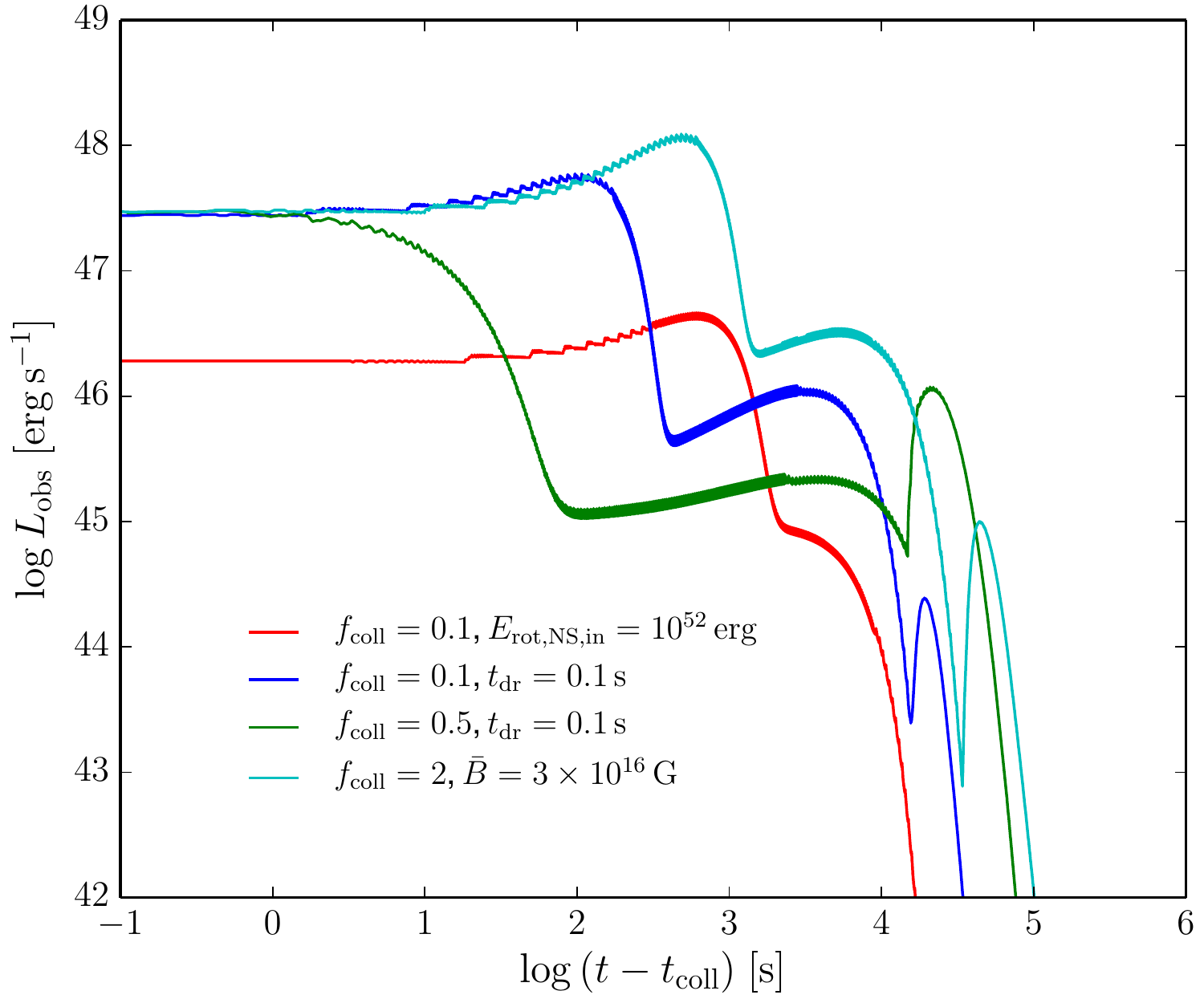}
\includegraphics[width=0.33
  \textwidth]{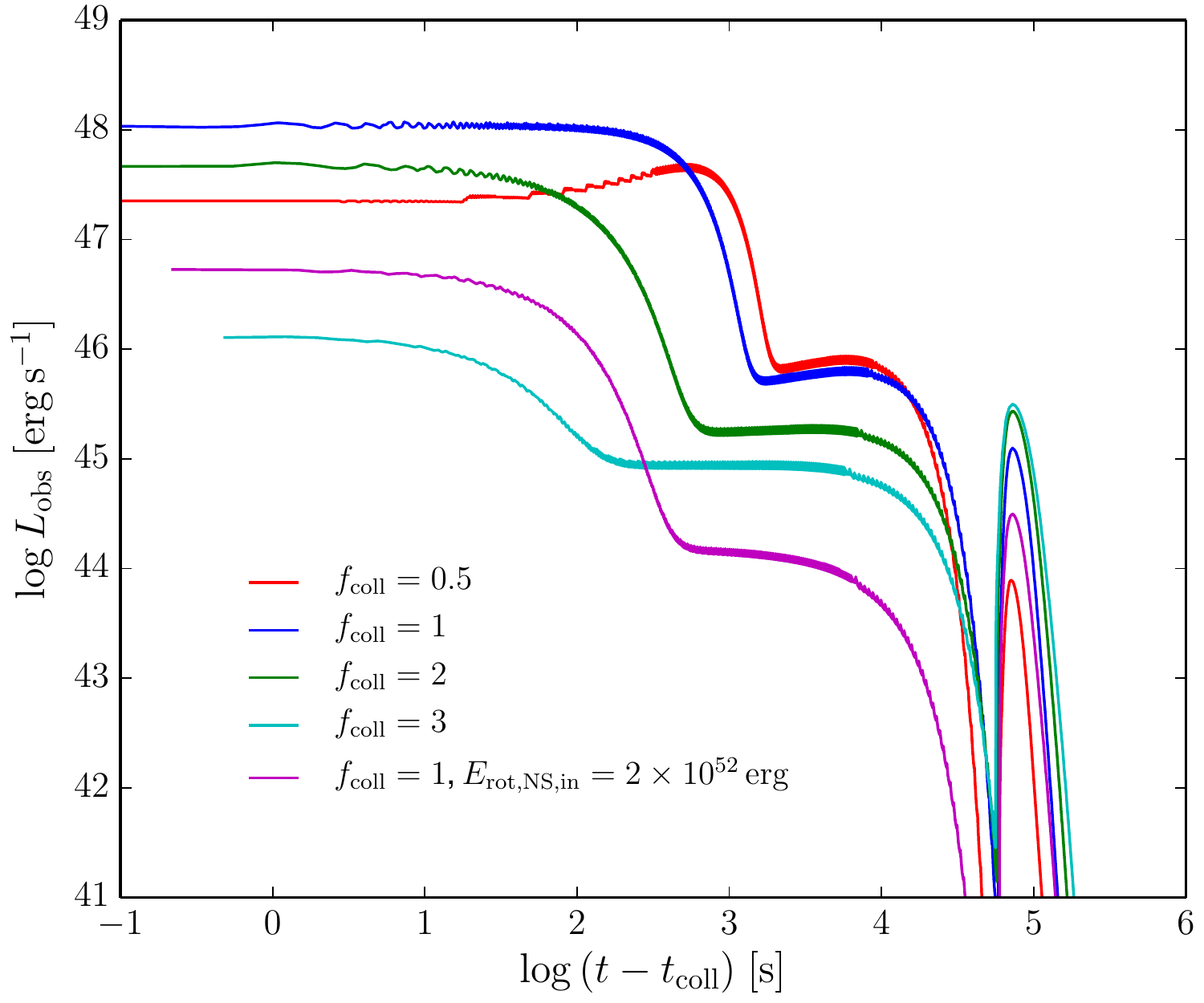}\\
\includegraphics[width=0.33
  \textwidth]{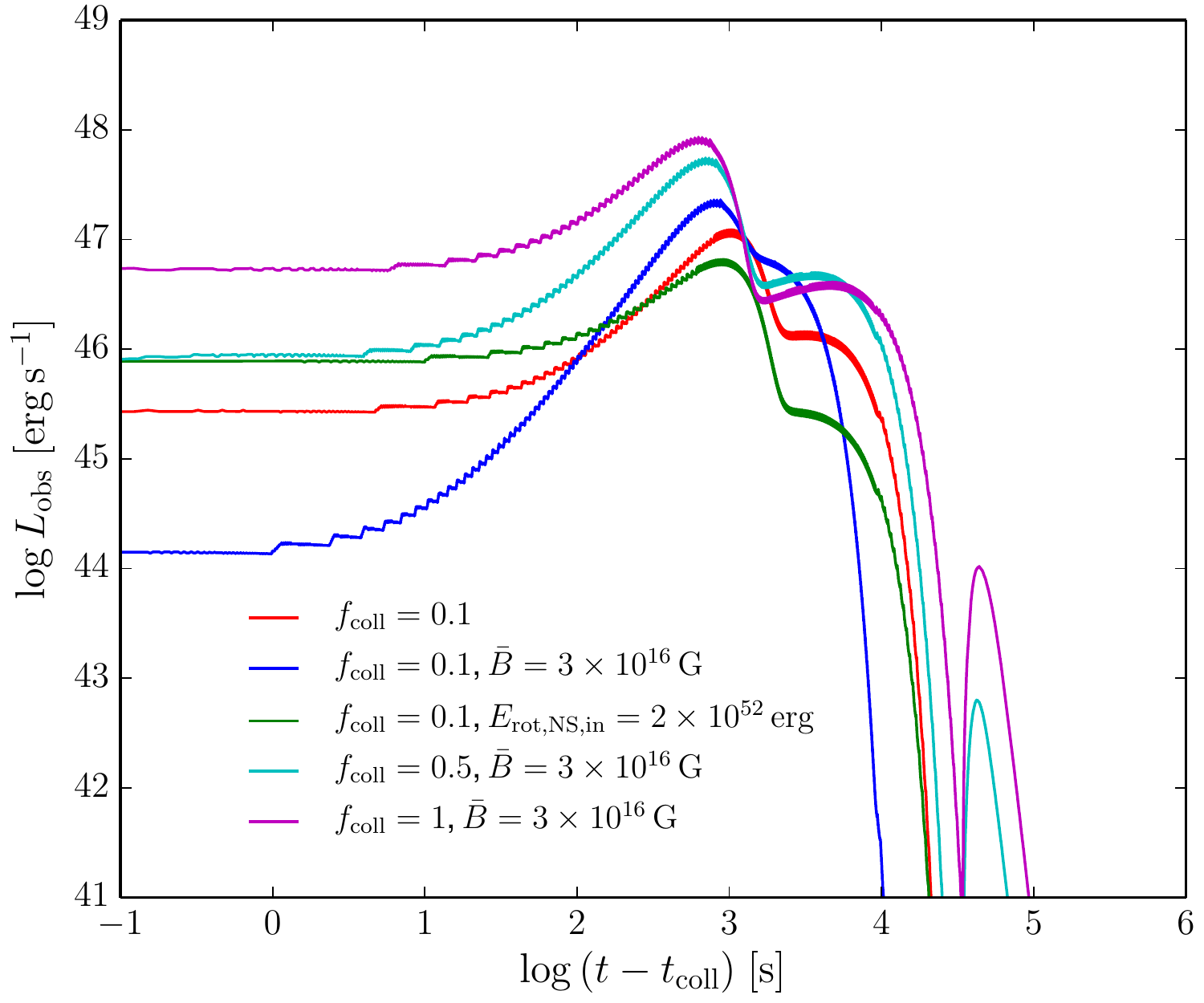}
\includegraphics[width=0.33
  \textwidth]{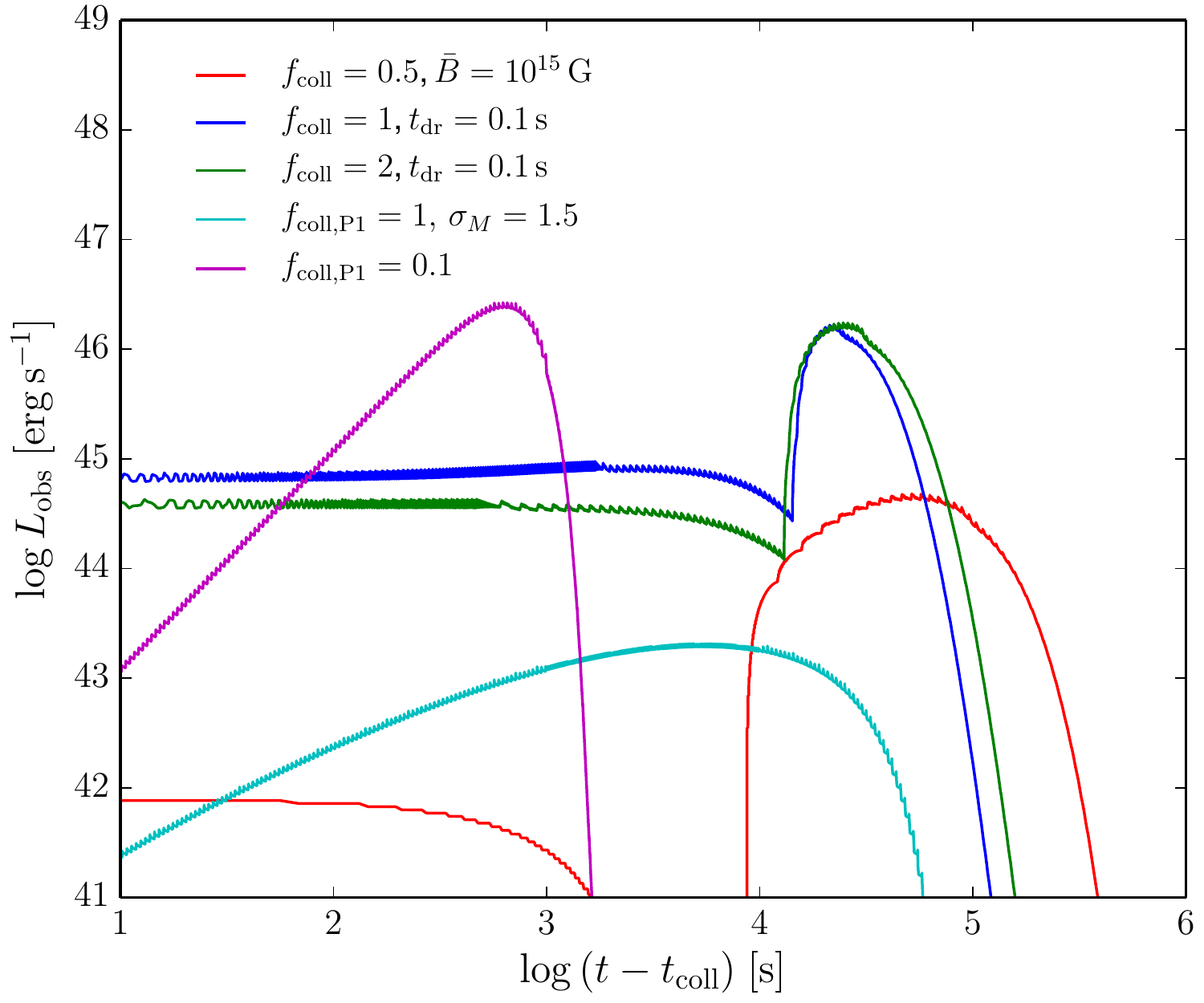}
  \caption{X-ray lightcurve morphologies (time-reversal scenario). Upper row (left to right):
    single-plateau lightcurves (SP), two-plateau lightcurves (TP),
    two-plateau lightcurves with a late-time flare (TPF). Bottom row
    (left to right): lightcurves with a rising plateau and abrupt
    decay (RP), lightcurves without (dominant/clear) plateaus
    (NP). All parameters are as in the fiducial setup
(cf.~Table~\ref{tab:model_parameters}), except for the ones specified
in the respective legend.}
  \label{fig:morphologies_TR}
\end{figure*}

Figure~\ref{fig:morphologies_TR} depicts examples for five different X-ray
afterglow lightcurve morphologies in the time-reversal scenario. In
contrast to the standard scenario (cf.~Section~\ref{sec:morphologies_noTR}) and
except for a few peculiar cases, these
lightcurves typically do not show a delayed onset. They are rather marked by
very high X-ray luminosities starting from the SGRB itself. This is
because at the times of collapse, the system is typically already
radiating at a very high luminosity level. We note that in the
time-reversal scenario, the X-ray luminosity predicted by our model
prior to the collapse as
discussed in Section~\ref{sec:morphologies_noTR} and
Sections~\ref{sec:fiducial_model} and \ref{sec:param_study} represents
a precursor signal that can be searched for by future X-ray
missions.

\paragraph{Cat.~SP} These are X-ray lightcurves characterized by a single plateau
lasting up to $10^4-10^5\,\text{s}$, followed by a rather steep
decay (upper left panel of Figure~\ref{fig:morphologies_TR}). The
radiation is purely non-thermal in the cases shown (broad-band synchrotron
plus inverse Compton, cf.~Section~\ref{sec:spectra}), as the collapse
of the NS occurs at late times when the ejecta shell is already
optically thin. This type of morphology seems to resemble some of the
X-ray plateaus discussed by \citet{Rowlinson2013}. Potential candidates
for this morphology could be, e.g., GRB 051221A, GRB
060313, GRB 061201, GRB 070809, GRB 090510,
GRB 111020A, 130603B, GRB 140903A. Some of these cases require a
contribution from the
standard X-ray afterglow of the SGRB jet itself in combination with
our prediction (see \citealt{Rowlinson2013} and Section~\ref{sec:discussion}).

\paragraph{Cat.~TP} Together with Cat. TPF this family of lightcurves
represents the most typical prototype for the time-reversal scenario,
as it shows the characteristic two-plateau morphology (see
Section~\ref{sec:col_p3}; middle panel of the upper row in
Figure~\ref{fig:morphologies_TR}). The plateau durations are
determined by the photon
diffusion times of the ejecta shell and the PWN (see
Section~\ref{sec:col_p3}). The luminosity levels and plateau durations
of typically
$\sim\!1-10^3\,\text{s}$ and $\sim\!10^3-10^4\,\text{s}$ for the first
and the second plateau, respectively, are in very good agreement with some
of the observed two-plateau structures discussed by
\citet{Gompertz2014}. Potential candidates for this morphology are GRB
051227, GRB
060614, GRB 070714B, GRB 070724A, GRB 071227, GRB 080123, GRB 111121A.

\paragraph{Cat.~TPF} Same as Cat. TP, but with a late-time `flare' of
non-thermal radiation from the nebula as the ejecta shell becomes
optically thin (upper right panel of
Figure~\ref{fig:morphologies_TR}). A potential candidate for this
morphology is GRB
050724, which shows such a flare at late times attached to the second
plateau (cf.~\citealt{Gompertz2014}). However, as these flares
are typically not well separated from the second plateau, many of such
events could be confused with Cat.~TP and some of the GRB candidates
mentioned there could also apply to this category and vice versa.

\paragraph{Cat.~RP} This family of lightcurves essentially shows a
single plateau, but with a rising luminosity level (lower left panel
of Figure~\ref{fig:morphologies_TR}). After the maximum
there is a very short second `plateau' up to typically
$\sim\!10^3-10^4\,\text{s}$, followed by a steep decay. The
radiation is thermal throughout. Candidate SGRBs for this category
could be, e.g., GRB 070714A, GRB 090607, GRB 131004A.

\paragraph{Cat.~NP} These lightcurves are again more peculiar examples
without a (dominant/characteristic) plateau (lower right panel of
Figure~\ref{fig:morphologies_TR}). This family is rather
marked by a single maximum or `bump' of thermal or purely non-thermal
radiation. A maximum at early times $\lesssim\!10^3\,\text{s}$ (as in
the case of a collapse during Phase I; cf.~the magenta curve and
Section~\ref{sec:col_p1}) is
typically characterized by thermal radiation and could
resemble some of the candidate GRBs noted in Cat.~M of
Section~\ref{sec:morphologies_noTR}. The other lightcurves with maxima
at later times ($\sim\! 10^4-10^5\,\text{s}$) are either thermal (cyan
curve; collapse during Phase I) or non-thermal (remaining curves;
collapse during Phase III). The absence of
strong X-ray radiation at early times in these cases is
either due to a high optical depth (collapse during Phase I) or a low
temperature of the ejecta
material (collapse during Phase III). These late-time maxima could be
consistent with the afterglow
rebrightening observed in some SGRBs, such as GRB 050724, GRB 080503,
and the r-process powered kilonova candidate GRB 130603B
(\citealt{Grupe2006,Perley2009,Fong2014}; see also the discussion of
Cat.~MM in Section~\ref{sec:morphologies_noTR}). The timescales for
rebrightening in the aforementioned cases of
$\sim\!10^4-10^5\,\text{s}$ are in good agreement
with the timescales for maximum brightness found here. This
rebrightening is
to be understood as an additional component to the fading standard X-ray
afterglow of the SGRB jet itself, as the early X-ray emission in the aforementioned
cases cannot be explained by the lightcurves such as those in the
lower right panel of Figure~\ref{fig:morphologies_TR}. Such an
interpretation is similar to \citet{Metzger2014b} and \citet{Gao2015},
who, according to their
models, noted that the observed X-ray excess at late times in some of
the aforementioned cases can be consistent with a magnetar driven
transient (or ``merger-nova'') in addition to the standard X-ray
afterglow.

\section{Discussion and Conclusion}
\label{sec:discussion}

In this paper, we have presented X-ray lightcurves and spectra
originating from the post-merger evolution of a BNS system, based on our
model proposed in the companion paper (Paper I). Given initial data
that can be read off from numerical relativity simulations of the
merger and early post-merger process, this model provides a
self-consistent evolution of the system over much longer time and length
scales inaccessible to those simulations. It thus bridges the gap
between such simulations and the timescales relevant for observations
of SGRB afterglows with satellite missions like \textit{Swift}.

\paragraph{Baryon pollution} Our model is based on the notion that a
long-lived NS is formed in
a large fraction of BNS merger events, which is either hypermassive,
supramassive, or stable. As we have argued (see Paper I), a generic
feature of such
a newly-formed object is strong baryon pollution in its surrounding
due to mass ejected dynamically during and shortly
($\sim\!\text{ms}$) after the merger (e.g.,
\citealt{Hotokezaka2013a,Oechslin2007,Bauswein2013,Kastaun2015a})
and subsequent neutrino-driven
and magnetically induced winds on longer timescales from the remnant NS
itself and from a potential accretion disk (e.g.,
\citealt{Dessart2009,Siegel2014a,Metzger2014c}). As we have shown,
these ejecta create
an optically thick environment which traps radiation on timescales of
interest, even for total ejecta masses as small as $M_\text{ej}\sim
5\times 10^{-4}\,\text{M}_\odot$. This is in contrast to typical
``magnetar models'' for SGRGs (e.g.,
\citealt{Rowlinson2013, Gompertz2013,Gompertz2014,Lue2015}), which
assume an optically thin environment such that the spin-down energy of
the newly-formed magnetar after the merger could immediately and
directly be
converted into X-ray radiation, $L_X(t)\propto L_\text{sd}(t)$ (by
some yet to be specified process). 

\paragraph{General properties of the X-ray signal} One
important consequence of our approach,
which takes such baryon pollution into account, is a `delayed onset'
of radiation from the post-merger system at $\sim\!1-10\,\text{s}$ after
merger. This is an intrinsic feature of baryon pollution that will be present
even if further potential sources for EM radiation not considered
here are included.\footnote{Additional sources of energy could be the
  gravitational, thermal, and EM energy of a short-lived
  accretion disk that might form around the remnant NS.} The radiation
typically peaks at $\sim\!10^2-10^4\,\text{s}$
after merger in the X-ray band with luminosities of
$\sim\!10^{46}-10^{49}\,\text{erg}\,\text{s}^{-1}$. Scanning the whole
parameter space (cf.~Table~\ref{tab:model_parameters}) we find that
these timescales and luminosities are a
robust prediction of the model, even when considering uncertainties in
the opacity of the ejecta material (cf.~Sections~\ref{sec:params}
and \ref{sec:non_coll_models}). These ranges are unlikely to change
even if a more detailed computation of opacities including bound-free
absorption due to only partial ionization of the material was
attempted. However, further acceleration of the ejected material in
Phase III (not considered here) may shift the times of maximum
brightening to somewhat earlier times (see below).

\paragraph{Electromagnetic counterpart to gravitational waves} We note
that according to our model, such a transient X-ray signal is
isotropic and it is expected for
all BNS mergers that lead to the formation of a long-lived ($\gtrsim$~tens\,ms) remnant NS, regardless of whether it is hypermassive,
supramassive, or stable. As we have pointed out (cf.~Section~1 of
Paper I), the fraction of such BNS
mergers should be very high, such that we can write the rate of these
X-ray transients $r_\text{X}$ in terms of the rate of BNS mergers
$r_\text{BNS}$ as
\begin{equation}
  r_\text{X} = f_\text{NS} r_\text{BNS},
\end{equation}
where $f_\text{NS}$ can be at least of the order of a few tens of
percent. This transient is also orders of magnitude
more luminous than other quasi-isotropic EM transients
expected to result from a BNS merger, such as r-process powered
kilonovae (or ``macronovae''), which are thought to emit mostly
in the IR to UV with typical luminosities of $L\sim
10^{41}-10^{42}\,\text{erg}\,\text{s}^{-1}$ (e.g.,
\citealt{Li1998,Kulkarni2005,Metzger2010,Piran2013}). We note again
that in contrast to such kilonovae, the X-ray transient found here
would be a clear sign of a long-lived NS and would thus distinguish
between a BNS and a NS--BH merger. In contrast to
these isotropic signals, the prompt SGRB radiation itself
(if associated with a BNS merger) is thought to be collimated and will
thus be beamed away from the observer in a large fraction
$1-f_\text{beam} = \cos\theta_\text{jet}$ of
events, where $\theta_\text{jet}$ denotes the beaming angle of the
jet (see \citealt{Berger2014} for an overview of observations to
date). Furthermore, the rate of potentially observable SGRBs,
\begin{equation}
  r_\text{SGRB}=f_\text{beam}f_\text{jet}r_\text{BNS},
\end{equation}
is further reduced by a factor $f_\text{jet}$ with respect to the BNS
merger rate $r_\text{BNS}$, since not all BNS mergers may generate a
jet to produce a SGRB. For instance, baryon pollution as described
above can choke jets (e.g.,
\citealt{Nagakura2014,Murguia-Berthier2014}) and it seriously
threatens the formation of a highly relativistic outflow at or shortly
after the time of merger. The combined product of the efficiency factors
is constrained to be 
\begin{equation}
  f_\text{beam}f_\text{jet}\lesssim 0.3\%
\end{equation}
if the likely local BNS merger rate of
$\sim\!10^{-6}\,\text{Mpc}^{-3}\text{yr}^{-1}$
(cf.~\citealt{Abadie2010} and references therein)\footnote{As BNS
  merger rates for the local universe are uncertain, we take here the
  ``realistic estimate'' of \citet{Abadie2010}.} is to be
reconciled with the local SGRB rate of
$\sim\!3\times 10^{-9}\,\text{Mpc}^{-3}\text{yr}^{-1}$
(\citealt{Wanderman2015} and references therein; cf.~also \citealt{Metzger2012})\footnote{Recent
  estimates for the current event rate of SGRBs essentially range within
  $(1-10)\times 10^{-9}\text{Mpc}^{-3}\text{yr}^{-1}$ (cf.~Table~4 of
  \citealt{Wanderman2015} and references given there; we have chosen a
  representative value of $3\times
  10^{-9}\,\text{Mpc}^{-3}\text{yr}^{-1}$).}. In particular, we note
that to date no SGRB event with known redshift has been observed
within the local sensitivity volume of the advanced LIGO/Virgo GW
detector network. Hence, for coincident EM and GW observations of BNS
mergers it is desirable to identify bright isotropic EM
counterparts that are formed in a high fraction of events. Furthermore,
the timescales for generating early-warning triggers for EM follow-up
observations of the GW signal is rather
challenging in the case of the SGRB itself \citep{Cannon2012}. During
the first
generation of ground-based laser interferometers such alerts were sent
out with typical latencies of 10--30 minutes
\citep{Evans2012,Cannon2012}. This is far too long compared to the
time frame of
at most 1--2 seconds to catch the prompt SGRB emission (assuming
that the SGRB is produced at the time of merger, see
below). While parts of this latency can, in principle, be
significantly reduced for the
advanced LIGO/Virgo setup \citep{Cannon2012}, sending out alerts to
search for the SGRB prompt emission still remains a
challenge. However, given a
timescale for maximum brightness of $\sim\!10^2-10^4\,\text{s}$, even the
latency of the first generation warning system should provide a
realistic chance to search for the X-ray transient predicted by our
model. Hence, the
first advanced LIGO/Virgo observing runs starting this year should be
able to trigger EM follow-up
observations to search for such X-ray transients. If
found, such an observation would represent a clear indication in
favor of our model. At the same time it would provide
strong evidence for the association of the GW
observation with a BNS merger (and not, e.g., a NS--BH binary merger)
and thus confirm the astrophysical
origin of the GW signal. In conclusion, all criteria mentioned in
Section~\ref{sec:introduction} are met: the high
luminosity, its isotropy, the long duration, the high fraction of
events, its ability to identify a BNS merger, and, additionally, a
realistic timescale for triggering EM follow-up
make the X-ray transient predicted by our model an ideal
EM counterpart to the GW signal of the
final inspiral and merger of a BNS system.

\paragraph{X-ray lightcurve morphologies and SGRBs} X-ray lightcurves arising
from the present model have been
discussed and classified according to their morphology
(cf.~Figures~\ref{fig:morphologies_noTR} and
\ref{fig:morphologies_TR}), both in the
standard and the
time-reversal scenario for SGRBs (prompt burst associated with the
time of merger or with the time of collapse of
the merger remnant). As a reference energy range we have employed the
\textit{Swift} XRT band. There is a wide spectrum of different
morphologies for both scenarios, although the variety is somewhat richer
in the time-reversal case. This can be a useful property if a single model
is to explain the observed wide variety of X-ray afterglow
lightcurves. In particular, our model is able to explain
single and two-plateau shaped lightcurves (cf.~Cat.~SP, TP, P) as,
e.g., reported by
\citet{Rowlinson2013}, \citet{Gompertz2013,Gompertz2014}, and
\citet{Lue2015}, with additional
late-time flares in some cases (cf.~Cat.~PF and TPF). The timescales
for such plateaus are given by the photon diffusion time of the ejecta
envelope and the PWN. We find that the
timescales ($\sim\!1-10^3\,\text{s}$ for a first and
$\sim\!10^3-10^4\,\text{s}$ for a second/single plateau)
and luminosity levels ($\sim\!10^{44}-10^{49}\,\text{erg}\,\text{s}^{-1}$) for such plateaus are in
excellent agreement with observations. Moreover, there are categories
showing a late-time rebrightening of the X-ray emission (cf.~Cat.~MM
and NP) similar to some
observed cases (GRB 050724, GRB 080503, GRB 130603B;
\citealt{Grupe2006,Perley2009,Fong2014,Metzger2014b,Gao2015}). Again,
the timescales and possible luminosity levels are broadly consistent
with these observations at late times. Many categories show
lightcurves with rather steep decays, which are due either to the ejecta material
cooling down and the peak energy of the spectrum moving out
of the XRT band or the PWN cooling down and consuming its residual
energy. Steep decays in X-ray lightcurves were previously interpreted as a
collapse of the remnant NS to a black hole (e.g.,
\citealt{Rowlinson2010,Rowlinson2013}). While our model allows for such
steep decays of the X-ray lightcurve, there might be problems in
explaining power-law decays with moderate slopes at late times. There
are a few observational cases that potentially challenge our model in
this regard; some of the events classified as ``external plateaus'' by
\citet{Lue2015}, which are consistent with a late-time temporal index
of $-1$ after a plateau phase, might belong to this category. However,
as the steepness of these decays is somewhat
dependent on the radiation transport and the cooling model (see, e.g.,
Section~4.4 of Paper I), a more detailed implementation of the thermal
and non-thermal emission in our model might resolve this issue (see
below). Furthermore, we
note that we have pointed out a few characteristic
observational examples for the respective X-ray lightcurve
morphologies in Section~\ref{sec:morphologies}. As a
general note of caution, we stress that
the present comparison
with observational data is only based on timescales and overall
luminosity levels and thus remains rather on a qualitative level. Routinely
fitting observational data by employing our model is beyond the scope of
the present paper, but will be attempted in future work.

\paragraph{Additional components to the X-ray lightcurve} The (X-ray)
afterglow lightcurves predicted by our model represent the intrinsic
contribution from the
post-merger system composed of a NS (or a black hole after
collapse), its PWN, and the confining ejecta envelope. They do not
include the X-ray signature of the prompt emission and a fading
tail. An early steep-decay component superimposed to our X-ray
lightcurves might be needed to explain the observed X-ray lightcurves
of some SGRB events (cf.~also \citealt{Rowlinson2013}). This initial
steep decay can represent a gradual decline of the X-ray
emission from the prompt burst itself. It can also be due to the
high-latitude effect if the prompt emission is `switched off' abruptly
\citep{Kumar2000,Kumar2015}. The latter phenomenon is essentially a
time-of-flight effect due to photons from different portions of the
jet arriving slightly retarded, depending on the angle with respect to
the observer direction. It fades out the prompt emission
as $\propto t^{-\alpha}$, where $\alpha = \beta + 2$, with $\beta$ the
spectral index of the prompt emission.

\paragraph{SGRBs with EE} \citet{Lue2015} have recently argued that
SGRBs with extended emission (EE) in the BAT band correspond to the same
phenomenon as SGRBs with an early plateau phase in the XRT
band, and that the detection of such early plateaus by BAT is just an
instrumental selection effect depending on brightness. EE in the
context of the magnetar
scenario has been interpreted as dissipation of kinetic energy of an early
modestly magnetized, neutrino-heated magnetar wind in internal shocks
that give rise to
synchrotron radiation \citep{Metzger2008a}, as the breakout of a magnetic jet
through the ejecta envelope of a PWN \citep{Bucciantini2012}, as mass
ejection via magnetic propellering of an accretion disk around the remnant NS
\citep{Gompertz2014}, or as magnetar wind leaking out from the
surrounding ejecta through a hole drilled out by the SGRB jet
\citep{Gao2015}. We note
that in the model presented here, such early X-ray plateaus and the
typical two-plateau structure evident in SGRBs with EE
\citep{Gompertz2014} are naturally explained in terms of photon
diffusion through the ejecta matter and the PWN (see Cat.~P, PF, TP,
TPF in Section~\ref{sec:morphologies} and
Section~\ref{sec:col_p3}). This does not exclude a possible further
contribution from magnetar wind leaking out from the
ejecta through the jet hole as suggested by
\citet{Gao2015}. However, such a component would only be visible along
the jet direction, while the radiation forming the early plateaus in
our model is, in general, isotropic.

\paragraph{Time-reversal scenario} The model presented here can
accommodate the recently proposed time-reversal scenario for SGRBs
\citep{Ciolfi2015a,Ciolfi2015b}, in which the SGRB is associated with
the collapse of the remnant NS. Detailed modeling in the present
paper quantitatively confirms the
estimates for the photon diffusion timescale of the nebula and
the ejecta shell and thus the duration of the X-ray afterglows
obtained earlier \citep{Ciolfi2015a}. Here, we have reported
for the first time detailed computations for the X-ray afterglow
lightcurves in this scenario, which show a wide range of
morphologies. In particular, this scenario gives rise to two-plateau
structures with plateau durations and luminosity levels compatible
with observations (see above). Additionally, the present study
predicts lightcurves for the X-ray radiation preceding the
prompt SGRB emission within this scenario. Such emission corresponds
to the part of
the X-ray lightcurve prior to the collapse of the NS, which is
characterized by a delayed onset after the BNS merger, reaching a maximum
brightness of $\sim\!10^{46}-10^{49}\,\text{erg}\,\text{s}^{-1}$ at a
timescale of minutes to hours after merger (see above). There
can be additional features not included in the lightcurve modeling
here, such as a short duration, strong non-thermal signal from the
shock front as it reaches the ejecta surface at the end of Phase II,
and possible flare-like irregularities as the SGRB jet drills through
the ejecta envelope. Such features could explain observed
precursors to the prompt SGRB emission as
early as $\sim\!100\,\text{s}$ before or as close as a fraction of a
second to the onset of the main burst \citep{Troja2010}. The
lightcurve calculations
presented here provide clear predictions for strong X-ray
emission prior to the SGRB that can be searched for by wide-field
X-ray telescopes. If found, it
would represent smoking-gun
evidence for the time-reversal scenario. As detailed above, the
timescales of the delayed onset makes the search for such isotropic X-ray
radiation prior to the SGRB a feasible and exciting task for
multi-messenger astronomy with joint GW and EM
observations. An observational strategy including GW
observations to trigger X-ray and $\gamma$-ray observations has thus
the potential to confirm the
association of SGRBs with BNS mergers, the time-reversal scenario, and
the astrophysical origin of the GW signal at the same
time. As the advanced
LIGO/Virgo detector network is starting
the first science runs later this year, such EM follow-up
campaigns become a very timely undertaking.

\paragraph{X-ray spectra} Our model predicts a characteristic
transition from thermal to non-thermal emission during the post-merger
evolution of the system (see Section~\ref{sec:spectra}). Only at late
times when the ejecta shell has expanded far
enough to become optically thin, genuine non-thermal radiation (mainly
broadband synchrotron radiation) from the PWN is emitted toward the
observer. Observationally, not much is known about the spectra
of SGRB afterglows and their temporal evolution. It is generally
assumed that they are of non-thermal nature, originating from
synchrotron radiation of a power-law distribution of electrons. In the
XRT band, spectra are typically fit by a power law
with a time-dependent power-law index (cf., e.g.,
\citealt{Rowlinson2013}). In our model, we assign a single
representative temperature to the ejecta shell, which reflects a high
degree of idealization. However, instead of a single temperature and
thus a pure Planck spectrum for the thermal radiation from the ejecta
shell, the ejecta material will be characterized by
spatial gradients in temperature. The actual emergent spectrum will
thus be a superposition of individual Planck spectra of different
effective temperatures that depending on the temperature gradient may
combine to give a power-law spectrum in the XRT band,
just as the individual synchrotron spectra of electrons following a power-law
distribution in the Lorentz factor superpose to give a power-law
spectrum. Moreover, as the ejecta material will be (at least
partially) ionized, inverse Compton scattering of thermal photons off
free electrons might create power-law-like high-energy tails for the
individual Planck spectra. Nevertheless, as long as the effective
temperature of the ejecta matter in our model resides within the XRT band,
the overall X-ray spectra should still capture the main dynamical
and energetical effects to reliably determine the X-ray lightcurve
morphology. Due to an enhanced high-energy
tail, a more detailed computation of
the emergent spectrum could lead to less steep decays of the X-ray
lightcurve as the ejecta is cooling down and the average temperature
is moving out of the XRT band. This could resolve the aforementioned
problem of describing more shallow decays of the X-ray radiation, in
particular at
the end of a (second) plateau phase.

\paragraph{Future improvements} Our modeling of the X-ray
lightcurves already reflects a high level of detail. Nevertheless, it
is built on simplifying assumptions. More
realistic X-ray spectra, their evolution and associated lightcurves
require a more detailed
spectral modeling of the thermal radiation from the ejecta shell as
discussed above. Moreover, a
time-dependent modeling of the radiative processes in the PWN is
required to include effects of a non-zero albedo of the ejecta matter
\citep{Metzger2014b} and to consider further acceleration of the
ejecta shell during Phase III. The latter aspect
may somewhat influence the timescales of the X-ray
lightcurves. Furthermore, such a time-dependent modeling of the
radiative processes in the PWN would also provide a more accurate
evolution of the PWN after the collapse and thus
provide a more realistic spectral evolution of photons inside the
PWN in the post-collapse phase. Finally, adding thermal Comptonization
to the radiative interactions in the PWN is desirable, as it might
affect the shape of the nebula X-ray spectrum. The main conclusions of
this work are, however, unlikely to be affected by such future improvements.

\acknowledgements
We thank B.~F.~Schutz, W.~Kastaun, B.~D.~Metzger, G.~Ghirlanda, and M.~G.~Bernadini for
valuable discussions. R.~C.~acknowledges support from MIUR FIR Grant No.~RBFR13QJYF. 

\newpage 


\bibliographystyle{apj}
\bibliography{aeireferences}

\end{document}